\documentclass[twocolumn,showpacs,preprintnumbers,amsmath,amssymb,aps,nofootinbib]{revtex4}

\usepackage{graphicx}
\usepackage{dcolumn}
\usepackage{bm}

\begin{document}

\preprint{APS/123-QED}

\title{Could Dark Energy be a Manifestation of Gravity?}
 
\author{Reva Kay Williams}
\affiliation{Department of Physics and Astronomy, The University of Toledo,
MS 111, Toledo, OH 43606-3390, USA}
\email{reva.williams@utoledo.edu}

\date{\today}

\begin{abstract}
It is shown that so-called dark energy 
could possible be a manifestation of the gravitational 
vortex 
producing the ``gravitomagnetic'' (GM) force 
field: associated with cosmic matter rotation and 
inertial spacetime frame dragging.  
The general relativistic G\"{o}del-Obukhov spacetime metric which 
incorporates expansion and rotation of the Universe is used to
evaluate this force.   This metric  
is expressed  
here in spherical  comoving coordinates. 
Through a cosmic time evolution, it is shown 
that cosmic acceleration is expected when the magnitude 
of the radial repulsive 
GM force exceeds that of the familiar or usual attractive 
gravitational ``gravitoelectric'' (GE) force:
associated with just cosmic matter and spacetime warping (or 
curvature).  
In general, this phenomenon of cosmic accelerated expansion 
appears to have occurred 
twice in 
the history of the Universe: the inflationary phase and the present-day
 acceleration phase.  It is suggested in this model that the two 
phases may or may not be related.  The cosmological model presented 
here is described in the context of Einstein's Theory of General 
Relativity in Riemann-Cartan spacetime (the ``generalized'' 
Einstein-Cartan theory of gravity), which includes cosmic 
rotation, its effect of spacetime torsion, and it being considered as
an intrinsic part of gravity.  
Also, an associated derived analytical
expression for the cosmic primordial magnetic field
is presented.  Evolving this magnetic field over cosmic time
shows it to be consistent with theory and observations.
In addition, it appears that the spin density of cosmic matter couples
this magnetic field to the GM field, and also couples this magnetic
field to the GE field.
\end{abstract}

\maketitle
\section{Introduction}
\label{sec:1}

Einstein's Theory of General Relativity together with ordinary matter, 
described by the standard model of particle physics, cannot fully 
explain the observational data from Type Ia supernovae 
\cite{Perlmutter1998, Perlmutter1999, Riess1998,
Riess2004}, the matter power spectrum
of large scale structure \cite{Tegmark2004},
and the anisotropy spectrum of the cosmic microwave background 
radiation \cite{Spergel2007}, with all these data suggesting the
presence of ``dark energy.'' 
From general relativity, assuming homogeneity  and isotropy,
the standard cosmological model is commonly described by the 
Friedmann, Lema\^{i}tre, Robertson, Walker
(FLRW) 1920s and 1930s solutions 
to the Einstein field equations for an expanding universe 
[Eq.~(\ref{GE6})].  According to this
standard cosmological model, the expansion of the 
Universe, if it contains only 
non-negative mass-energy density $\rho$ and pressure $p$, 
decelerates, as expected on the grounds that gravity is attractive
and the cosmological constant $\Lambda$ is zero.
The recent observations of cosmic acceleration, first discovered 
and confirmed by Perlmutter et al. \cite{Perlmutter1998}
 and Riess et al. \cite{Riess1998}
from type Ia supernovae, can only be 
explained by considering repulsive gravity. In the  standard 
cosmological model, this is achieved by models 
introducing matter with negative pressure and/or $\Lambda\neq 0$
(see \cite{Perivolaropoulos2006}, and references therein).  
Observations suggest that the alleged cosmic acceleration can only
be a very recent phenomenon and must have set in during the late
stages of the mass dominated expansion of the Universe.  Subsequent
observations by Riess et al. \cite{Riess2004} identify the 
transition from a 
decelerating to an accelerating universe to be at $z=0.46\pm 0.13$.   
Now, among suffering from the coincidence problem (e.g.,Why is 
the energy density of matter and radiation nearly equal to the dark 
energy density today?) and the cosmological constant problem (e.g., How 
could  $\Lambda $ have been so large during inflation but so 
incredibly small today?), the  standard model does not 
explain why the acceleration has started in the recent past.  

To avoid resorting to anthropic principle arguments to gain acceptance 
of the above mentioned models, of $p<0$ and/or $\Lambda\ne 0$,
 which are constrained by the so-called standard
cosmological model,  perhaps we should seek a wider understanding 
of a general relativistic cosmology, not constrained by non-rotation, 
a spacetime being an extension of the standard cosmological model.
This ``new'' standard cosmological model then should take
into account cosmic rotation as well as cosmic expansion: two 
degrees of freedom.  When this 
is done, we find that a repulsive force of  gravity is a natural
occurrence and could possibly provide an explanation for 
the recent acceleration phase 
of the present-day
Universe, and possibly shed light on our understanding of
the physics of inflation in 
the early Universe.      

In this paper, the nature of so-called dark energy is investigated.
The aim is to answer the question, Could dark energy be a manifestation
of gravity? i.e., Could it be that component of gravity, the 
so-called
 gravitomagnetic (GM) force field,  associated 
with cosmic rotation
and inertial spacetime frame dragging?
  Inertial is used here 
in the general dynamical sense. 
It seems reasonable to refer to the cosmic expansion frame
as an inertial (or ``flat'') spacetime frame in
a general relativistic dynamical sense because it
appears to have inertial force properties as well as inertial
motion properties.
The G\"{o}del-Obukhov metric \cite{Obukhov2000, Jain2007} 
which incorporates expansion and rotation
of the Universe, derived from general relativity, is used to
define spacetime separation (or distance), in this quest to 
answer the above 
question. Importantly, the G\"{o}del-Obukhov metric or geodesic 
line element can ensure the absence of closed timelike curves,
 making it completely causal, different from the originally proposed 
G\"{o}del metric \cite{Godel1949}.  The 
G\"{o}del-Obukhov cosmological model \cite{Obukhov2000} 
contains parameters which smoothly interpolate between 
this cosmology and the 
standard FLRW cosmology 
(which describes an isotropic and homogeneous universe filled
with matter: commonly represented by an ideal fluid). 
Note, the independent nature of vorticity as
associated with shear of a fluid and pure
rotation does not allow limits on cosmic rotation to be placed by
limits on vorticity \cite{Obukhov2000, Obukhov2002, Jain2007}.
Namely, the G\"{o}del-Obukhov spacetime metric is shear-free but the 
vorticity and expansion  are nontrivial.  It is not vorticity of pure cosmic 
rotation that would lead to anisotropy of the microwave background 
radiation temperature distribution, but effects of vorticity associated 
with a shearing force.

The G\"{o}del-Obukhov  model (sometimes referred to as a 
G\"{o}del-type model
with rotation and expansion) does not conflict with any known
cosmological observations.  
The G\"{o}del-Obukhov  cosmological model is a 
Bianchi  type III, which means that the metric of Eq.~(\ref{metric_cc})
is shear free, spatially
homogeneous, and  isotropic in
the cosmic microwave background (CMB) radiation (like
the standard FLRW cosmology) for any moment of
cosmological time $t$ \cite{Obukhov2000}.  Importantly,
the G\"{o}del-Obukhov model is not the Bianchi type $\rm VII_h$.
The Bianchi type $\rm VII_h$ has shear and is anisotropic in 
the CMB radiation: of which 
WMAP \cite{WMAP2013} and Plank observations 
\cite{PLANCK2015} constrain the vorticity at 
$ {(\omega/ H)_0}<8.6\times 10^{-10}$
and $ {(\omega/ H)_0}<7.6\times 10^{-10}$,
respectively.
Further, and in summary, the cosmological model of
 Eq.~(\ref{metric_cc}), with rotation
and expansion, does not suffer from the three major problems
associated in the past with cosmic rotation. This cosmological
model is causal, isotropic in the CMB radiation, and
parallax free; and thus,
the limits on
the cosmic rotation, obtained earlier from the study of CMB radiation
and of the parallaxes in a rotating world, are not true for
the class of cosmologies in which the G\"{o}del-Obukhov metric 
is a member \cite{Obukhov2000}.

Among distinctive predictions of the G\"{o}del-Obukhov
cosmological model are effects on the propagation of 
light \cite{Jain2007}.
Cosmic rotation affects
a polarization of radiation which propagates in this curved spacetime, 
resulting in some observable anisotropy \cite{Obukhov2000}.
The plane of polarization of electromagnetic waves 
is expected to rotate in the same direction as the cosmic 	
matter; this being caused by the angular momentum 
of the gravitating matter \cite{Balazs1958}, and, thus, 
inertial frame dragging.	
This anisotropy in the polarizations of radio galaxies appears 
to have been confirmed
\cite{Birch1982_1983, Kendall1984,
Nodland1997a, Nodland1997b,  Jain2006}.
Observational tests have been
done that do not require redshift information, by Jain \& Ralston
\cite{Jain1999}.  They found significant signals
of anisotropy in a large sample of data.  Several other observations of
radiation propagating on cosmological scales have been found to
indicate a preferred direction, all of which are aligned along the same
axis  (e.g., \cite{Ralston2004,
 Hutsemekers2005, Perivolaropoulos2011}).
  The origin
of  these effects, however,  may be independent of gravitation and
restricted to
modifications of the electromagnetic sector in which polarization
observations are exquisitely sensitive \cite{Jain2007}.
Perhaps the preliminary evidence for alignment of 
handedness of spiral galaxies indicating a preferred
axis \cite{Longo2011, Perivolaropoulos2011} and  
the model presented in this present manuscript will lend
support to the possibility that such
effects may indeed be gravity related.

Dark energy is the popular motivation to consider models beyond the
standard Friedmann-Robertson-Walker spacetime metric,
such as the the G\"{o}del-Obukhov model.  
Early observational data \cite{Jarosik2011, Tegmark2004} appear to
 fit a flat cosmology with $\Omega_{\rm mat}\sim 0.27$ and 
$\Omega_\Lambda\sim 0.73$ for matter 
and dark energy density
parameters,
respectively, in the popular lambda cold dark matter ($\Lambda$CDM)
cosmological model, which assumes negative pressure in a FLRW 
cosmology.  These fits assume an isotropic universe, while, at face 
value, the data used in the fits substantially contradicts isotropy 
\cite{Jain2007},
at least it appears so from the observational tests mentioned above.
Jain et al. \cite{Jain2007} used large redshift type Ia Supernova data 
(see \cite{Jain2007}, and 
references therein)  and related magnitudes, to place constraints 
upon parameters appearing in the G\"{o}del-Obukhov metric 
(which does not have the
restriction of an isotropic universe). This is
done by obtaining bounds on an anisotropic redshift  versus 
magnitude relationship and on accompanying parameters of the  
G\"{o}del-Obukhov metric.
They found that the outcome depends on what are 
used for the host galaxy extinctions.  The most reasonable fits do not 
show any signals requiring anisotropy.  Yet, the existence of some 
small anisotropy cannot be ruled out.  It appears that their findings 
are consistent with present-day observations, and it might be 
reasonable to investigate models that perhaps yield some anisotropy,
particularly the G\"{o}del-Obukhov model.

The G\"{o}del-Obukhov metric, with exact general relativistic
solutions as expressed by 
the cosmic scale factor 
$R=R(t)$ and its derivatives \cite{Obukhov2000}, 
describing the evolution of $R$,
just as commonly done for the
Friedmann-Robertson-Walker metric,
avoids the principal difficulties of
old cosmological models with rotation, where $R$ describes the
expansion of physical spatial distances. 
For example, the G\"{o}del-Obukhov metric 
is consistent with isotropy of
the microwave background radiation, like the standard cosmology,
 and it produces
no parallax effects.
The final state
according to this
metric depends on the values of two cosmological 
coupling constants (discussed below) in which torsion can 
cause the Universe to
either accelerate or decelerate 
(${\ddot{R}/ R}>0$ or ${\ddot{R}/ R}<0$) or prevent 
cosmological collapse (${\ddot{R}/ R}\approx 0$),
with these constants playing a similar role to that played by the 
elusive cosmological constant, $\Lambda$, in the FLRW
spacetime cosmology.  However, the origin and the physics of these 
spin-torsion cosmological coupling constants  \cite{Obukhov2000} 	
can be readily identified.
  
Using the G\"{o}del-Obukhov metric to define separation of 
spacetime events,
we find that cosmic acceleration
is expected when the radial repulsive GM force (associated
with rotational energy) exceeds the familiar 
attractive radial ``gravitoelectric'' (GE) force (associated with 
rest mass energy or mass-energy).\footnote{The 
terms gravitomagnetic and gravitoelectric  
defined in this manuscript are not the same as those defined in  
so-called gravitoelectromagnetism (sometimes loosely referred
to as gravitomagnetism), which is a 
mathematical analogy between weak gravity and 
 Maxwell's equations for electromagnetism (see Mashhoon, B., 2003,
arxiv:gr-qc/0311030).}  
This appears to be somewhat 
the idea behind 
Einstein's introduction of the 
cosmological constant $\Lambda$, when he introduced it in his 
Theory of General Relativity to explain 
how the Universe could resist collapse under the inward force of 
gravity. However it was later thrown out by Einstein as his greatest 
blunder.  
It seems that he did not see any use for
$\Lambda$ in 
a universe already expanding according to the Hubble law,
 before discarding it.
In any case, maybe $\Lambda$, in a general sense,  is like 
the Hubble parameter $H$ that 
changes over time with the age of the Universe. That is, just as 
the Hubble parameter relates to cosmic expansion of the Universe, 
and, as we shall see, cosmic rotation, perhaps $\Lambda$ relates 
only to the effects of cosmic rotation.  
In the G\"{o}del-Obukhov spacetime cosmology, presented here, two 
cosmological coupling constants ($\lambda_1$ and $\lambda_3$ 
\cite{Obukhov2000}) due to rotation (or spin) and torsion 
(as related to general relativistic frame 
dragging or the so-called Lense-Thirring 
\cite{Lense-Thirring1918} effect)
in curved spacetime take on the role of 
$\Lambda$.  Moreover, we find that the G\"{o}del-Obukhov
spacetime cosmology appears not only to explain recent 
observations of the accelerated expansion, but suggests a dynamical
description of the Universe over time that is a natural general
relativistic extension of the standard FLRW cosmology.
In this sense the answer to the question posed in the title
appears to be yes.

The organization of this paper is as follows.  In Sec.~\ref{sec:2},
a detailed description is presented of the model used here to explain 
the present epoch acceleration of the Universe: as being a 
gravitational-rotational-inertial phenomenon.  A formalism 
containing the 
astrophysical and mathematical descriptions of the components to
validate the model's claims is presented in Sec.~\ref{sec:3}.
It includes analytical derivations of the G\"{o}del-Obukhov 
metric in spherical comoving
coordinates, the cosmic radial GM and 
GE force fields, and the density
of the Universe, where the GE and GM  fields are the 
gravitational analogues of electric and magnetic force fields, 
respectively. 
In general, the GE and GM fields relate 
directly to the total mass
and rotation, respectively, of a gravitating system 
(see, e.g., Refs.~\cite{Thorne1986,Williams2002}). 
 Also, included are 
the cosmological 
parameters used, which includes the cosmic rotational (or angular) 
velocity, the scale
factor, and the Hubble constant. 
The numerical results from evolving the analytical expressions
for the GE and GM accelerations over time are presented
in Sec.~\ref{sec:4}.  The
Discussion is presented
in Sec.~\ref{sec:5}.  
Included in Sec.~\ref{sec:5}, an analytical expression for 
the cosmic magnetic field [Eq.~(\ref{D15})] given by 
the G\"{o}del-Obukhov metric
in terms of the spin density is evolved over cosmic time and 
compared with observations and theory. 
This expression
suggests how the magnetic field, the mass density, and the GM
field might be related through the spin density.   
Also included in Sec.~\ref{sec:5}, the equation of state from 
Obukhov \cite{Obukhov2000} is used to test the validity of these present 
model calculations.
In Sec.~\ref{sec:5.7} a summary is given of the individual Discussion
sections.
Conclusions are presented in 
Sec.~\ref{sec:6}.  

\section{Model Description}
\label{sec:2}

It seems reasonable to assume that at $t\approx 0$ the Universe 
had, at least, two degrees of freedom: translational, associated 
with the expansion and collapse (or infall), in the 
${\bf \hat e}_r$-direction and 
rotational, associated with  cosmic rotation, in the 
${\bf \hat e}_\phi$-direction
 where  ${\bf \hat e}_r$ and ${\bf \hat e}_\phi$ are global unit 
vectors of the cosmological spacetime continuum.  
Rotation at $t\approx 0$ is consistent with the G\"{o}del-Obukhov 
 spacetime metric \cite{Obukhov2000}, which allows for rotation and
 expansion.  

Note, here we will assume that the only different between the FLRW
cosmology and the G\"{o}del-Obukhov \cite{Obukhov2000} cosmology is the 
rotation (producing inertial frame dragging), meaning that the 
properties that apply to the mass-energy (producing the warping 
or curvature of 
spacetime) apply to both cosmologies.  For example, the general form 
of the mass-energy critical density $\rho_c$ and the form of the 
solution of the scale factor $R(t)$ [compare Eqs.~(\ref{GE7}),
(\ref{D9}), and~(\ref{CP6})] are of the same form in both cosmologies,
save for the difference due to the effects of cosmic rotation.  
This appears to be a valid assumption.

For the initial conditions of the Universe, although speculative but 
consistent with the G\"{o}del-Obukhov cosmology,
 we will assume that a 
gravitationally unbound ``hot'' energetic rotating and expanding 
dense plasma existed at 
$t\simeq 5.4^{-44}$~s, the 
Planck time, with the observable universe 
corresponding to the Planck length, 
$l_P \simeq 1.6\times 10^{-33}$~cm.  We assume that this rotating 
matter is ``embedded'' in 
an inertially expanding spacetime coordinated frame, inertia as 
analogous to Newton's first law
of motion.  This inertially expanding frame, however, can be 
associated with an inertial force field that wants to expand
the cosmic spacetime matter out of
rotation, while the cosmic matter wants to drag (or torque) the 
inertially expanding frame into rotation.
The inertial frame dragging angular velocity oriented along the 
global $z$-axis (or symmetry axis) is $\omega_{\rm FD}<0$ 
[Eq.~(\ref{FD})], with 
frame dragging in
the direction of the cosmic rotation. 
It seems that the cosmic rotation is
coupled with gravity and the inertial expansion is associated
with the initial force that ``ignited'' the Big Bang.   

Now, let us go further to assume that the Big Bang was perhaps due to 
this inertial force
``stretching''  the
cosmic matter apart as spacetime expands 
(like splitting the nucleus of an atom)
with a cosmic cataclysmic quantum-gravitational, 
$E_{\rm U}=M_{\rm U} c^2$,
type  explosion that caused infinitely dense
matter to expand relativistically outward due
to a force with a strength similar to that of the quantum-gravitational
singularity
that gives rise to the event horizon of
a black hole, where $M_{\rm U}$ 
is the total mass of the Universe at $t=0$, and $E_{\rm U}$ is the 
total energy of the Universe.  It seems that this  inertial force
wants to ``flatten'' spacetime wherein world lines will have straight
instead of curved geodesics.  This inertial force appears to
play an intrinsic part in the process from whence our Universe is 
expanded.  Perhaps its origin was the initial force, as mentioned
above,  needed to release 
the infinitely large binding energy trapped in pre-existing Big Bang 
conditions in which, at least, the physical
forces of nature were unified.

The proposed existence of this cosmic 
inertial force field appears to be 
similar to the idea that Einstein had in
proposing, in his Theory of General Relativity
(see Ref.~\cite{deSitter1916}),
 that ``world-matter'' (the matter he postulated to be the
origin of the energy of inertia) was  located at the boundary of
the Universe, and controlled and provided energy for the whole Universe,
from so-called supernatural masses.
It appears that de Sitter \cite{deSitter1917} later persuaded
Einstein to adopt a new hypothesis with the world-matter not at
the boundary of the Universe, but distributed over the whole
Universe, proposing the Universe to be finite, 
though unlimited (a sphere or an ellipse).
 In this new hypothesis
inertia is produced by the whole of world-matter, and gravitation
is produced by local deviations from homogeneity.   
Thus, in terms of modern general relativity, the mass-energy 
density $\rho(t)$ regulates the warping of the spacetime 
continuum, 
while supplying the small scale
inhomogeneity to the original homogeneously expanding Universe.
It appears that the cosmic expansion frame is inherently a ``flat'' 
spacetime continuum; and that the spacetime continuum has 
field properties, analogous to the electromagnetic field 
\cite{Einstein1956}.  The spacetime continuum expansion, it seems, has 
the inertial property of electricity and light 
(or electromagnetic radiation), and the 
gravitational property of general relativity wherein the 
spacetime continuum (or so-called world-matter) can 
be warped (or dragged) by the presence of mass-energy 
and momentum.

Now if we constitute Einstein's new hypothesis with
properties of the proposed inertial field, then, based on this 
constitution,
the global gravitational
field of the rotating cosmic matter is a deviation 
from uniformity
of the cosmic inertial expansion.  If the energy of
inertia is to keep the Universe expanding,
then deviations
from this due to gravity and rotation would have negligible 
effects
locally, i.e., on small scales,  but on large scales, say
globally, any significant deviation
may have an effect on the expansion rate. This has been realized in
the standard cosmological Big Bang model described by the
Friedmann-Robertson-Walker metric, and incorporated in the deceleration 
parameter:
\begin{equation}
q=q(t)\equiv q_t={-{\frac{\ddot{R} R}{\dot{R}^2}}},
\label{dec}
\end{equation} 
which depends 
on the Hubble parameter $H(t)$ [compare Eq.~(\ref{hubble})], 
as related to the mass 
(or mass-energy) density $\rho(t)$ of the  
Universe [see Eqs.~(\ref{GE7}) and~(\ref{GE10})], 
and the scale factor $R$ 
and its derivatives, with 
$\Lambda=p=0$ in Eq.~(\ref{GE6}), where, when the subscript is $0$,
it  means 
$t=t_0$, the 
present epoch.  

A further example of affecting 
the expansion is the 
recently observed cosmic 
acceleration.  Now, gravitating systems
 are associated with 
mass and, it seems safe to say, rotation (through
conservation of angular momentum). Since 
the mass density has an effect on the spacetime expansion, thus 
determining the geometry in the
standard cosmological model, then cosmic rotation (appearing to be a
property of gravity through
conserved angular momentum) may also have an
effect on the expansion.  The apparent tendency of the
inertial spacetime continuum force field (or so-called world-matter) 
to stay expanding 
as a flat spacetime continuum and the dragging of this inertial 
spacetime frame by the rotating 
cosmic matter
are the physical mechanisms 
proposed here to be the origin of recently observed  
acceleration of the comic expansion.

Consistent with what is allowed by the G\"{o}del-Obukhov
 spacetime metric, the frame dragging angular 
velocity, $\omega_{\rm FD}$,
 can be $<0$ or $>0$, with the $<0$ expression 
naturally being chosen if we express $\omega_{\rm FD}$ in the 
usually sense,
as we shall see in the following section. 
To understand physically what is meant by $\omega_{\rm FD}<0$, 
we use the analogy of a rotating black hole.  As mentioned above we 
assume that the Universe 
has two major degrees of 
freedom, rotation and expansion (which includes infall).  Now, in the 
case of a massive rotating black hole, 
the gravitational force of the black hole drags inertial 
frames into rotation, with 
$\omega_{\rm FD}>0$, in the direction of the rotating black hole.  
But in the case of the 
cosmic matter of the Universe,
undergoing what one might call an ``anti-gravitational'' expansion 
(i.e., a 
reversed gravitational collapse), the inertial spacetime frame 
of the expansion force appears to be dragged (or torqued) by 
the cosmic matter, into rotation, with  
$\omega_{\rm FD}<0$,  in the direction of the rotating universe.  
In both cases, a relativistic fictitious force is  
produced, which we refer to as the GM force, 
that acts on 
any moving matter or particle in the dragged frame.  It is like say
 the Coriolis force acting on moving matter in a rotating frame, 
and analogous to the Lorentz force acting on a charged particle in a 
magnetic field: thence came the word GM (see, e.g., 
Ref.~\cite{Thorne1986}).
  However, importantly, in the case 
of the Universe, with   $\omega_{\rm FD}<0$, this GM force is of a 
repulsive nature, and could very well be associated with the 
present-day observed cosmic acceleration.
Note, in the case of the black hole, with $\omega_{\rm FD}>0$, the 
GM force is also of a repulsive nature (see Refs.~\cite{Williams2002,
Williams2004,Williams2005}).

Now, consistent with the degrees of freedom stated above, 
we will assume that at $t\sim 10^{-43}$~s, 
at least three large-scale forces, producing
the following magnitudes of acceleration, were present to act on the 
mass-energy of the Universe: 
(1)  the 
GE acceleration
 of gravity $g_{\rm GE}$ associated with $\rho(t)$; 
(2) the GM acceleration of gravity $g_{\rm GM}$ 
associated with cosmic rotational velocity $\omega_{\rm rot}$, 
with direction $\bm{\omega}_{\rm rot}<0$ \cite{Obukhov2000}; and
(3) the initial acceleration of the expansion $a_{\rm I}$
associated with the inertial spacetime-expanding
coordinate frame (indicated by subscript $I$).

Next, we consider the 
following scenario: If we assume that the energy, or the work 
done by the force, of the
Big Bang at $t=0$
was enough to overcome the infinitely large binding force 
 associated with $g_{\rm GE}$
(due to the mass-energy that initially warped spacetime closed), 
then we will have expansion, with $a_{\rm I}\agt g_{\rm GE}$,
 implying a flat or open universe.
We assume that at the event of the Big Bang the Universe became
gravitationally unbound, with matter transforming to 
relativistic particle expansional and rotational energy.  The Universe,
 in general,
will expand with the expansion velocity given to it by the force
of expansion, $\bm{F}_{I}=\bm{F}_{I}(t)$, 
which appears to be expressed 
 by $\bm{F}_{I}\sim {E_{U}c^{-2}}H^2 \bm{r}$, where, again,   
$H=H(t)$ is the Hubble parameter, and $\bm{r}=\bm{r}(t)$ is the 
spacetime separation between events [Eq.~(\ref{GE11})]. 
Since in the Big Bang ``explosion'' matter was converted
entirely into energy, then the expansion velocity $v_I$,
 at $t\approx 0$,
was at least  $\approx  c$, the speed of light.  
The acceleration of the expansion,
$a_{\rm I}\sim  H^2 r$,  will decrease over time.
If the scenario ended here, add inflation, and exclude rotation,
this would be a universe explained by the standard FLRW cosmology,
more or less.  But  with the existence
of $g_{\rm GM}$, associated with the non-inertial rotating frame,
 the expanding mass-energy of the 
Universe will experience an additional acceleration, perhaps
one related to the recently observed cosmic acceleration. 
 Now, this is where the FLRW cosmology develops the well known 
problems pointed out in Sec.~\ref{sec:1}, i.e., 
when attempting to explain 
the physics of the accelerated 
expansion (or cosmic acceleration) we observe to exist in the 
present-day Universe. 

It appears that the
inconsistencies in the standard cosmological model of the 
Friedmann-Robertson-Walker
spacetime metric might be due to our lack of considering the effects of 
cosmic
rotation, which requires the use of a rotating and expanding 
cosmological spacetime metric, like that employed in this present paper.    
In the following sections
the physics we need to further discuss the model described above, and 
to test its validity with observations, is devised.

\section{Formalism}
\label{sec:3}

\subsection{The G\"{o}del-Obukhov Spacetime Metric in Spherical 
Coordinates}
\label{sec:3.1}

The G\"{o}del-Obukhov \cite{Obukhov1997,Obukhov2000, 
Jain2007}  shear free
and spatially homogeneous
 spacetime metric, defining separations of events in Cartesian 
comoving coordinates, is given by \cite{Obukhov2000,Jain2007}
\begin{eqnarray}
{\rm d}\tau^2&=&{\rm d}t^2-2\sqrt{\sigma}R(t){\rm e}^{mx}
{\rm d}t{\rm d}y\nonumber\\
&&-R^2(t)
     ({\rm d}x^2+k {\rm e}^{2mx}{\rm d}y^2+{\rm d}z^2),
\label{metric_cc}
\end{eqnarray}
with rotation  directed along $z$-axis 
and acceleration along 
$y$-axis \cite{Obukhov1990},
 where ${\rm d}\tau$ is the proper time interval, 
$\sigma\equiv\sigma(t)$ (Sec.~\ref{sec:3.5}), 
$m$, and $k$ are 
related  geometrical
parameters; $R=R(t)$ is a time dependent scale factor, and 
$k>0$ ensures 
absence of closed timelike curves (note, $k$ is not the 
spatial curvature index unless noted otherwise); 
with $c=1$ unless noted otherwise. 
Clearly, $\sigma(t)$ must be $ > 0$, 
and for 
definiteness, we choose $m>0$ \cite{Obukhov2000}.
 According to Eq.~(\ref{metric_cc}) the Universe 
is spatially homogeneous, rotating,
 and expanding.
Note, the usual G\"{o}del \cite{Godel1949} 
metric that suffers from the presence of closed timelike curves 
is obtained by setting
\[
R(t)=1,~~ \sigma(t)=1, ~~m=1, ~~k=-{\frac{1}{2}}
\]
in Eq.~(\ref{metric_cc}).  The magnitude of the global cosmic 
rotational velocity 
$\omega_{\rm rot}$ oriented along the $z$-axis is \cite{Obukhov2000,
Jain2007}
\begin{eqnarray}
\omega_{\rm rot}=\sqrt{\omega_{\mu\nu}\omega^{\mu\nu}}
={\frac{m}{2 R}}\sqrt{{\frac{\sigma}{k+\sigma}}}\geq 0,
\label{metric_1}
\end{eqnarray}
with
\begin{eqnarray}
m= 2 R \omega_{\rm rot}\sqrt{\frac{k+\sigma}{ \sigma}}
\label{metric_2}
\end{eqnarray}
(see  Eq.~\ref{metric_cc}), where, recall, $R=R(t)$ and 
$\sigma\equiv\sigma(t)$.
Thus, we see that, vanishing of $m$ and/or $\sigma(t)$ yields
zero vorticity.

Upon assuming a spinning fluid of intrinsic angular momentum along the
global $z$-axis with 
electromagnetic dynamical
characteristics in a Riemann-Cartan spacetime \cite{Obukhov2000, 
Obukhov1987,Minkevich2006},
Obukhov \cite{Obukhov2000} gives an exact solution to Einstein's field 
equations: an equation of motion describing
the evolution  of the scale factor $R$.  From Obukhov 
\cite{Obukhov2000}, after 
some algebraic manipulations and substitutions, we can
show that 
\begin{eqnarray}
\frac{\ddot R}{R}&=&-H^2 + {\frac{\omega_{\rm rot}^2}{{3 k\sigma}}}
(k+\sigma)(3\sigma+4k)\nonumber\\ 
 &&+ {\frac{1}{\omega_{\rm rot}^2}}
\biggl({\frac{k+\sigma}{144k}} \biggr)
(4\lambda_3^2 - \lambda_1^2){\frac{B^4}{R^8}}\nonumber\\
&&+ {\frac{8\pi G}{3c^2}} \biggl(\frac{k+\sigma}{k}\biggr)
\biggl(c^2\rho -p-{\frac{B^2}{R^4}}\biggr),
\label{Exact_1} 
\end{eqnarray}
where the variables $H$, $R$, $\omega_{\rm rot}$, $B$ (the cosmic 
magnetic field strength), $\rho$, and $p$ 
are all functions of time; $\lambda_1$ and $\lambda_3$ are 
cosmological coupling 
constants of the spin and torsion tensors \cite{Obukhov2000},
mentioned in Sec.~\ref{sec:1}.
It appears that the parameters $\sigma$ and $k$, in a sense, 
determine the magnitude of acceleration of a fluid element due to 
rotation of the Universe \cite{Obukhov1990,Jain2007}; 
we shall see more evidence of
this in Sec.~\ref{sec:3.2}. 
Note, $B$ is related to the spin density (angular momentum per
unit volume), as we shall see in Sec.~\ref{sec:5.5}. 
Thus, one can see the repulsive nature of the above equation of motion
for the scale factor $R$, as found, it 
appears, independently by Obukhov 
\cite{Obukhov2000} and Minkevich \cite{Minkevich2005}, from an 
adaptation of general relativity to
a Riemann-Cartan spacetime.  
Minkevich, Garkun, \& Kudin \cite{Minkevich2007} have found 
this repulsive characteristic not only in the extreme conditions
of the early Universe, but also at sufficiently small energy 
densities of later times.  Minkevich et al. \cite{Minkevich2007}
 conclude that the 
effect of 
the accelerated cosmological expansion, even of today,
 is geometrical in nature and
is connected with the geometrical structure of spacetime. 
Indeed this might be the case:
it appears that these authors are finding the effect that
frame dragging, producing the GM field, has on the geometry of 
spacetime.    In this present paper, with the author's model proposed 
independently and unaware of concluding remark  by Minkevich et al. 
\cite{Minkevich2007}, it is  
shown that the recently observed cosmic acceleration  
may be the effect of the frame dragging 
nature of cosmic rotation, interacting
with an inertially expanding spacetime geometry.  
 Note, in deriving Eq.~(\ref{Exact_1}), 
for a specific epoch time $t$,
\begin{equation}
H=H(t)\equiv H_t={\frac{\dot{R}}{R}},
\label{hubble}
\end{equation}
 the Hubble parameter, was used.  In addition, one cannot help but notice 
that the first term on the 
right-hand side of Eq.~(\ref{Exact_1}) 
is the same as that in the standard FLRW model: wherein the equation 
of motion of the cosmic scale factor
reduces exactly to this term when $\Lambda=p=k~
(\text{spatial curvature index})=0$ and 
$q=1$ [see Eq.~(\ref{GE6}) along with Eq.~(\ref{GE7})].
We will return to this and similar comparisons in Sec.~\ref{sec:5.4}.

For the
G\"{o}del-type universe of Eq.~(\ref{metric_cc}), the
evolution of the scale factor
reveals several possible stages of the Universe as pointed
out by Obukhov \cite{Obukhov2000}, and elaborated on
here, in this present paper,
based on
Eq.~({\ref{Exact_1}}).
  The first stage is short and occurs in the vicinity $t=0$.
There is no initial cosmological singularity due to the dominating
spin contribution, a characteristic of Einstein's gravitational
theory in Riemann-Cartan spacetime \cite{Minkevich2006}, in which
$R(t=0)\neq 0$ implies a regular, as opposed to a singular, spacetime
metric in the transition from compression (pre-existing Big Bang
 conditions) to
cosmological expansion.  The duration of this first
stage is $\ll 1$~s,
since the spin
term quickly decreases with the growth of the scale factor
\cite{Obukhov2000}. Compare Eq.~({\ref{metric_1}}) 
and the second term on the
right-hand side of Eq.~({\ref{Exact_1}}).
But, importantly,
notice that $\omega_{\rm rot}$ in the denominator  of
third term on the
right-hand side of Eq.~({\ref{Exact_1}}) will cause this
accelerating term to increase over time in some degree
as $\omega_{\rm rot}
\longrightarrow 0$.	
Now, at
this stage ($\ll 1$~s) the fluid source describing the material of the
Universe can be characterized by the approximate stiff matter
 equation of state
\cite{Obukhov2000}:  
\begin{equation}
p\approx \biggl({\frac{\lambda_1-4\lambda_3}{6}}\biggr) 
{\frac{\tau^2}{R^6}},
\label{press1}
\end{equation}
where, in general,  as found from Obukhov \cite{Obukhov2000},
\begin{equation}
p=\biggl({\frac{\lambda_1-4\lambda_3}{3}}\biggr) 
{\frac{\tau^2}{R^6}}-\epsilon +{\frac{2B^2}{R^4}};
\label{press2}
\end{equation}
 $\tau=\tau(t)$ is the spin density
(discussed in Sec.~\ref{sec:5.5}); and 
$\epsilon=\epsilon(t)=\rho c^2$ is the
internal energy density of matter and radiation, 
assuming the Big Bang had 
a relativistic mass-energy origin
as mentioned in Sec.~\ref{sec:2}. At some point in this
stage, perhaps at $t\alt 10^{-36}$~s,
the equation of
state, possibly being that of a gravitational repulsive ``false vacuum''
might drive cosmic inflation. 
Upon substitution of $p$ from Eq.~(\ref{press1}) or~Eq.~(\ref{press2}) 
into the forth term on the right-hand side of Eq.~(\ref{Exact_1}),
which appears to be related to the inertial spacetime
expansion and cosmic rotation, it can be shown from
the results of this present investigation that 
for $\vert 4\lambda_3\vert
\gg \vert\lambda_1\vert$
and $\lambda_3>0$, the rapid increase
of the scale factor at the onset of inflation
will produce a large repulsive acceleration;
one that might at least assists in cosmic inflation
\cite{Williams2014}.
Moreover, during inflation it is commonly accepted that
the scale factor increases by a factor
$\sim{\rm e}^{H\Delta{t}}$ 
 (as discussed in Sec.~\ref{sec:3.5}).
If the Big Bang consisted of some sort of 
``explosive'' expansion with spin, 
such initial conditions at $t\approx 0$ could possibly be
associated with inflation, at least the initial condition of the 
scale factor would be satisfied (see the following paragraph).  
This speculation would have to be investigated further.
Next comes the  stage when the scale factor 
increases like 
$R(t)\propto t^{1/2}$, while the equation of state is of the
radiation type, $p\approx c^2\rho/3$. This ``hot universe'' 
expansion lasts until the Universe becomes mass dominated.
  After this the 
``modern'' stage
starts with the effective dust equation of state 
$p\approx 0$ and, it can be shown from \cite{Obukhov2000},
\begin{equation}
\epsilon\approx {\frac{2 B^2}{R^4}}.
\label{energy1}
\end{equation}
The scale factor still increases, now like $R(t)\propto t^{2/3}$,
but the deceleration of the expansion takes place.  The final 
stage depends on the value of the third term on the right-hand
side of Eq.~(\ref{Exact_1}) referred to as
the cosmological term by Obukhov \cite{Obukhov2000}, 
containing $\lambda_1$ 
and $\lambda_3$,
which specifically are made up of coupling constants relating
spacetime curvature, spin, and torsion, where torsion can 
either accelerate the
expansion or prevent
cosmological collapse.  
Notice the striking similarity of this cosmological
term and accelerated expansion to the 
popular view of the cosmological constant $\Lambda$ 
as the source of the present-day observed 
accelerated expansion.

The above stages are consistent with the model description  proposed 
in Sec.~\ref{sec:2}, which includes being consistent with a  
Big Bang cosmology.  This could mean that 
if expansion is part of
the conditions occurring around $t=10^{-43}$~s, then rotation, which
appears to be a natural  phenomenon associated with gravitation, could
very well be a part also. So, avoiding the initial 
singularity that  exists  at  $R(t=0)=0$ for the standard FLRW cosmology
 suggests that 
the G\"{o}del-Obukhov metric allows us to get somewhat closer
to conditions existing at $t=0$,
with $R(t=0)=1$ \cite{Obukhov2000} or $R(t=0)=e^0=1$, consistent
with inflation (see above).  That is, perhaps 
cosmic rotation and
cosmic expansion are intrinsic parts left over from the earlier 
quantum-gravitational spacetime makeup 
of the primordial matter of the Universe at $t\simeq 0$. Imagine that
cosmic expansion, and deceleration of cosmic rotation, of the Universe 
are like a reversed process of 
gravitational contraction (or collapse) and conservation of angular
momentum. This helps one to conceive  the strong possibility of how 
the two: rotation and expansion, cannot, it appears,  be 
separated  in the physics to describe the Universe, as commonly done
by assuming $\omega_{\rm rot}=0$.

Importantly, it appears that macroscopic torsion of spacetime might 
be directly related to inertial frame dragging (a Lense-Thirring effect), 
and, thus, the GM force field.
In support of this, the characteristics of torsion given
by Mao et al. \cite{Mao2007}, that {\it a rotating body also generates
torsion through its rotational angular momentum, and the torsion
in turn affects the motion of spinning objects such as gyroscopes},
are exactly those of the GM field (see, e.g., Refs.~\cite{Thorne1986,
Williams2002}).  In general, torsion in the Einstein-Cartan theory 
appears to be produced by any intrinsic
spin density (angular momentum per unit volume) of mass-energy  
that torques (or drags)
the spacetime continuum whether it be of 
microscope or macroscopic 
origin: from the intrinsic spin of elementary particles to that of compact 
objects (stars, planets and centers of galaxies) to that of global cosmic 
matter rotation of the Universe as a whole. This generality has the potential
 to set to rest the controversy 
surrounding the above claim by Mao et al. (see Ref.~\cite{Hehl2007}).  
In this present model the cosmic rotation (or spin) is intrinsic to
the matter and has a spin density in which the comoving observer is 
 within the source.
Therefore, it seems reasonable to associate the dominant repulsive
nature of
Eq.~({\ref{Exact_1}}) with that of the GM field.
In this paper, we
derive the GM field associated with cosmic rotation to see what
role  it may have in
the recently observed cosmological accelerated expansion.
Further analysis of
Eq.~({\ref{Exact_1}}) will allow us to identify, as we shall see
see in Sec.~\ref{sec:5.4}, the
suspected GM acceleration and other terms one would expect to be
measured
by a rotating and expanding comoving frame observer.

Since the
G\"{o}del-Obukhov metric or line element of Eq.~(\ref{metric_cc})
 is inconvenient for our
present application,  we transform to
spherical  (polar) coordinates for convenience.  
Taking spatial homogeneity
of Eq.~(\ref{metric_cc}) into account, we 
assume that the observer's coordinates are 
$P=(t=t_0,~x=0,~y=0,~z=0)$ at the local infinitesimal point $P$,
where $t_0$ is the present epoch observer  
\cite{ Korotkii1991}.
The comoving observer is in free
fall, which naturally, through the Equivalence Principle,
 makes him locally an inertial frame observer whose
unit four vector is an orthonormal tetrad 
\cite{Hartley1995}.  
In other words, the comoving observer is the 
accelerated observer whose frame is inertial at $t=t_0$, and whose  
Riemann-Cartan geometry  is Euclidean at the point $P$ 
\cite{Hehl2007}. 
Therefore,
it is appropriate to use the Euclidean space transformation
from local Cartesian coordinates ($x,~y,~z$) to local
spherical coordinates ($r,~\theta,~\phi$)  
centered on the comoving observer at point
$P=(t=t_0,~r=0)$:
 \begin{eqnarray}
x&=&r\sin\theta\cos\phi, \nonumber\\
y&=&r\sin\theta\sin\phi, \nonumber\\
z&=&r\cos\theta; 
\label{sph_transformation1}
\end{eqnarray}
and derivatives:
 \begin{eqnarray}
dx&=&\sin\theta\cos\phi dr+ r\cos\phi\cos\theta d\theta
-r\sin\theta\sin\phi d\phi, \nonumber\\
dy&=&\sin\theta\sin\phi dr+ r\sin\phi\cos\theta d\theta
+r\sin\theta\cos\phi d\phi, \nonumber\\
dz&=&\cos\theta dr-r\sin\theta d\theta; 
\\ \nonumber
\label{sph_transformation2}
\end{eqnarray}
with $t$ being invariant, i.e., $t=t^\prime$ 
[see Eq.~(\ref{sph_transformation4})].
Applying the above transformations to Eq.~(\ref{metric_cc})
yields 
\begin{widetext}
\begin{eqnarray}
{\rm d}\tau^2&=&{\rm d}t^2-2\sqrt{\sigma(t)}
           R(t){\rm e}^{mr\sin\theta\cos\phi}(\sin\theta\sin\phi
          {\rm d}t{\rm d}r+r\sin\phi\cos\theta{\rm d}t{\rm d}\theta
           +r\sin\theta\cos\phi{\rm d}t{\rm d}\phi)\nonumber\\
&&   -R^2(t)[(\sin^2\theta\cos^2\phi
     +k {\rm e}^{2mr\sin\theta\cos\phi}\sin^2\theta\sin^2\phi
     +\cos^2\theta){\rm d}r^2\nonumber\\
&&   +(\cos^2\theta\cos^2\phi
     +k {\rm e}^{2mr\sin\theta\cos\phi}\cos^2\theta\sin^2\phi
     +\sin^2\theta)r^2 {\rm d}\theta^2\nonumber\\ 
&&   +(\sin^2\phi+ k {\rm e}^{2mr\sin\theta\cos\phi}\cos^2\phi)
         r^2\sin^2\theta {\rm d}\phi^2\nonumber\\
&&   +2(\cos^2\phi+ k {\rm e}^{2mr\sin\theta\cos\phi}\sin^2\phi-1)
         r\sin\theta\cos\theta{\rm d}r{\rm d}\theta\nonumber\\
&&   +2(k {\rm e}^{2mr\sin\theta\cos\phi}-1)
       r\sin^2\theta\cos\phi\sin\phi{\rm d}r {\rm d}\phi\nonumber\\
&&   +2(k {\rm e}^{2mr\sin\theta\cos\phi}-1)
       r\sin^2\theta\cos\theta\cos\phi\sin\phi
       {\rm d}\theta{\rm d}\phi], 
\label{sph_transformation3}
\end{eqnarray}
\end{widetext}
as approved by Obukhov \cite{Obukhov2014}. 
Note, the transformed metric of Eq.~(\ref{sph_transformation3})
can also be obtained from the metric tensor transformation law at
any given point 
\cite{Weinberg1972}:
\begin{equation}
g^\prime_{\mu\nu}={{\frac{\partial x^\rho}{\partial x^{\prime\mu}}}
{\frac{\partial x^\sigma}{\partial x^{\prime\nu}}}}g_{\rho\sigma}.
\label{sph_transformation4}
\end{equation}

Now, for
simplicity, it seems appropriate to assume polar axisymmetry
of spacetime for this  local comoving observer.   If we assume
such axisymmetry for the comoving observer,
the metric coefficients must be independent of
the azimuthal $\phi$ coordinate [i.e., $g_{\mu\nu}\equiv
g_{\mu\nu}(t,r,\theta)$].   This is a valid assumption according
to the Killing vector isometries associated with the G\"{o}del-Obukhov
 metric of Eq.~(\ref{metric_cc}) \cite{Obukhov2000}.  
A Killing vector field is
one that preserves the metric.  This means that the
Lie derivative of the metric tensor in the direction of a Killing 
vector vanishes: $\mathcal{L}_{\bm{\xi}}g_{\mu\nu}=0$. 
 For $k>0$
[see Eqs.~(\ref{metric_cc})
and~(\ref{CP2})], the three Killings vector fields that provide spatial
homogeneity of the $t={\rm constant}$ hypersurfaces are
\begin{eqnarray}
&&\xi_{(1)}=\frac{\partial}{\partial y};
~~\xi_{(2)}=\frac{\partial}{\partial z}; 
~~\xi_{(3)}=\frac{\partial}{\partial \phi}
=\frac{1}{m}{\frac{\partial}{\partial x}}
-y{\frac{\partial}{\partial y}}. \nonumber\\
&&
\label{killing_vectors}
\end{eqnarray}
These vector fields 
indicate that
Eq.~(\ref{metric_cc}) has symmetry along $y$-axis,  $z$-axis,
and in coordinate $\phi$ direction, i.e., azimuthal
direction
\cite{Obukhov1990, Obukhov2000}.  Recall that the spacetime metric
of Eq.~(\ref{metric_cc}) has rotation directed along $z$-axis
and acceleration along $y$-axis.
Thus, the above clearly means that the $z$-axis has symmetry 
along and axisymmetry about the axis, validating the above 
assumption of polar axisymmetry. 

The axial symmetry about the $z$-axis Killing vector $\xi_{(3)}$ 
above allows us to show below that the usual
Euclidean transformation equations from Cartesian
to spherical coordinates yield axisymmetrical spherical
coordinates if $\cos\phi\longrightarrow 1$, which means that
we set  $\phi=0, 2\pi, 4\pi,\ldots, 2n\pi$, for integer $n$,
in the equations above.   This defines the infinitesimal 
($x=r\sin\theta,z=r\cos\theta$) planes of 
axisymmetry about the local $z$-axis.  
In other words, this
choice of $\phi$ respects the rotation symmetry of spacetime
about the $z$-axis \cite{Obukhov1990} of the comoving 
observer and, thus, simplifies the mathematical description.
For example, 
in Eqs.~(\ref{sph_transformation1}) and~(11),  
the principal value $\phi=0$ gives
\begin{eqnarray}
x&=&r\sin\theta, \nonumber\\
y&=&0, \nonumber\\
z&=&r\cos\theta; 
\label{sph_transformation5}
\end{eqnarray}
and derivatives:
\begin{eqnarray}
dx&=&\sin\theta dr+ r\cos\theta d\theta, \nonumber\\
dy&=&r\sin\theta d\phi, \nonumber\\
dz&=&\cos\theta dr-r\sin\theta d\theta. 
\label{sph_transformation6}
\end{eqnarray}
In general, if we transform the usual flat 3-dimensional 
spatially isotropic metric:
\begin{eqnarray}
{\rm d}s^2=
     {\rm d}x^2+ {\rm d}y^2+{\rm d}z^2,
\label{sph_transformation7}
\end{eqnarray}
in Cartesian coordinates 
 to spherical  coordinates 
using either the transformations of
Eqs.~(\ref{sph_transformation1}) and~(11) 
or Eqs.~(\ref{sph_transformation5}) 
and~(\ref{sph_transformation6}), in both cases, we find the 
 geometry or metric of a hyperspace ($t=\text{constant}$) 
sphere of radius 
$r=(x^2+y^2+z^2)^{1/2} $, 
with polar-axis symmetry (i.e., axisymmetry about the $z$-axis), 
and, in this case, spherical symmetry as well, 
surrounding an 
observer at point
$P$:
\begin{eqnarray}
{\rm d}s^2=
     {\rm d}r^2+ r^2{\rm d}\theta^2+ r^2\sin^2\theta {\rm d}\phi^2.
\label{sph_transformation8}
\end{eqnarray}
Similarly, letting $\cos\phi\longrightarrow 1$, implying substituting  
the   principle value $\phi=0$ into the transformed 
metric [Eq.~(\ref{sph_transformation3})] or
transforming Eq.~(\ref{metric_cc}) directly from 
Eqs.~(\ref{sph_transformation5}) and~(\ref{sph_transformation6}),
the G\"{o}del-Obukhov metric of
   Eq.~(\ref{metric_cc})  in spherical coordinates for
non-stationary and polar-axisymmetric characteristics of 
local spacetime
for a comoving observer is given by
\begin{eqnarray}
{\rm d}\tau^2&=&{\rm d}t^2-2\sqrt{\sigma(t)}
           R(t){\rm e}^{mr\sin\theta}r\sin\theta
          {\rm d}t{\rm d}\phi\nonumber\\
&&   -R^2(t)({\rm d}r^2+r^2 {\rm d}\theta^2 + k {\rm e}^{2mr\sin\theta}
         r^2\sin^2\theta {\rm d}\phi^2)\nonumber\\
&&
\label{metric_sph}
\end{eqnarray}
[compare Eq.~(\ref{metric_cc})].  
Note, local, in this
context, appears to mean the causally-connected region 
about the point $P$ in which the comoving observer
measures proper distances.

Now, we know that the metric of Eq.~(\ref{metric_cc}) is spatially
 homogeneous \cite{Obukhov2000};
 then we need the transformed metric [Eq.~(\ref{metric_sph})]
 to be homogeneous also.  To show
that it is indeed homogeneous, Eq.~(\ref{metric_sph})
 must have a maximally symmetrical hyperspace or subspace
\cite{Obukhov2014, Weinberg1972}.  This means that the space is
homogeneous on
each hypersurface of constant time or subspace of constant
radius.  Mathematically,
homogeneity means all points are equivalent, i.e., there exist
infinitesimal isometries (rotations and translations)
that carry or can map any
given point $P$ into any other point in its immediate (or local)
neighborhood. We compare the G\"{o}del-Obukhov hyperspace
($t=\text{constant}$) of Eq.~(\ref{metric_cc}):
\begin{eqnarray}
-{\rm d}\tau^2&=&R^2
     ({\rm d}x^2+k {\rm e}^{2mx}{\rm d}y^2+{\rm d}z^2) \nonumber\\
&=&R^2({\rm d}x^2+ {\rm d}z^2+ k {\rm e}^{2mr\sin\theta}
         r^2\sin^2\theta {\rm d}\phi^2)\nonumber\\
&&
\label{sph_transformation9}
\end{eqnarray}
to that
of Eq.~(\ref{metric_sph}):
\begin{eqnarray}
-{\rm d}\tau^2&=&R^2({\rm d}r^2+r^2 {\rm d}\theta^2
+ k {\rm e}^{2mr\sin\theta}
         r^2\sin^2\theta {\rm d}\phi^2)\nonumber\\
&=&   R^2({\rm d}x^2+ {\rm d}z^2 + k {\rm e}^{2mr\sin\theta}
         r^2\sin^2\theta {\rm d}\phi^2),\nonumber\\
&&
\label{sph_transformation10}
\end{eqnarray}
 where the last steps in Eqs.~(\ref{sph_transformation9})
and~(\ref{sph_transformation10}) are given by 
Eqs.~(\ref{sph_transformation5})
and~(\ref{sph_transformation6}), where
\begin{equation}
{\rm d}x^2+{\rm d}z^2= {\rm d}r^2+r^2 {\rm d}\theta^2,
\label{sph_transformation11}
\end{equation}
\begin{equation}
{\rm d}y^2=   r^2\sin^2\theta{\rm d}\phi^2.
\label{sph_transformation12}
\end{equation}
We find that the spacetime metrics of Eqs.~(\ref{metric_cc})
and~(\ref{metric_sph}) have
identical maximally symmetric 3-dimensional subspaces 
($r= \text{constant}$), whose metrics $-{\rm d}\tau^2={\rm d}s^2$:
\begin{eqnarray}
{\rm d}s^2&=&R^2 K^{-1}( {\rm d}\theta^2
+ k {\rm e}^{2mr\sin\theta}
         \sin^2\theta {\rm d}\phi^2),
\label{sph_transformation13}
\end{eqnarray}
have positive eigenvalues and  a {\it constant of curvature}  
$K=1/r^2$, describing
the surface on a 2-sphere of radius $r$, centered on the origin, 
guaranteeing
homogeneity of spacetime \cite{Weinberg1972}.

It appears that the choice of $\cos\phi\longrightarrow 1$,
giving rise to the ``unique'' transformations of
Eqs.~(\ref{sph_transformation5})
and~(\ref{sph_transformation6}), satisfies the
uniqueness theorem that {\it
given two maximally symmetric metrics with the same $K$ and the same
number of eigenvalues of each sign, it will always be possible to find
a coordinate transformation that carries one metric into another}
\cite{Weinberg1972}, as
in the case of Eqs.~(\ref{metric_cc})  and~(\ref{metric_sph}).

Moreover, again, and in summary, as found by Obuklov 
\cite{Obukhov1990, Obukhov2000},
according to the Killing vector fields [Eq.~(\ref{killing_vectors})]
and displayed in the spacetime matrices of 
Eqs.~(\ref{metric_cc}) and~(\ref{metric_sph}), a comoving observer 
 will observe
symmetry along  $y$-axis and $z$-axis and 
coordinate $\phi$ direction, acceleration 
along $y$-axis, and rotation directed along 
$z$-axis.

Next, we identify the following metric coefficients for our convenience:
\begin{eqnarray}
g_{tt}&=&1, \nonumber\\
g_{t\phi}&=& -\sqrt{\sigma(t)}R(t){\rm e}^{m r\sin\theta}r\sin\theta
            = g_{\phi t},  \nonumber\\
g_{rr}&=&-R^2(t), \nonumber\\
g_{\theta\theta}&=&-R^2(t)r^2,  \nonumber\\
g_{\phi\phi}&=&-R^2(t)k {\rm e}^{2mr\sin\theta}r^2\sin^2\theta.
\label{metric_sph_1}
\end{eqnarray}
Further, for our convenience, we find the corresponding inverse
metric components:
\begin{eqnarray}
g^{tt}&=&{\frac{k}{k+\sigma(t)}}, \nonumber\\
g^{t\phi}&=& -{\frac{\sqrt{\sigma(t)}}{R(t){\rm e}^{m r\sin\theta}
                r\sin\theta[k+\sigma(t)]}}
            = g^{\phi t},  \nonumber\\
g^{rr}&=&-{\frac{1}{R^2(t)}}, \nonumber\\
g^{\theta\theta}&=&-{\frac{1}{R^2(t)r^2}},  \nonumber\\
g^{\phi\phi}&=&-{\frac{1}{R^2(t) {\rm e}^{2mr\sin\theta}r^2\sin^2\theta
                       [k+\sigma(t)]}}.
\label{metric_sph_2}
\end{eqnarray}
 
The frame dragging angular velocity for the G\"{o}del-Obukhov metric
of Eq.~(\ref{metric_sph}), with diagonal component signature
 (+, -, -, -, -), is 
given by
\begin{eqnarray}
\omega_{\rm FD}&=&-{\frac{g_{\phi t}}{g_{\phi\phi}}} \nonumber\\
            &=&-{\frac{\sqrt{\sigma(t)}}{R(t)k 
               {\rm e}^{m r \sin\theta}r\sin\theta}},
\label{FD}
\end{eqnarray}
where we are assuming it to be given by the general
geometrical expression
of Bardeen, Press, \& Teukolsky \cite{Bardeen1972} for the frame
dragging angular velocity in the Kerr \cite{Kerr1963} 
metric of a spinning mass.
This appears to be a valid assumption.  In Eq.~(\ref{FD}), we see 
that the frame dragging angular velocity
parallel to the symmetry axis,
with frame dragging tangential velocity in 
the global azimuthal coordinate direction, can be either
 positive or negative because of 
the square root. But $\omega_{\rm FD}<0$ occurs naturally it
seems,  being consistent with the cosmic angular 
(or rotational) velocity axial vector 
$\bm{\omega}_{\rm rot}<0$ \cite{Obukhov2000}; compare 
the magnitude  [Eq.~(\ref{metric_1})]. 

\subsection{The Cosmic ``Gravitomagnetic'' (GM) Force}
\label{sec:3.2}

We now derive an expression for the GM force,
$\bm{F}_{\rm GM}$, exerted on a
test particle (or an object such as a galaxy) of cosmic space
momentum $\bm{P}$.
Apparently, using the analogy of a rotating compact object 
\cite{Thorne1986, Williams2002}, 
for a rotating general
relativistic system such as our Universe, the general
invariant for the GM force measured by
an arbitrary comoving observer can be expressed by 
\begin{equation}
\Biggl(\Bigg\lgroup{\frac{{\rm d}{\bm{P}}}{{\rm d}\tau}}
\Bigg\rgroup_{\rm GM}\Biggr)_{i}
=H_{ij}P^{j}, \;\;{\rm i.e.,}\;
\Bigg\lgroup{\frac{{\rm d}{\bm{P}}}{{\rm d}\tau}}\Bigg\rgroup_{\rm GM}=
{\tensor{\bf H} \bm{\cdot} \bm{P}}
\label{GM1}
\end{equation}
(${\rm d}\tau$ is the proper time interval),
with
\begin{displaymath}
H_{ij}={\rm e}^{-\nu}(\beta_{_{\rm GM}})_{_{j\vert i}}
\end{displaymath}\
(the vertical line indicates the covariant derivative in
3-dimensional absolute space), where, like in the case of a rotating
black hole \cite{Thorne1986},
\begin{equation} 
(\beta_{_{\rm GM}})^r = (\beta_{_{\rm GM}})^\theta=0, \;\;
(\beta_{_{\rm GM}})^\phi = -\omega_{\rm FD};
\label{GM2}
\end{equation}
$\omega_{\rm FD}$ is given by Eq.~(\ref{FD}).
The field $\tensor{\bf H}$
 is called the
GM tensor field, and
$\bm{\beta}_{{_{\rm GM}}}$ is sometimes called the GM potential.
We perform the metric component operations in Eq.~(\ref{GM1}) to
give the following expression for the 
GM force exerted, $\bm{F}_{{\rm GM}}$:
\begin{eqnarray}
\Bigg\lgroup{\frac{{\rm d}{\bm{P}}}{{\rm d}\tau}}\Bigg\rgroup_{\rm GM}
\equiv\bm{F}_{{\rm GM}}
&=&[(F_{\rm GM})_r,(F_{\rm GM})_\theta,(F_{\rm GM})_\phi]
            \nonumber\\
&=&(H_{r\theta} P^\theta+H_{r\phi}P^\phi){\bf\hat{e}_r} \nonumber\\ 
   &&+(H_{\theta r}P^r+H_{\theta\phi}P^\phi){\bf\hat{e}_\theta} 
\nonumber\\
&&+(H_{\phi r}P^r+H_{\phi\theta}P^\theta){\bf\hat{e}_\phi}.
\label{GM3}
\end{eqnarray}
Here we are only interested in the radial component of
Eq.~(\ref{GM3}), where we are 
assuming that the other components are not important in 
explaining  cosmological acceleration along the
line-of-sight of the observer.  Therefore, we need to determine
the force in the radial direction:
\begin{eqnarray}
(F_{\rm GM})_r&=&H_{r\theta}P^\theta+H_{r\phi}P^\phi \nonumber\\
              &=&H_{r\theta}g^{\theta\theta}P_\theta
                +H_{r\phi}g^{\phi\phi}P_\phi,
\label{GM4}
\end{eqnarray}
where we have used $P^\mu=g^{\mu\nu}P_\nu$.  
[Note, Eq.~(\ref{GM3}) is a general expression, existing for 
any gravitating and rotating system.]

From $H_{ij}$ of 
Eq.~(\ref{GM1}) we can identify the so-called blueshift factor
 ${\rm e}^{-\nu}$ [$\equiv \sqrt{g^{tt}}$ for a metric with a signature of 
diagonal components of the type as in 
Eq.~(\ref{metric_sph})].  Thus, from Eqs.~(\ref{metric_sph_2}),
\begin{eqnarray}
{\rm e}^{-\nu}=\sqrt{\frac{k}{k+\sigma(t)}}.
\label{GM5}
\end{eqnarray}

Next we determine the relevant GM tensor components: $H_{r \theta}$ 
and $H_{r\phi}$, to be substituted into Eq.~(\ref{GM4}). These 
components are given by $H_{ij}$ of 
Eq.~(\ref{GM1}): 
\begin{equation}
H_{ij}=\sqrt{g^{tt}}\,(\beta_{\rm GM})_{j\vert i},
\label{GM6}
\end{equation}
where we have
used the definition of Eq.~(\ref{GM5}).
In general the covariant 
derivative in 3-dimensional absolute space is given by  
\begin{equation}
\beta_{j\vert i}=\beta _{j,i}-\Gamma_{j i}^k\beta_k
\label{GM7}
\end{equation}
(repeated indices of $i,~j,~k$ sum over $r,~\theta,~\phi$);
so that
\begin{eqnarray}
H_{r\theta}&=&\sqrt{g^{tt}}\,(\beta_{\rm GM})_{\theta\vert r}
\nonumber\\
&=&-\sqrt{g^{tt}}\,\Gamma_{\theta r}^\phi(\beta_{\rm GM})_\phi,
\label{GM8}
\end{eqnarray}
since, as given from Eq.~(\ref{GM2}), $(\beta_{\rm GM})_r=
(\beta_{\rm GM})_\theta=0$ and  $(\beta_{\rm GM})_\phi\neq 0$, 
where
\begin{eqnarray}
(\beta_{\rm GM})_\phi&=&g_{\phi\phi}(\beta_{\rm GM})^\phi
                          \nonumber\\
                     &=&g_{\phi\phi}(-\omega_{\rm FD})
                        \nonumber\\
&=& -\sqrt{\sigma(t)}\,R(t) {\rm e}^{mr\sin\theta}r\sin\theta,
\label{GM9}
\end{eqnarray} 
upon substitutions from  Eqs.~(\ref{metric_sph_1}) and~(\ref{FD}).
Note, $(\beta_{\rm GM})_\phi$ is measured
in units of 
length and $(\beta_{\rm GM})^\phi$ in units of per length.
Note also that $\omega_{\rm FD} < 0$, yielding 
$(\beta_{\rm GM})_\phi<0$ and GM potential 
$(\beta_{\rm GM})^\phi>0$ [Eq.~(\ref{GM2})], is consistent with the
description of the model
presented in Sec.~\ref{sec:2}: proposing that the frame dragging,
$\omega_{\rm FD}$, tends to drag (or torque) the inertially
expanding spacetime frame into rotation.
Now, similarly, 
\begin{eqnarray}
H_{r\phi}&=&\sqrt{g^{tt}}\,(\beta_{\rm GM})_{\phi\vert  r}
\nonumber\\
         &=   &\sqrt{g^{tt}}\,\bigl[(\beta_{\rm GM})_{\phi, r}
            -\Gamma_{\phi r}^\phi(\beta_{\rm GM})_\phi\bigr]
\nonumber\\
&=&\sqrt{g^{tt}}\biggl[ {\frac{\partial (\beta_{\rm GM})_\phi}
   {\partial r}}-\Gamma_{\phi r}^\phi (\beta_{\rm GM})_\phi
\biggr].
\label{GM10}
\end{eqnarray}

It can be shown that $H_{r\theta}=0$, from 
Eqs.~(\ref{metric_sph_1}),~(\ref{metric_sph_2}), 
(\ref{GM8}), and~(\ref{GM13}); therefore, 
the GM force in the radial direction relative to an arbitrary 
comoving observer, given by Eq.~(\ref{GM4}), reduces to
\begin{equation}
(F_{\rm GM})_r= H_{r\phi}g^{\phi\phi}P_\phi.
\label{GM11}
\end{equation}
We want to simplify and analyze the above vector component to see 
under what 
conditions it may contribute to a repulsive accelerating force, i.e.,
we want the GM radial force component              
to be repulsive ($>0$).
We first evaluate the partial derivative of Eq.~(\ref{GM10}). 
From Eq.~(\ref{GM9}),
\begin{eqnarray}
{\frac{\partial (\beta_{\rm GM})_\phi}{\partial r}}&=
&-\sqrt{\sigma(t)}\,R(t)\sin\theta\, {\frac{\partial}
     {\partial r}}(r {\rm e}^{m r\sin\theta}) \nonumber\\
&=&-\sqrt{\sigma(t)}\,R(t)\sin\theta \,{\rm e}^{m r\sin\theta}
(m r \sin\theta + 1). \nonumber\\
&&
\label{GM12}
\end{eqnarray}
We next evaluate $\Gamma_{\phi r}^\phi$ of Eq.~(\ref{GM10}). In 
general the Christoffel symbol (or affine connection) is given by
\begin{equation}
\Gamma_{\mu\nu}^\lambda={\frac{1}{2}} g^{\lambda\kappa}
\biggl({\frac{\partial g_{\kappa\nu}}{\partial x^\mu}}
+{\frac{\partial g_{\kappa\mu}}{\partial x^\nu}}
-{\frac{\partial g_{\mu\nu}}{\partial x^\kappa}}
\biggr).
\label{GM13}
\end{equation}
So, with $g_{t r}=g_{\phi r}=g^{\phi r}=g^{\phi\theta}=0$,
\begin{eqnarray}
\Gamma_{\phi r}^\phi&=&{\frac{1}{2}}g^{\phi t}
 \biggl({\frac{\partial g_{t\phi}}{\partial r}}\biggr)
    +{\frac{1}{2}}g^{\phi\phi} \biggl({\frac{\partial g_{\phi\phi}}
     {\partial r}}\biggr)
       \nonumber\\
&=&{\frac{m r\sin\theta+1}{2r}}\biggl[{\frac{\sigma(t)+2k}
   {k+\sigma(t)}} \biggr],
\label{GM14}
\end{eqnarray}
upon substitution of nonzero metric components from 
Eqs.~(\ref{metric_sph_1}) 
and~(\ref{metric_sph_2}) into Eq.~(\ref{GM13}). 
Now we substitute from Eqs.~(\ref{metric_sph_2}), 
(\ref{GM12}), (\ref{GM14}), and~(\ref{GM9})
into Eq.~(\ref{GM10}) yielding
\begin{eqnarray}
H_{r\phi}&=&\biggl[{\frac{k\sigma(t)}{k+\sigma(t)}}\biggr]
           ^{1/2}R(t)\sin\theta\,{\rm e}^{m r\sin\theta}
           (m r \sin\theta+1) \nonumber\\ 
&&\times\biggl\lbrace {\frac{\sigma(t)+2k}
         {2[k+\sigma(t)]}}-1\biggr\rbrace.
\label{GM15}
\end{eqnarray}
Then, substitution of Eq.~(\ref{GM15}) 
and from Eqs.~(\ref{metric_sph_2}) into Eq.~(\ref{GM11}) yields the 
following for the cosmic GM radial force component along 
the line-of-sight:
\begin{eqnarray}
(F_{\rm GM})_r&=&\biggl\lbrace{\frac{k\sigma(t)}
{[k+\sigma(t)]^3}}\biggr\rbrace^{1/2}
\biggl[{\frac{m r\sin\theta+1}{R(t) r^2\sin\theta 
{\rm e}^{m r \sin\theta}}}
\biggr] \nonumber\\  
&&\times\biggl\lbrace 1-{\frac{\sigma(t)+2k}
{2[k+\sigma(t)]}}\biggr\rbrace P_\phi,
\label{GM16} 
\end{eqnarray}
in geometrical units ($G=c=1$), 
where $\sigma(t)$, $R(t)$, and $k$ are dimensionless; and $m$ 
has unit of per length.  

Next, we want to find an expression for the covariant component 
of the azimuthal coordinate angular momentum 
$P_\phi$, in Eq.~(\ref{GM16}), for a test particle (or object)
 moving in spacetime as measured by
an arbitrary comoving observer.
Globally relative to the center of a rotating gravitational system, 
in general relativity, the covariant component 
of the azimuthal coordinate angular momentum 
$P_\phi$ of the energy-momentum four vector
of an object  of mass $M$ equals the component
of the angular momentum $L$ parallel to the symmetry axis.
So, we need the global angular momentum $L$ of an object (say, galaxy)
as measured by an arbitrary comoving observer at the proper
distance $r$.
The proper distance given by the vector $\bm{r}$ only measures 
the relative position vector, 
$\bm{r}=\bm{r}_{\rm gal}-\bm{r}_{\rm co}$, 
between the global position
 vector of the  galaxy, $\bm{r}_{\rm gal}$,  and
global position vector  of the observer, $\bm{r}_{\rm co}$, with 
respect to the global
``center'' of the Universe, in spherical coordinates.  In order to 
determine $L$ and, 
thus,  $P_\phi$, we need $\bm{r}_{\rm gal}=\bm{r}
+\bm{r}_{\rm co}$.
Since we do not yet, if ever,  know
$\bm{r}_{\rm co}$ (the arbitrary comoving observer's distance from
the global center), for simplicity
 we set $\bm{r}_{\rm co}=0$, assuming that the results, at least, 
qualitatively, will not change.  This means placing the 
G\"{o}del-Obukhov metric
[Eq.~(\ref{metric_sph})] at the global center of the Universe,  
in this particular case, for simplicity, 
allowing  us to derive $L$ relative to the global center
out to proper distance $r_{\rm gal}=r$.
This configuration appears permissible to give reasonable results 
since Eq.~(\ref{Exact_1}) applies to
the global system.
Now, in general, as measured by a comoving observer located at the 
global center 
\begin{equation}
\bm{L}=\bm{r}\bm{\times}\bm{p},
\label{ang_mom1}
\end{equation} 
where $\bm{p}=M\bm{v}$ is the global linear momentum in 
spherical coordinates; $\bm{v}$ is the linear velocity tangent 
to the 
trajectory of the mass $M$.  It can be shown
in spherical coordinates that
\begin{eqnarray}
\bm{v}&=& v_r{\bf \hat e}_r +v_\theta {\bf \hat e}_\theta
          + v_\phi{\bf \hat e}_\phi \nonumber\\
      &=&\dot{r} {\bf \hat e}_r - r\sin\theta\dot{\phi} 
{\bf \hat e}_\phi,
\label{velocity}
\end{eqnarray}  
where the dot represents differentiation with respect to time,
and $v_\theta=r\dot{\theta} =0$, consistent with cosmic rotation 
about the global symmetry axis.  
Therefore, upon substitution and evaluation of the cross 
product in Eq.~(\ref{ang_mom1}), we find that 
\begin{equation}
\bm{L}=M  \omega_{\rm rot}r^2 \sin\theta{\bf \hat e}_\theta.
\label{ang_mom2}
\end{equation}
We immediately identify Eq.~(\ref{ang_mom2}) as the 
component of the angular momentum along the global $z$-axis
(i.e., parallel to the symmetry axis), with magnitude
\begin{equation}
{L}=M  \omega_{\rm rot}r^2 \sin\theta=P_\phi.
\label{ang_mom3}
\end{equation}

Finally, upon substitution of Eq.~(\ref{ang_mom3}), 
into Eq.~(\ref{GM16}),  the cosmic GM radial force
acting on the galaxy of mass $M$, moving with angular velocity
$\omega_{\rm rot}$, at the distance $r$, as 
measured by a comoving observer over time,
expressed in non-geometrical units, 
becomes
\begin{widetext}
\begin{eqnarray}
(F_{\rm GM})_r&\sim&\biggl\lbrace{\frac{k\sigma(t)}
{[k+\sigma(t)]^3}}\biggr\rbrace^{1/2}
\biggl[{\frac{(m/c) r\sin\theta+1}{R(t) {\rm e}^{(m/c) r
\sin\theta}}}
\biggr]
\biggl\lbrace 1-{\frac{\sigma(t)+2k}
 {2[k+\sigma(t)]}}\biggr\rbrace M\omega_{\rm rot}c \nonumber\\
&\sim&\biggl\lbrace{\frac{k\sigma^3(t)}
{4[k+\sigma(t)]^5}}\biggr\rbrace^{1/2}
\biggl[{\frac{(m/c) r\sin\theta+1}{R(t) {\rm e}^{(m/c) r
\sin\theta}}} 
\biggr] M\omega_{\rm rot}c,
\label{GM17} 
\end{eqnarray}
\end{widetext}
repulsive, i.e., $(F_{\rm GM})_r> 0$, 
where, again, $\omega_{\rm rot}$ is the magnitude of the
global cosmic angular (or rotational) velocity, and $m$ 
is given by Eq.~(\ref{metric_2}).
This derived GM force due to frame dragging behaves similar
to the torsion term in Eq.~(\ref{Exact_1}) (the third term on the 
right-hand side) of which the final
state, as observed or predicted by Obukhov \cite{Obukhov2000},  can
either accelerate or prevent cosmological collapse.
This will be discussed further in the
following sections, where
we shall see in
Secs.~\ref{sec:4} and~\ref{sec:5.5}, how Eq.~(\ref{GM17}),
expressed as the GM force
per unit mass [Eq.~(\ref{NMR1})] and subsequently expressed 
as the GM force per unit mass per unit length  
[Eq.~(\ref{D13})],
can be compared to the torsion term in the equation of motion of
 the cosmic scale factor [Eq.~(\ref{Exact_1})].

\subsection{The Cosmic ``Gravitoelectric'' (GE) Acceleration}
\label{sec:3.3}

We now calculate the familiar or usual cosmic gravitational 
acceleration or
the GE force per unit of mass
throughout an
assumed axisymmetrical expanding  universe (of
infinite extent relative to an arbitrary comoving observer).
The magnitude of this negative GE acceleration will be compared
to the magnitude of the positive GM acceleration of 
Eq.~(\ref{GM17})
over time to see if and when acceleration of the
cosmic expansion occurs.
We will assume that the
scale factor of the
FLRW cosmological model
is still at least approximately valid in the G\"{o}del-Obukhov
cosmology. Support of this assumption is that the rotating and
expanding G\"{o}del-Obukhov metric [Eq.~(\ref{metric_cc})],
in the limit of large times and
nearby distances, reduces to the open metric of Friedmann
\cite{Carneiro2002}.  Moreover, we will also assume that the derivation
of the GE force per unit mass using
spherical axisymmetric comoving coordinates is not much
different from the FLRW cosmology using
 spherical symmetric comoving coordinates.
The result will allow us to test
the validity of this assumption, once the exact GE term can
be identified in Eq.~(\ref{Exact_1}).

The gravitational potential inside the Universe is assumed to be 
given by the post-Newtonian approximation, 
 \begin{eqnarray}
\Phi(\bm{r}) \approx - G\int {\rm d}^3r^\prime 
{\frac{T^{00}(\bm{r}^{\prime})}
{\vert\bm{r}-\bm{r}^{\prime}\vert}},
\label{GE1} 
\end{eqnarray}
for a system of particles (or galaxies) that are bound together
by their mutual gravitational attraction,
where, $\bm{r}-\bm{r}^{\prime}$ is the relative position
vector of the ``source'' point $\bm{r}^{\prime}$ with 
respect to the ``field'' point $\bm{r}$ between 
comoving arbitrary 
observers; and where
\begin{equation}
T^{00}=\sum\limits_{n} m_n\delta^3(\bm{r}-\bm{r}^{\prime})
\label{GE2} 
\end{equation}
for a gravitational bound system of masses $m_n$.
The component $T^{00}$ is the rest-mass density, or commonly referred 
to as the mass density, of the energy-momentum
tensor, $T^{\mu\nu}$, which serves as the source of the gravitational
field.  For nonrelativistic matter Eq.~(\ref{GE2}) 
can be set equal to
the mass density $\rho(\bm{r}^{\prime})$.
Again, the G\"{o}del-Obukhov spacetime metric has spatial homogeneity, 
and isotropy in 
the CMB radiation only, i.e., no spatial isotropy (as one would 
expect in a rotating universe).
In the FLRW model, the
Cosmological Principle of spatial homogeneity and spatial isotropy is 
assumed, which is 
consistent with CMB temperature measurements (save for the puzzling 
anomalies found 
in the Wilkinson Microwave Anisotropy Probe temperature maps 
that are not expected from gaussian 
fluctuations \cite{Chiang2003, 
deOliveira-Costa2004}), and consistent with 
large-scale structure observations (save for the large-scale 
asymmetries that are equally unexpected in an isotropic, homogeneous
space \cite{Eriken2004, Land2005}).  
These measurements and observations 
confirm however,  to a strong degree, the Cosmological 
Principle.  Perhaps the small 
anomalies and asymmetries are effects
predicted by cosmic rotation and do not conform to the standard 
FLRW cosmological model. 

Not considering the topology \cite{Liddle2003} of the Universe,
it seems reasonable to assume that at any given time the
causally-connected observable Universe, $r=r_H$ (Sec.~\ref{sec:5.2}),
surrounding an arbitrary comoving observer can
be represented  by a sphere of homogeneous expanding medium
of average mass
density $\rho=\rho(t)\equiv\rho_t$, such that for spacetime expanding
from a Big Bang origin,
the distance the Universe has expanded from its initial ``point''
(or state) is equal to the coordinate separation between galaxies (or
protogalaxies). The above reasoning allows use of Eq.~(\ref{GE1})
 to derive the GE acceleration (i.e., the familiar
attractive gravitational acceleration
experienced by all galaxies throughout a spacetime-expanding
universe independent of cosmic rotation).  The requirement
must be that
for a ``freely falling,'' locally flat spacetime 
observer this gravitational
acceleration is approximately zero, according to the Equivalence
Principle.
So, for a point inside a sphere of radii $r\leq r_H$, with the average
(or uniform) mass density existing throughout the Universe, 
$\rho_t$,
 for any given epoch, Eq.~(\ref{GE1}) yields
\begin{eqnarray}
\Phi(\bm{r})&\approx&-4\pi G\int_0^{r_H} {\frac{\rho(\bm{r}^{\prime}) }
{\vert\bm{r}-\bm{r}^{\prime}\vert}}{r}^{\prime 2}
dr^{\prime}\nonumber\\
&\approx& -{\frac{4\pi G\rho_t}{r}} \int_0^r  \bm{r}^{\prime 2}
dr^{\prime} - 4\pi G\rho_t \int_r^{r_H} \bm{r}^{\prime}
dr^{\prime} \nonumber\\
&\approx&-{\frac{2\pi G\rho_t}{3}}(3r_H^2-r^2),
\label{GE3}
\end{eqnarray}
where for the field  outside the source we set 
$\vert\bm{r}-\bm{r}^{\prime}\vert=\vert\bm{r}\vert$ 
(i.e., $\bm{r}^{\prime}=0$) for $\bm{r}^{\prime}<\bm{r}$ 
and for the field inside the source we set
 $\vert\bm{r}-\bm{r}^{\prime}\vert=
\vert -\bm{r}^\prime\vert$ ~(i.e., $\bm{r}=0$)
for $\bm{r}<\bm{r}^{\prime}$
in the first and second integrals, respectively.
Then from the 
relationship between the gravitational force per 
unit mass or acceleration $\bm{g}$ and the gravitational potential 
$\Phi$, $\bm{g}=-\bm{\nabla}\Phi$, 
requiring $\bm{\nabla\times g}=0$, Eq.~(\ref{GE3}) gives 
the radial 
component of the gravitational (i.e., GE) acceleration
\begin{eqnarray}
(g_{\rm GE})_r &\approx& -{\frac{4}{3}}\pi G \rho r,
\label{NMR2} 
\end{eqnarray}
  assumed to be
that measured by an arbitrary 
comoving observer
at a coordinate separation distance $r$.
 With 
$r$ being a measure of spacetime separation, it can be 
identified as the same $r$ as that in the spacetime metric 
of Eq.~(\ref{metric_sph}). 
 Equation~(\ref{NMR2}) satisfies the
above requirement that the gravitational acceleration goes to
zero as $r\longrightarrow 0$, as measured by the comoving local
inertial spacetime observer. 
The validation of the above reasoning
used in deriving
Eq.~(\ref{NMR2}) will be given in the following section.
Importantly, we shall see that $\rho_t$ is just the mass density
in the equation of motion of the scale factor in the standard
FLRW cosmology when $\Lambda=p=0$ [see Eq.~(\ref{GE6})].
Notice that the GE acceleration or force per unit mass given by
Eq.~(\ref{NMR2}) is negative and opposite the sign of the
radial component of the GM force
given by  Eq.~(\ref{GM17}).  The
repulsive nature of the GM force is consistent
with it acting to accelerate the cosmic expansion of the
Universe.  To test this claim of consistency, in
Secs.~\ref{sec:4} and~\ref{sec:5}, we will compare the magnitudes
of the accelerations produced by the GM and GE forces at
redshift $z\sim 0.5$, to see which is dominant.

To summarize, we are assuming that the Universe can be represented
locally by a spherical axisymmetric cosmology 
[Eq.~(\ref{metric_sph})].  The force per unit of mass 
$\bm{g}_{\rm GE}$
of Eq.~(\ref{NMR2}) expresses the gravitational 
acceleration (i.e., deceleration), due to the average mass 
density $\rho_t$, acting on say a galaxy  of mass $M$ at 
a distant $r$, as measured by an arbitrary comoving observer,
where, for this observer,  $r\longrightarrow  0$, which means 
that $\bm{g}_{\rm GE}
\longrightarrow  0$, as it should locally, in accordance with the 
Equivalence Principle, and, therefore, satisfying the requirement above.  
Notice, however, the same is not true
for $(F_{\rm GM})_r$ of Eq.~(\ref{GM17}), i.e., 
$\bm{F}_{\rm GM}$ does not go to zero at the observer, where
$r\longrightarrow  0$, because $\bm{F}_{\rm GM}$ exerts a force  
on {\it moving} inertial frames; then only if 
$\omega_{\rm rot}
\longrightarrow  0$ will $\bm{F}_{\rm GM}\longrightarrow  0$.
In other words, the GM force in general acts on the momentum
of a test particle (or galaxy) 
in a rotating frame [compare Eq.~(\ref{GM1})].  
 
\subsection{The Density of the Universe}
\label{sec:3.4}

We now derive an expression for the mass density $\rho(t)$ of
the Universe, which includes any contribution from radiation. 
We assume that the standard
FLRW cosmological model is
approximately correct.  The 
Friedmann-Lema\^{i}tre's solutions to Einstein's gravitational field 
equations yield the following acceleration equation for the cosmic
scale factor $R(t)$:
\begin{equation}
{\frac{\ddot R}{R}}={\frac{\Lambda }{3}}-{\frac{4\pi G}{3}}
\biggl( \rho + {\frac{3p}{c^2}}\biggr),
\label{GE6} 
\end{equation}
where the Robertson-Walker metric was used and 
$k~ (\text{spatial curvature index})=0$. Then the 
general expression for the time-dependent critical mass density is
given by
\begin{equation}
\rho_c(t)={\frac{3 q H^2(t)}{4\pi G}},
\label{GE7} 
\end{equation}
with $\Lambda=p=0$,  implying specifically a
Friedmann cosmology, where  
we have used Eqs.~(\ref{dec}) and~(\ref{hubble}); $\rho_c$ is
the density needed to make the Universe flat.
Note, the way in which $\rho_c$ of Eq.~(\ref{GE7}) was derived,
from the standard FLRW cosmology [Eq.~(\ref{GE6})], does not
rule out contribution from  dark energy and its relation
to gravity, but only
sets  $\Lambda =0$, for a matter
dominated ($p=0$) universe.  This does, however, suggests 
that the standard FLRW cosmological model cannot
adequately account for the presence of dark energy.
Thus, the critical mass density $\rho_c$
in terms of the measured present epoch cosmological 
parameters, is given by
\begin{equation}
\rho(t=t_0)_c\equiv \rho_c(t_0)={\frac{3 q_0 H_0^2 }{4\pi G}}, 
\label{GE8} 
\end{equation}
where we  find that
\begin{equation}
\rho_c(t_0)\approx 9.5\times 10^{-30}~{\rm {g}\,cm^{-3}}
\label{GE9} 
\end{equation}
for  the currently suggested  values of 
$H_0 \simeq 71$~$\rm km \,s^{-1}\,Mpc^{-1}$ 
and $q_0\simeq{1/2}$.  
This value of deceleration
parameter $q_0$ indicates a flat universe, which implies that
$\Omega\equiv {\rho(t_0)/\rho_c(t_0)}=1$, consistent with
observational data finding that $\Omega\simeq 1$
\cite{Tegmark2004, Jarosik2011}, if we assume that 
$\rho\equiv\rho_{\rm mat}+\rho_{\Lambda}$; then division 
by $\rho_c$ yields $\Omega=\Omega_{\rm mat}+\Omega_{\Lambda}\simeq 1$,
where observations suggest that $\Omega_{\rm mat}\simeq 0.27$ and
 $\Omega_{\Lambda}\simeq 0.73$, requiring  
$\Lambda\neq 0$ and $p<0$ in the
standard FLRW cosmology, specifically defining  the $\Lambda$CDM
cosmological model.  This model, however, fails to tell the true 
nature of so-called dark energy,  
leaving the subject open to speculation.  Nevertheless,
Eq.~(\ref{GE7}) can be identified as the source of the
universal gravitational field of attraction and 
 will be used in the GE acceleration given by
Eq.~(\ref{NMR2}), which 
gives the GE force per unit of mass for
different epochs, where evaluated using Eq.~(\ref{GE8}) gives the
present strength.

Now we return to give validity to the expression for the 
attractive universal 
gravitational force per unit mass
  [Eq.~(\ref{NMR2})], 
and, thus, to the reasoning that led to its derivation.  
Upon substitution of
the critical density of the Universe [Eq.~(\ref{GE7})]
 into Eq.~(\ref{NMR2}), we obtain the GE acceleration 
[i.e., the gravitational
force per unit mass $(g_{\rm GE})_r {\bf \hat e}_r=\ddot{\bm{r}}\,$].
We can express this as a deceleration of the scale factor $R$
by dividing through by $\bm{r}$, the proper distance: 
\begin{eqnarray}
  \ddot{\bm{r}}&\approx &- qH^2\bm{r},\nonumber\\
 {\frac{\ddot{R}\bm{\chi}}{R\bm{\chi}}}&\approx&-q H^2,\nonumber\\
       {\frac{\ddot{R}}{R}}&\approx&-qH^2,
\label{GE10}
\end{eqnarray}
which is,  as would be expected, the same as that of
the standard FLRW cosmology,
when $\Lambda =p=0$ and Eq.~(\ref{GE7}) is substituted 
into Eq.~(\ref{GE6}),
where  
\begin{equation}
\bm{r}(t)=R(t)\bm{\chi}, 
\label{GE11}
\end{equation}
relating the physical distance $\bm{r}$ to
the comoving coordinate distance $\bm{\chi}$, and its 
derivatives have been used. 
 The vector $\bm{\chi}$ 
comoves with the cosmic expansion. One can think of 
Eq.~(\ref{GE11}) as a coordinate grid  which expands with time.  
Galaxies remain at fixed locations in the  
$\bm{\chi}$ coordinate system.  
The scale factor $R(t)$ then tells how physical separations are growing 
with time, since the coordinate distances 
$\bm{\chi}$ are by definition fixed.  Further, solving for 
$q$, we can identify Eq.~(\ref{GE10}) as 
that of Eq.~(\ref{dec}), with $H$ given by 
Eq.~(\ref{hubble}), i.e., we identify the 
deceleration parameter as defined in the standard FLRW model. 
Again, this 
is what one would expect for the behavior of the GE acceleration of 
Eq.~(\ref{NMR2}), as it relates to the standard 
model, and, thus, this can serve to validate the 
reasoning behind assumptions 
made in its derivation.  
Importantly, note, Eq.~(\ref{GE10}) is
exactly equal to the first term on the right-hand side of
Eq.~(\ref{Exact_1}) with $q=1$.  Therefore, this term can 
be identified
as the GE deceleration of the scale factor in the G\"{o}del-Obukhov 
spacetime (we will return to this discussion in Sec.~\ref{sec:5.4}).
So, in summary, the validity of the derivation leading to
Eq.~(\ref{NMR2}), which can be used to express the GE 
acceleration  approximately in both the G\"{o}del-Obukhov and FLRW 
cosmologies,  has been established. 
The assumption that the 
derivation of the GE acceleration for the spherical axisymmetric 
case is not 
much different from that of the spherical symmetric (FLRW) case
has been validated, at least qualitatively; and it appears from
Eqs.~(\ref{Exact_1}) and~(\ref{GE10}) that the strengths 
will differ quantitatively by a 
factor of $q$.

\subsection{Cosmological Parameters}
\label{sec:3.5}

For a qualitative and somewhat quantitative analysis of the model 
described in this paper we
choose the following parameters of Eqs.~(\ref{metric_1}), 
(\ref{metric_2}), and~(\ref{metric_sph}): $\sigma $, 
$m$,  $k$, and 
$\omega_{\rm rot}$, based on observations and, of course, 
on theoretical 
insight.  A possible way to express evolution of 
the force $\bm{F}_{\rm GM}$
of Eq.~(\ref{GM17})  over time is 
to let
\begin{equation}
\sigma\equiv\sigma(t)\equiv {\rm e}^{c_1 t/ t_0},   
\label{CP1} 
\end{equation}
and let $k$ be defined as a function of  $\sigma(t)$ by 
Obukhov's \cite{Obukhov2000}  model relation
\begin{equation}
k=c_2 \sigma(t),   
\label{CP2} 
\end{equation}
where, when estimated from the G\"{o}del-Obukhov metric 
and general relativity, $c_2\approx 71$, using  $q_0=0.01$,
$(\omega_{\rm rot})_0= 0.1 H_0$,  
$H_0\sim 70$ ~$\rm km\, s^{-1}\,Mpc^{-1}$ (again, the $0$ subscripts
indicate the present epoch), with $\omega_{\rm rot}=
\omega_{\rm rot}(t)\equiv
(\omega_{\rm rot})_t$.
Note, although Obukhov \cite{Obukhov1990} 
and Korotkii and Obukhov \cite{Korotkii1991}  formulated
 the G\"{o}del-type metric of Eq.~(\ref{metric_cc}) for a 
constant value of the unknown 
parameter $\sigma$, Carneiro \cite{Carneiro2002}
 claims that  Eq.~(\ref{metric_cc}) 
remains valid 
when $\sigma$ is a function of time; and Obukhov \cite{Obukhov2014}
stated that formally, i.e., in essence or correctly,
it can be considered as a function of time.  Therefore,
 we will assume  that $\sigma$  is a function of time. 
Then we must further assume that the deviation from isotropy of the CMB 
radiation and the existence of parallax effects will
continue to be negligible, thus, 
making the so-called G\"{o}del-Obukhov metric at least approximately 
valid, which sounds reasonable \cite{Obukhov2014}. 
Proof of the latter above assumption is beyond the scope of this 
present manuscript. That is, validation of this
assumption must await an 
analysis of constraints imposed by observations of anisotropy in CMB 
radiation and parallax effects on the parameters of a rotating and 
expanding shear-free universe. To date it appears that no such study has
been done \cite{Obukhov2014}.
However, the choice of $\sigma$  [Eq.~(59)] can  
be validated theoretically as we shall see in Sec.~\ref{sec:5.5}.
So, it follows that, in these present calculations we are assuming that
$\sigma~[\equiv\sigma(t)]$  is
constant only for a specific epoch or hypersurface (where
$t=\text{constant}$), like the Hubble
parameter $H(t)$ and the scale factor R(t), for example.  Recall, the 
parameter $\sigma$ determines
the magnitude of acceleration of a fluid element due to
rotation of the Universe. 
So it is not unreasonable to expect $\sigma$ to 
be a function of cosmic time.

Now, reasonable choices for the constants $c_1$ and $c_2$ 
appear to be as follows: $c_1\approx -115$.
The validity of this choice is confirmed in Sec.~\ref{sec:5.5}.
The value of $c_1$ is related to the magnitude of the force
$\bm{F}_{\rm GM}$ of Eq.~(\ref{GM17});
for example, upon changing from
$c_1=-105$ to $c_1=-115$, in these model calculations, the magnitude
of the force, for a typical case, increases
by about two orders of magnitude.
We chose to
use the value  $c_2\approx 71$, like that of \cite{Obukhov2000}
since making it relatively
larger or smaller appears to have little effect on the model
outcome.
Subsequently, the chosen expressions for $\sigma$ and $k$ are approximately
within the limit of negligible or some small large-scale
spatial anisotropy
(see Ref.~\cite{Jain2007}).
Note, at $t=0$, with such choices above,
$\sigma(t=0)=1$ and $k\approx 71$,
consistent with the $k\ge 0$
requirement for causality.

Moreover, concerning the derivation of Eq.~(\ref{Exact_1}), we will
assume that the additional terms with derivatives with respect to
time of the unknown
parameter $\sigma(t)$
in the gravitational field equations are trivial when $\sigma(t)$
and $k$ are defined in terms of the parameters 
used in this present
manuscript [see Eqs.~(\ref{CP1}) and~(\ref{CP2})].
Details of the validation of this assumption of triviality 
can be found in Appendix.
This validation includes the following:
\begin{enumerate}
\item In the local Lorentz connection 
$ \tilde{\Gamma}_{b\mu}^a$ \cite{Obukhov2000},
used to derive the gravitational field equations, it is
shown that the first-time derivative of $\sigma$  
reduces to a trivial constant term 
that goes to zero when the time derivative
is taken in the Riemann-Christoffel curvature
tensor [Eq.~(\ref{Riemann})] and its associated Ricci tensor
[Eq.~(\ref{Ricci_tensor})].
This means that the derivatives of $\sigma$ cannot produce 
an acceleration (or force) over time that would  affect  the 
expansion rate like $\ddot R$ does in the
equation of motion of the scale factor [Eq.~(\ref{Exact_1})].
In fact, there will be no time derivatives
of the parameter $\sigma$ in the gravitational field equations.
\item The energy-momentum tensor of \cite{ Obukhov2000}
does not contain derivatives of the components $g_{\mu\nu}$
of  the spacetime metric Eq.~(\ref{metric_cc}); therefore, the 
energy-momentum will be the same for $\sigma=\sigma(t)$
and $\sigma=\rm constant$.
\end{enumerate}
In addition, the validation of our choice of $\sigma$ can be
found in Sec.~\ref{sec:5.5}.

Next, we will assume an analytical expression [Eq.~(\ref{NMR7})]
consistent with the more recent estimate
 for the ratio of the
magnitude of cosmic rotation to the Hubble constant $H_0$, where
observations of anisotropy in electromagnetic
propagation from distant radio sources, expected typically 
of cosmic
rotation, are used to determine the estimate given below
\cite{Obukhov1997, Nodland1997a, Obukhov2000}:
\begin{eqnarray}
{\frac{(\omega_{\rm rot})_0}{H_0}}=6.5\pm 0.5,
\label{CP3} 
\end{eqnarray}
with galactic coordinate direction
$l=50^\circ\pm 20^\circ$, $b=-30^\circ\pm 25^\circ$.
This value is larger than a previous estimate \cite{Obukhov1992}:
\begin{eqnarray}
{\frac{(\omega_{\rm rot})_0}{H_0}}=1.8\pm 0.8,
\label{CP4} 
\end{eqnarray}
with direction 
$l=295^\circ\pm 25^\circ$, $b=24^\circ\pm 20^\circ$,
obtained from Birch's \cite{Birch1982_1983} 
data. Moreover, recent analysis of the large-scale 
distribution of galaxies
\cite{Broadhurst1990}
has revealed an apparently periodic structure of 
the number of sources as a function of red shift. From this we get
yet
another estimate of the rotational velocity which appears
necessary to produced this observed  periodicity effect.
This estimate gives the 
largest  ratio of the three \cite{Korotky1994,Obukhov2000}:
\begin{eqnarray}
{\frac{(\omega_{\rm rot})_0}{H_0}}\approx 74
\label{CP5} 
\end{eqnarray}
[compare Eqs.~(\ref{CP3}) and~(\ref{CP4})].
It is clear from above that further careful observations and 
statistical analyzes will
be extremely important in overcoming the inconsistencies, in 
establishing the true value of the cosmic 
rotation (or vorticity), which may result from too few 
empirical data.

Next, we use the scale factor $R(t)$ to relate the
 Hubble parameter  with  time.
In general, with the usually power-law solution
for the scale factor as a function of time ($R\propto t^n$)
according to the
FLRW cosmological model, assumed to be applicable here 
(Sec.~\ref{sec:3.3}),
\begin{equation}
R(t)\equiv \Biggl({\frac{t}{t_0}}\Biggr)^{n},
\label{CP6} 
\end{equation}
normalized at the present epoch $t=t_0$.
Using Eqs.~(\ref{dec}), (\ref{hubble}), and~(\ref{CP6}),
 we get the general
expressions
\begin{equation}
H_t = n^\prime t^{-1}, 
\label{CP6_1}
\end{equation}
and 
\begin{equation}
q=-{\frac{(n-1)}{n}},
\label{CP6_2}
\end{equation}
where, in Eq.~(\ref{CP6_1}), $n\equiv n^\prime$,
which 
gives for $n^\prime=n={2/ 3}$ an age of the Universe 
($\simeq 9.2\times 10^9$~yr)
 too low to be consistent with recent
 observational estimates of $H_0\simeq 71~{\rm km\,s^{-1}\,Mpc^{-1}}$,
with $q={1/ 2}$ according to Eq.~(\ref{CP6_2}).
On the other hand, the expression $H_t=t^{-1}$, with $n^\prime=1$, 
gives an 
age ($13.80\times 10^9$~yr), which is consistent with recent 
observational estimates, with cosmic acceleration \cite{Lineweaver1999}, 
and without acceleration, in the absence of  
deceleration \cite{Freedman2001}. So, it 
seems reasonable
to assume the following limits for the present age $t_0$:
\begin{eqnarray}
{\frac{2}{3H_0}}< t_0\alt {\frac{1}{H_0}},
\label{CP7} 
\end{eqnarray}
i.e., ${2/ 3}<n^\prime\alt 1$.  Note, with $n=1$, 
according to Eq.~(\ref{CP6_2}) $q=0$, 
implying an open universe in the standard FLRW cosmology 
[compare Eq.~(\ref{dec})]. 
Specifically, for concreteness, it appears appropriate to choose
$t_0\equiv H_0^{-1}$ ($n^\prime =1$) for the present epoch, but 
with $n={2/ 3}$ in Eq.~(\ref{CP6}).  Note, 
with observations suggesting that the age of the Universe is closer
to the Hubble time ($H_0^{-1}$), instead of that given by the 
standard FLRW cosmological model [${(2/ 3)}H_0^{-1}$)] implies 
that the Universe has at least not decelerated continuously.
The discrepancies leading to the limits above 
can possibly be attributed to the 
evolution of $R(t)$, i.e., how it might
change as the Universe undergoes phase changes, thus reflecting how 
the value of $n$ might change, where $n={2/ 3}$, recall, is also 
the starting scale factor at  $t\sim 0$ of  the Einstein-Lema\^{i}tre
\cite{1931ab_Lang1980} expanding 
cosmological model.  

Moreover, for completion, reference, and review, during inflation 
(indicated by the 
subscript ``infl'') it appears that 
\begin{eqnarray}
{\frac{[R(t)]_{\rm f}}{[R(t)]_{\rm in}}}\approx {\rm e}^{H\int {\rm d}t}
\approx {\rm e}^{H_{\rm infl}\Delta{t}}
\label{CP8} 
\end{eqnarray}
(i.e., ${\rm e}^{\int H {\rm d}t}\approx {\rm e}^{H\int {\rm d}t}$),
where $[R(t)]_{\rm in}$ and $[R(t)]_{\rm f}$ are the 
initial (subscript ``in'') and final (subscript ``f'') scale factors 
before and after inflation; 
$H_{\rm infl}=1/t_{\rm infl}$ is the Hubble parameter at 
the onset of inflation, which remains approximately constant 
during inflation; and
$\Delta{t}=t_{\rm f}-t_{\rm in}$, with $t_{\rm in}$ and
$t_{\rm f}$ indicating the beginning and ending
times of inflation,
respectively. For example, assuming that inflation occurs between
$10^{-36}~{\rm s} \alt t \alt 10^{-34}~{\rm s}$,
we find that the scale factor by which the Universe increased
during inflation is $[R(t)]_{\rm f} \sim {\rm e}^{99}
[R(t)]_{\rm in}\sim 10^{43} [R(t)]_{\rm in}$.  Now, 
whether or not inflation occurred as we know it or its origin, 
we do not know for 
certain, but we do know that the Universe,
early on in its history, appears to have  increased or inflated  
by a factor of  
$\sim 10^{43}$ from a small causally-connected  comoving region of 
spacetime $r_{\rm in}\sim 10^{-43}r_{\rm f}$, according to 
Eq.~(\ref{GE11}) and the above relationship 
between the initial and 
final scale factors.  
It appears a false vacuum or the release of  a
type of quantized-gravity binding-like  energy,
resulting from symmetry
braking of  at least three of the fundamental forces (strong,
gravitational,
electromagnetic), drove inflation (see also Sec.~\ref{sec:3.1}).
The details as to what initiated 
inflation are yet to be understood; at present we can only speculate. 
Nevertheless, and importantly, it appears that cosmic vorticity 
enhances the inflation, i.e.,
when the vorticity is large, the inflation rate is much bigger 
than in the vorticity-free case \cite{Obukhov2002}.

\section{Numerical Model Results}
\label{sec:4}

For comparison and completion, plotted in Fig.~1 is the cosmic scale 
factor $R(t)$. Figure~1(a) displays a schematic plot of  the  scale 
factor given by Eq.~(\ref{CP6}), with $n={1/ 2}$ or 
$ n={2/ 3} $, or given by Eq.~(\ref{CP8}) over a
period from when the age of the Universe was
$\sim 10^{-43}$~s to the present estimated age of 
$t_0=13.8\times 10^9$~yr.  The lower time limit corresponds 
to the Planck era. 
Immediately following the Planck era we believe that 
the Universe was  at least a thermal causally-connected
spacetime  gaseous plasma.   We assume that  $ n={2/ 3}$ in 
Eq.~(\ref{CP6}) at $t\sim 10^{-43}$~s, with this value lasting up 
to the beginning of the inflationary phase at which $R(t)$ is given by
Eq.~(\ref{CP8}),
as indicated in Fig.~1(a). 
Here also we are assuming, as usually assumed in the standard model, 
that 
after inflation, during the radiation 
dominated era, from when the age of the Universe was $t\sim 10^{-34}$~s 
up to $t=t_{\rm eq}\sim 1.7\times 10^{12}$~s $\simeq 54,000$~yr, 
 $n={1/ 2}$ in Eq.~(\ref{CP6}), where  
$t_{\rm eq}$ is the time of matter and
radiation equality \cite{Liddle2003}.   Beyond $t_{\rm eq}$ we set
$n={2/ 3}$, indicating mass dominance, producing the
step-like feature clearly seen in Fig.~1(b) 
at $t=t_{\rm eq}\sim 4\times 10^{-6}\,t_0$.  Before this time 
relativistic particles dominated. 
As the Universe continues in a mass dominated phase after 
recombination, at $t\sim 350,000 $~yr after the Big Bang,
in Eq.~(\ref{CP6}) we still have $ n={2/ 3}$ up to the
present epoch. 
Note, this expression for $R(t)$ indicates a flat,
decelerating universe [i.e., $q={1/ 2}>0$,
using Eq.~(\ref{CP6_2})], as would be expected in a FLRW
expanding cosmology.
Yet, this is only somewhat consistent with observations,
because
recent observations appear to indicate an
open, accelerating universe, at least for the present epoch,
with $q<0$ according to the standard FLRW cosmological model.
This would cause $R(t)$ to have a somewhat steeper incline 
 (or slope)  near the present epoch than that displayed.
For example, for $q=-0.2$, Eq.~(\ref{CP6_2}) gives $n=1.25$.
Now, Fig.~1(b) displays $R(t)$ of Eq.~(\ref{CP6}) over the time 
($138~{\rm yr}\le t \le 13.8\times 10^9 ~{\rm yr}$) that the 
GM and the GE gravitational accelerations are calculated,
 as we shall see below.
The lower time limit is set here by the computational
capacity of the computer in the units used in calculating the
GM acceleration from Eq.~(\ref{GM17}). This limit will, however,
 be overcome in Sec.~\ref{sec:5} using an approximate analytic
expression.  
Note, the step-like feature is an ``artifact''
indicative of where $n = 1/2$ 
changes to n = 2/3, at
the radiation-mass equilibrium time
[compare Eq.~(\ref{CP6})].  Realistically, the change would be
more gradual.
\begin{figure*}
\includegraphics[width=10cm,height=16cm,angle=-90]{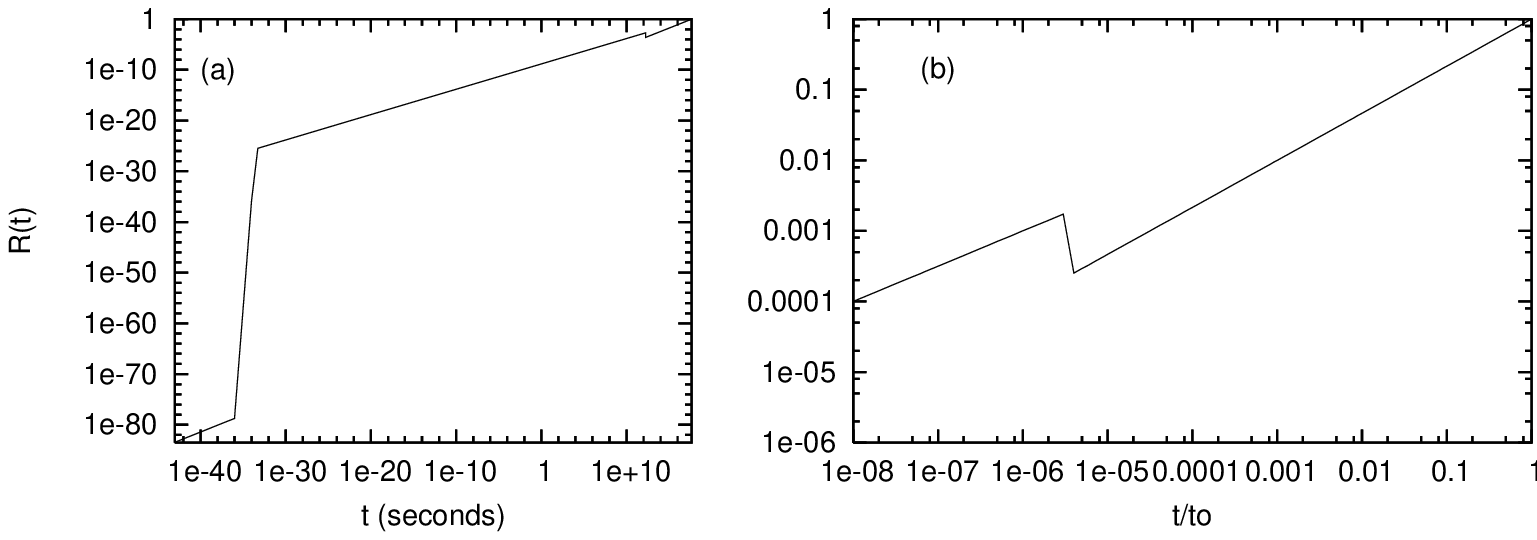}
\vspace{-4.85cm}
\caption{The scale factor $R(t)=(t/t_0)^n$ versus time up to
$t=t_0 =13.8\times 10^9$~yr, with $n=2/3$ or $n=1/2$. (a) Schematic
plot of $R(t)$ vs. $t$
in seconds, from the Planck time ($\sim 10^{-43}$~s) to $t_0$.
Inflation is indicated by
the steep rise in the curve at $10^{-36}~{\rm s}\le t\le 10^{-34}$~s,
where $R(t)$
increases by a factor $\sim {\rm e}^{99}$ from its initial value given by
$R(t)=(t/t_0)^n$, with $n=2/3$ before inflation (see text).
(b) $R(t)=(t/t_0)^n$ vs.
$t/t_0$ for $138~{\rm yr}\le t\le t_0$, where $t=10^{-8}t_0=138.0$~yr.  
The step-like feature
indicates where $n=1/2$ (just after inflation)  changes to $n=2/3$,
at the radiation-mass equilibrium time
$t_{\rm eq}\sim 1.7\times 10^{12}$~s
(see text); this can also be seen in (a).
}
\label{figone}
\end{figure*}

Displayed in Figs.~2(a) through~2(f) are the evolutions of the 
magnitudes of the cosmic gravitational 
accelerations (force per unit mass), $(g_{\rm GM})_r$ 
from Eq.~(\ref{GM17}) and $(g_{\rm GE})_r$ of 
Eq.~(\ref{NMR2}), versus $t/t_0$,
where, upon dividing Eq.~(\ref{GM17}) by $M$,
\begin{eqnarray}
(g_{\rm GM})_r&\sim &\biggl\lbrace{\frac{k\sigma^3(t)}
{4[k+\sigma(t)]^5}}\biggr\rbrace^{1/2}
\biggl[{\frac{(m/c) r\sin\theta+1}{R(t) {\rm e}^{(m/c) r
\sin\theta}}}
\biggr] 
\omega_{\rm rot}c. \nonumber\\
&&
\label{NMR1} 
\end{eqnarray}
In these calculations, we set $\theta={\pi/ 2}$ in 
Eq.~(\ref{NMR1}) for simplicity. 
The radial gravitational accelerations, $(g_{\rm GM})_r$ 
and $(g_{\rm GE})_r$, of Eqs.~(\ref{NMR1}) and~(\ref{NMR2}), 
respectively, are measured at a coordinate separation distance $r$ 
(corresponding to a particular redshift $z$) by a 
comoving observer, 
as this distance expands over time, while the 
gravitational accelerations at that  distance 
evolve over time, from
$t=10^{-8}t_0=138.0$~yr after the Big Bang to the present 
estimated age of the Universe:
$t_0=13.8\times 10^9$~yr (for $H_0=71$~$\rm km  \,s^{-1}
\,Mpc^{-1}$). 
Note, dividing $(g_{\rm GM})_r$  
and $(g_{\rm GE})_r$  
by $r$ convert these gravitational accelerations into an
acceleration of the scale factor ($\ddot R/R$), as done in 
Eqs.~(\ref{D13}) and~(\ref{GE10}), respectively.
\begin{figure*}
\includegraphics[width=14cm,height=17cm,angle=-90]{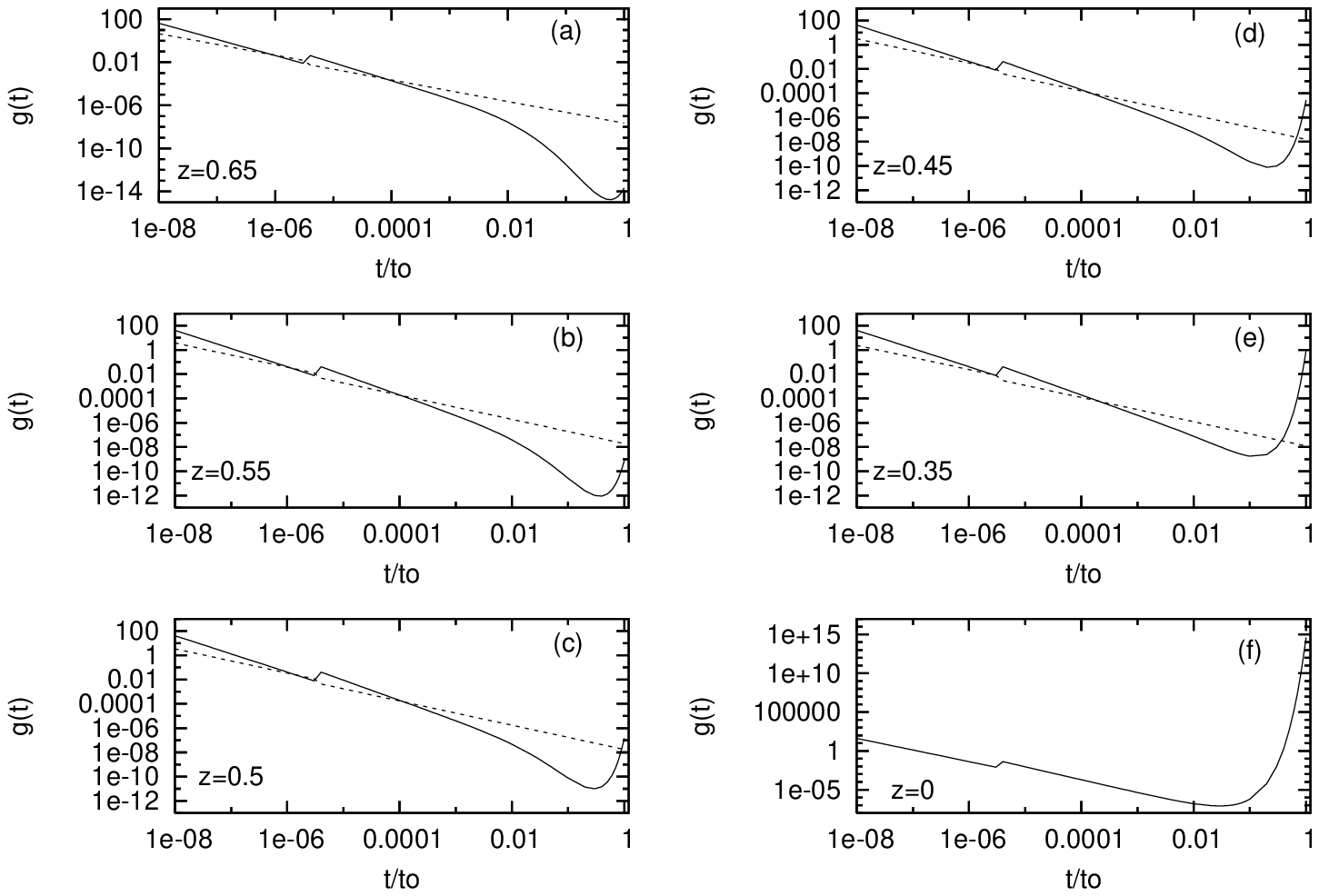}
\caption{The magnitudes of the radial accelerations
$(g_{\rm GM})_r$ (solid curve)
and $(g_{\rm GE})_r$ (short-dashed curve) produced by the
gravitomagnetic (GM) and
gravitoelectric (GE) forces, respectively, in cgs units, versus $t/t_0$,
representing the evolution from $t=138$~yr 
after the Big Bang up to
the present time $t=t_0=13.8\times 10^9$~yr:
(a) Evolution of accelerations at a
distance with $z=0.65$ (see text).
(b)  Evolution of accelerations at a distance with $z=0.55$.
(c)  Evolution of accelerations at a distance with
$z=0.5$.
(d)  Evolution of accelerations at
a distance with  $z=0.45$.
 (e)  Evolution of accelerations at a distance with $z=0.35$.
(f) Evolution of the accelerations at $z=0$, where
$(g_{\rm GE})_r\longrightarrow 0$.
Notice that $(g_{\rm GM})_r$ reaches a maximum finite magnitude at
$z=r=0$, as measured at the comoving observer (see text).
(Note, the step-like feature is an artifact of the computer
simulation, indicating the change from radiation dominance to mass
dominance at $t=t_{\rm eq}\sim 54,000$~yr, where $n=1/2$
changes to $n=2/3$.)
}
\label{figtwo}
\end{figure*}

In these calculations we step through the independent variable $t$
by assuming the following:
\begin{equation}
t= f_i t_0,
\label{NMR2a}
\end{equation}
where $f_i$ is the fraction of the total time that we step through, 
normalized to equal one 
 at the present epoch, i.e., $t=t_0$; and subscript $i$ indicates a step
 size.  The Hubble parameter then evolves as 
\begin{equation}
H(t) = {\frac{H_0}{f_i}}.
\label{NMR2b}
\end{equation}
The evolving mass density $\rho=\rho(t)$ of Eq.~(\ref{NMR2}) 
is thus given by Eq.~(\ref{GE7}) for
a flat universe according to the standard FLRW cosmology. 
Similarly, the evolving distance $r=r(t)$ is assumed to be given by 
\begin{eqnarray}
r(t)\simeq{\frac{c z}{H(t)}}={\frac{cz}{H_0}}f_i,
\label{NMR2c}
\end{eqnarray}
for recession velocities $\ll c$,  which is just the 
nonrelativistic Hubble law. 
Note, for $z> 1$, the relativistically corrected
Hubble law [Eq.~(\ref{D2})] must
be used for accuracy; this will be discussed further in 
Sec.~\ref{sec:5.2}.
  
The above evolving distance $r(t)$  is for a 
specific $z$, measured by a present-day observer, indicating
how a specific coordinate point in spacetime has evolved. 
Substitution of
the evolving variables: $r(t)$, $\sigma(t)$ [Eq.~(\ref{CP1})], 
$R(t)$ [Eq.~(\ref{CP6})], and $\rho(t)$ [Eq.~(\ref{GE7})] into  
$(g_{\rm GM})_r$ and $(g_{\rm GE})_r$ [Eqs.~(\ref{NMR1}) 
and~(\ref{NMR2}), respectively]  
allows us to see how these 
cosmic gravitational
accelerations have evolved at that specific comoving coordinate point, 
indicated by $z$, as measured by a present epoch observer.
Note, it shall be interesting to see what
happens to $(g_{\rm GM})_r$ as $t$ approaches zero and what 
role it may play in the cosmic inflationary era.  
In Sec.~\ref{sec:5}, we shall see what role, if any, it may play,
where we will
attempt to go back in time as far as theoretically 
possible using a valid
approximation to the GM acceleration of Eq.~(\ref{NMR1}).  
However, for the present, we find that for 
$z\longrightarrow 0$ (i.e., $r\longrightarrow  0$) 
in Eq.~(\ref{NMR1}), 
$(g_{\rm GM})_r$ reaches a finite
maximum value of $(g_{\rm GM})_r\sim 4
\times 10^{14}$~$\rm cm\, s^{-2}$
at the comoving observer as shown in Fig.~2(f).
Now, by Eq.~(\ref{NMR1}) and 
conservation of angular momentum, as $t$ decreases, in Fig.~2, 
for $t<0.01 \,t_0$ (or $< 1.38\times 10^8$~yr after the Big Bang),   
$(g_{\rm GM})_r$  increases with increasing $\omega_{\rm rot}$
as would be expected.  Yet, on the other hand, as $t$ increases,
for $t>0.01 \,t_0$, $(g_{\rm GM})_r$ first decreases as would be 
expected, but then as 
$\omega_{\rm rot}$ gets smaller and smaller, $(g_{\rm GM})_r$ once 
again increases, at least for the $z$ values shown 
(compare Fig.~2; see also Fig.~3). 
[The behavior
of $(g_{\rm GM})_r$ for larger values of $z$ will be discussed in
Sec.~\ref{sec:5.2}].  
The above
behavior of $(g_{\rm GM})_r$ appears to be  
consistent with the
third term on the right-hand side of
Eq.~(\ref{Exact_1}); and, importantly, in Fig.~2(c),
the magnitude of $(g_{\rm GM})_r$ overtakes that of
$(g_{\rm GE})_r$, indicating a net positive acceleration
 or repulsive force per unit mass.
  We shall return
to this discussion in the following section.
\begin{figure}
\centerline{\includegraphics[width=55mm,height=80mm,angle=-90.]
{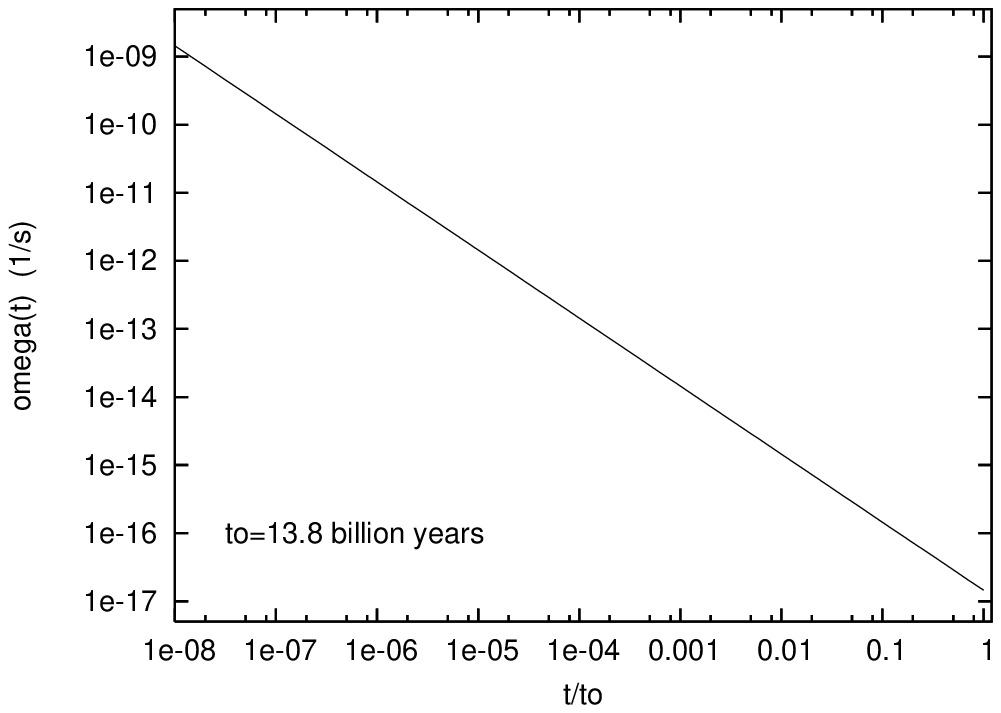}}
\caption{(a) Evolution of the magnitude of the 
cosmic rotational velocity
$(\omega_{\rm rot})_t$
over time (see text), from $t=138$~yr to the
present $t_0=13.8\times 10^9$~yr for $H_0=71$~$\rm km  \,s^{-1}
\,Mpc^{-1}$.
}
\label{figthree}
\end{figure}

Figure~3 displays 
how the cosmic rotational velocity decreases over time:
 Specifically plotted,
as we shall see below, is a derived analytical expression 
consistent with
Eq.~(\ref{CP3}).  We shall see that these model calculations 
suggest that the magnitude of the rotational 
velocity  $\omega_{\rm rot}$ has a value $\sim 6.3 H$, which
is consistent with present-day observations.
So,  in what follows, an analytical expression is derived for the 
cosmic rotational
velocity $\omega_{\rm rot}$ of Eq.~(\ref{NMR1}), whose
numerical value is consistence with observations.  As usual the
magnitude of the angular velocity (or the angular frequency) 
for circular motion is given by
\begin{eqnarray}
\omega&=&{\frac{{\rm d}\phi}{{\rm d}t}}\nonumber\\
& \approx&{\frac{2\pi}{t}},
\label{NMR3} 
\end{eqnarray}
assuming simple harmonic-like motion.
From Eq.~(\ref{NMR3}), it seems reasonable to 
 relate the Hubble parameter, the rate of cosmic
expansion, to $\omega_{\rm rot}$, the rate of cosmic rotation,
 by 
\begin{eqnarray}
\omega_{\rm rot}&\sim&2\pi H\nonumber\\
& \sim&6.3 H,
\label{NMR7} 
\end{eqnarray}
where we have use $ H\sim 
1/t$.
Importantly, this expression is consistence with observations;
 compare Eq.~(\ref{CP3}).
Equation~(\ref{NMR7}) is plotted in Fig.~3.

\section{Discussion}
\label{sec:5}

In the following sections, we analyze further and discuss the results 
above.  We will look at the long term behavior of the GM acceleration
$(g_{\rm GM})_r$ and the GE acceleration $(g_{\rm GE})_r$ 
over time.   Because 
of the large interval of time covered, we do this in two separate 
epochs for computational simplicity. 
However, since 
these calculations are computations of analytical equations, none of 
the physics
is lost because the values and times converge at the ``interface,''
indicative of a specific cosmological time. 
Note, assuming that spacetime torsion and spacetime
frame dragging (producing the GM acceleration) 
are one in the same (as we shall see in
Sec.~\ref{sec:5.5} that this is a valid assumption), in
this analysis,
we are essentially comparing the first term (GE acceleration)
and third term (torsion) on the right-hand
side of Eq.~(\ref{Exact_1}); see also Eq.~(\ref{GE10}).
The other terms of
Eq.~(\ref{Exact_1}), which
are compared in Williams \cite{Williams2014}, 
appear not to be important in the later
Universe where the recently observed 
acceleration of the expansion or so-called dark energy arises.
That is, although these other terms appear to be important in the
overall expansion
and deceleration of the early Universe and perhaps during inflation,
the torsion term overtakes these
terms in the later Universe
resulting in the acceleration of the expansion.
Included in Sec.~\ref{sec:5.5} is an analysis and discussion of 
the G\"{o}del-Obukhov associated cosmic magnetic field.

\subsection{The Gravitomagnetic (GM) and the Gravitoelectric (GE)  
Accelerations: From 138 Years to the Present}
\label{sec:5.1}

We first analyze the GM and the GE accelerations over the time
for which we have the exact analytical expression for the GM
acceleration [Eq.~(\ref{NMR1})], with the GE acceleration given by
Eq.~(\ref{NMR2}). As displayed in Fig.~2, this time is 
$138~{\rm yr}\le t \le 13.8\times 10^9 ~{\rm yr}$ after the Big
Bang.  Note, Eq.~(\ref{NMR1}) is referred to as the 
exact in comparison to the
approximation to this equation we will use in the following section to 
find its value in the  early Universe ($t< 138$~yr). 
As mentioned in Sec.~\ref{sec:4}, the evolution of the GM and GE 
accelerations over time at 
a distance $r(t)$
as measured by a  present epoch comoving 
observer for a specific $z$ [see Eq.~(\ref{NMR2c})] 
as related to the cosmological distance given by
Eq.~(\ref{GE11}) is plotted in Fig.~2. 
These model calculations show that the GM acceleration 
[Eq.~(\ref{NMR1})] goes to zero at $z\agt  0.8$ 
as measured by a comoving 
present  observer.  This implies that the Universe was in a
decelerating  phase for $z$ at least greater than
$\sim 0.8$, i.e., at an earlier
cosmic time; this also is consistent with observations
\cite{Riess2001}.    Figures~2(a) and~2(b) seem to show
that  the Universe starts to decelerate at a slower and then an even 
slower rate at $z=0.65$ and $0.55$, respectively.
This is because according to Fig.~2(a),
 $(g_{\rm GM})_r$ begins to increase at 
$\sim 8 \times 10^9$~yr 
after the Big Bang, from a minimum value $\sim 2\times 
10^{-15}~{\rm cm \,s^{-2}}$, at the spacetime coordinate point, $r$, 
associated with $z=0.65$ (as measured by a present epoch 
observer); and according to Fig.~2(b), $(g_{\rm GM})_r$ 
begins to increase at 
$\sim 6 \times 10^9$~yr  after the Big Bang, from a minimum 
value $\sim 9\times 
10^{-13}~{\rm cm\, s^{-2}}$, at the spacetime coordinate point 
associated with $z=0.55$.
Importantly, we see that Fig.~2(c) is consistent with recent
observations that suggest that the Universe entered into an accelerating
phase at $z\sim 0.5$, with $(g_{\rm GM})_r>\vert (g_{\rm GE})_r\vert$.
Figure~2(d)
shows how  as $z$ gets smaller, which means that the distance from 
the comoving observer is getting smaller, the GM acceleration 
gets larger and
 larger; and we find that, as $z\longrightarrow 0$, the
GM acceleration will continue to get larger  until a maximum 
value of $(g_{\rm GM})_r\sim 4
\times 10^{14}$~$\rm cm \,s^{-2}$  is reached at the 
comoving observer, as
mentioned in Sec.~\ref{sec:4}, and as can be seen in Fig.~2(f).
So, overall, and importantly, Fig.~2 is consistent with observations
that suggest the cosmic acceleration of the expansion
 started about $4.5\times 10^9$
years ago, i.e., at $z=0.46\pm 0.13$  \cite{Riess2004}.

\subsection{The GM and the GE
Accelerations at the Hubble Radius: 
From Time of Planck Scale through Inflation 
to the Present}
\label{sec:5.2}

Upon substitution of
Eq.~(\ref{NMR7}) into Eq.~(\ref{NMR1}), setting $r\approx 0$,
$\theta={\pi/ 2}$, and using model parameters defined in
Sec.~\ref{sec:3.5} [Eq.~(\ref{CP6}), $\sigma=\exp[{-115(t/ t_0)}]$,
$k=71\sigma$, $H_t\sim{1/t}$], we can derive an 
approximate analytical
expression for the GM cosmic acceleration that appears to be valid at
early times in the Universe as well as later times where the 
scale factor of Eq.~(\ref{CP6}) is normalized
at the present epoch [i.e., where $f_i$ of Eqs.~(\ref{NMR2a})
to~(\ref{NMR2c}) equals 1].  Thus, we find that
\begin{eqnarray}
 (g_{\rm GM})_r &\sim&\biggl\lbrace{\frac{k\sigma^3(t)}
{4[k+\sigma(t)]^5}}\biggr\rbrace^{1/2} {\frac{\omega_{\rm rot}c}
{R(t)}}
 \nonumber\\
&\sim& 9.6\times 10^{-5}\exp{\Biggl({\frac{57.5 t}{t_0}}\Biggr)}
{\frac{\omega_{\rm rot}c}{R(t)}}.
\label{D1} 
\end{eqnarray}
For $t=t_0$
(i.e., $H=H_0$), Eq.~(\ref{D1}) gives the same value 
as that calculated
using the exact analytical expression for the GM acceleration 
[Eq.~(\ref{NMR1})], 
measured at a comoving present epoch
observer: $(g_{\rm GM})_r\sim 4\times 10^{14}$~$\rm cm \,s^{-2}$, for
$H_0=71$~$\rm km \,s^{-1}\,Mpc^{-1}$, by letting $z\longrightarrow 0$ 
[i.e., $r\longrightarrow 0$; compare Eq.~(\ref{NMR2c}) and Fig.~2(f)].  
This appears to
validate our use of Eq.~(\ref{D1}) at earlier times for which 
$r\approx 0$ and $t\ne t_0$.
\begin{figure*}
\includegraphics[width=10cm,height=16cm,angle=-90]{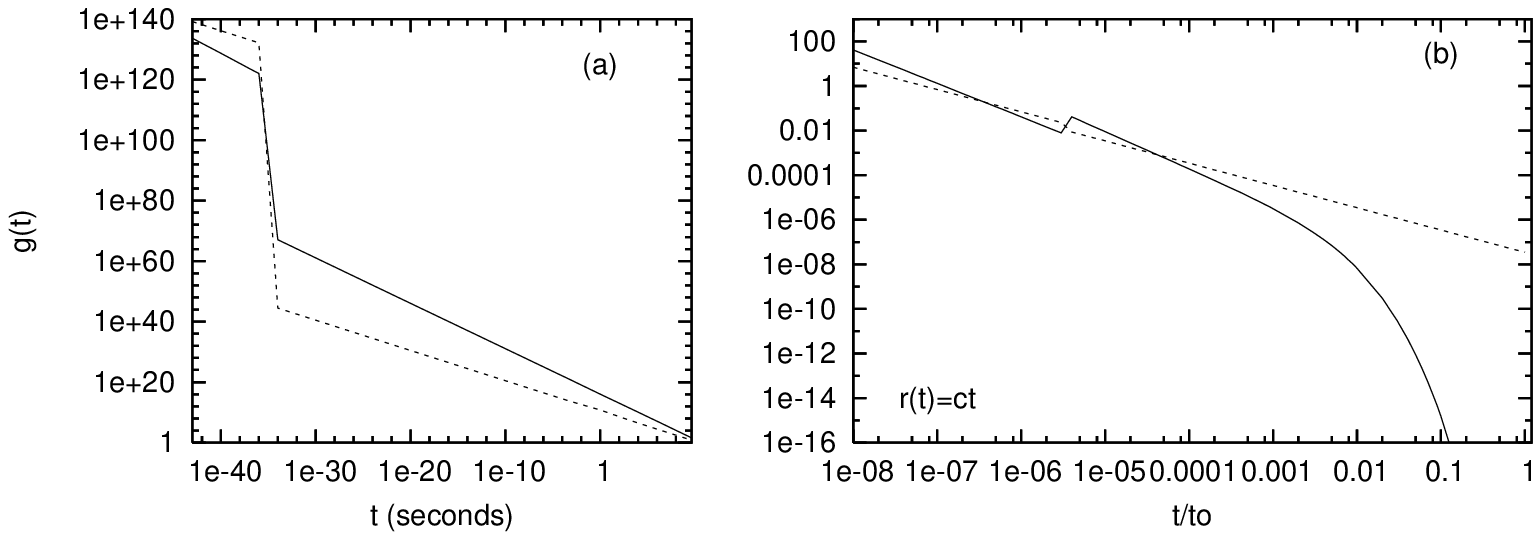}
\vspace{-4.85cm}
\caption{Magnitudes of gravitational accelerations,
 $(g_{\rm GM})_r$ and $(g_{\rm GE})_r$, over time:
(a) From the
Planck time ($t\simeq 5.4\times 10^{-44}$~s) up to $t\simeq
4.4\times 10^9$~s ($\simeq 138$~yr$=10^{-8} t_0$). Solid curve is for
$(g_{\rm GM})_r$, $n={2/3}$ before inflation at
$t\alt 10^{-36}$~s
and  $n=1/2$ after inflation at $t\agt 10^{-34}$ (see text);
short-dashed curve is for $(g_{\rm GE})_r$, $n=2/3 $
(implying $q={1/ 2}$)
before inflation and $n=1/2$ (implying $q=1$) after inflation;
$n={2/3}$ is for mass dominance and $n={1/2}$
is for radiation dominance (see text). (b) From
$t=10^{-8}\, t_0$ ($=138$~yr) up to 
$t_0=13.8\times 10^9$~yr. Solid
curve is for $(g_{\rm GM})_r$, short-dashed curve for
$ (g_{\rm GE})_r$,
evaluated  at the Hubble radius, being consistent
with (a). Not shown, $(g_{\rm GM})_r\longrightarrow 0$ at
$t\sim 0.8 \,t_0$.
 (Note, the step-like feature is an artifact; see note on Fig.~2.)
}
\label{figfour}
\end{figure*}

Displayed in Fig.~4 are the evolutions of the magnitudes of 
the GM and the GE accelerations $(g_{\rm GM})_r$  and
$(g_{\rm GE})_r$, respectively.
Plotted in Fig.~4(a) are the magnitudes of $(g_{\rm GM})_r$
[Eq.~(\ref{D1})]  and
$(g_{\rm GE})_r $ [Eq.~(\ref{NMR2})], from the Planck time
($t_P\simeq 5.4\times 10^{-44}$~s) up to
 $t\simeq 4.4 \times 10^9$~s
($\simeq 138$~yr $=10^{-8}\,t_0$).  
Note, for any given epoch, after 
inflation, the magnitudes of $(g_{\rm GM})_r$
and $(g_{\rm GE})_r $ are evaluated at the Hubble radius 
($r_H=c t\equiv cH^{-1}$), i.e.,
the limit of the causally-connected observable Universe for
any comoving observer.  Before inflation, we are assuming that 
this spacetime region of the Universe was $\sim 10^{-43}r_H$,
and, thus, in the quantum-gravity regime.
If we assume that $n={2/ 3}$ (implying mass dominance), 
 being consistent with
the Einstein-Lema\^{i}tre \cite{1931ab_Lang1980} 
early cosmology, then from the Planck time up to the time
of inflation ($\sim 10^{-36}$~s),  $(g_{\rm GM})_r$ goes from
about five to ten orders of magnitude, respectively, 
smaller  than	
$\vert(g_{\rm GE})_r\vert$; this can be seen in Fig.~4(a).
Yet, after inflation, with $n={1/2}$ (implying radiation dominance),  
$(g_{\rm GM})_r$ 
has decreased by a factor $\sim 10^{-43}$ according to 
Eq.~(\ref{D1}) and statements made in the last paragraph of
Sec.~\ref{sec:3.5} concerning the scale factor $R(t)$ and
proper distance $r$; 
$\vert(g_{\rm GE})_r\vert$ has decreased by a factor $\sim 10^{-86}$,
being now smaller than $(g_{\rm GM})_r$ by a factor $\sim 10^{-22}$, 
where we have used Eqs.~(\ref{NMR2}) and~(\ref{D21}).
(Note, this does not necessarily mean that the Universe 
would be in an accelerating phase, since it can be shown from these
calculations that the 
fourth term on the
right-hand side of Eq.~\ref{Exact_1} will be negative 
and its absolute value greater
than $(g_{\rm GM})_r$ at this point in spacetime, particularly
the acceleration term produced by $B$, where $B$ will be discussed in
Sec.~\ref{sec:5.5}.)
So, based on the above and Fig.~4(a),  
it appears that $(g_{\rm GM})_r$, resulting from spacetime
frame dragging (or torsion), does not directly
 cause inflation.  However, considering the fourth term on the 
right-hand side of Eq.~(\ref{Exact_1}) and the pressure 
of Eq.~(\ref{press1}) or~(\ref{press2}), for 
spin-torsion
cosmological coupling constant relations
 $4\lambda_3\gg\lambda_1$ and $\lambda_3>0$, as mentioned in 
Sec.~\ref{sec:3.1},
the net acceleration of the expansion might 
at least contribute to inflate
the initially very small quantum-gravity region of the Universe 
to the causally-connected region given by the
Hubble radius $r_H$, consistence with
what we assume to have
occurred
in our standard FLRW cosmology with inflation.
The above statement requires an investigation
to find out specifically the contributions from all the terms in 
Eq.~(\ref{Exact_1}).  This is investigated elsewhere
\cite{Williams2014}.
Now, notice in Fig.~4(a), even up to age
$t\sim 1$~s, the proposed limit for inflation to have occurred to
be consistent with nucleosynthesis  \cite{Liddle2003},
$(g_{\rm GM})_r$ is still
$\sim 10^{11}$ times larger than
$\vert(g_{\rm GE})_r\vert$.   Since, however, we want this
present model to be consistent with the standard cosmological
model
and with the assumption that $(g_{\rm GM})_r$ is related to
spacetime torsion,
we must keep in mind that the negative terms in the
last acceleration component
on the right-hand side of Eq.~(\ref{Exact_1}) could very well come
into play to keep the expansion
of the Universe at a
rate consistent with the standard model \cite{Williams2014}.
Moreover, importantly, at	
$t=138$~yr ($\simeq 4.4\times 10^9$~s),  
$(g_{\rm GM})_r$ has fallen
significantly:
$(g_{\rm GM})_r\sim 41.5~{\rm cm \,s^{-2}}$ and
$\vert(g_{\rm GE})_r\vert\sim 1.7~{\rm cm\,s^{-2}}$, for $n={1/ 2}$,
$n^\prime =1$ [compare Eqs.~(\ref{CP6}) and~(\ref{CP6_1})].

In Fig.~4(b), we continue the calculations of cosmic gravitational 
accelerations of the GM [Eq.~(\ref{NMR1})] and 
GE [Eq.~(\ref{NMR2})] radial 
components over time from $t\ge 138$~yr up to the present, 
still evaluated at
the Hubble radius: $r_H(t)=cH_t^{-1}$,  the causally-connected region
surrounding a comoving observer who is at the center
($r=0$) in the metric of Eq.~(\ref{metric_sph}), 
where this local inertial 
center can be any point in this homogeneous spacetime, and
where this comoving observer measures relative distances (or positions)
 with respect to
the global center,  as mentioned 
in Sec.~\ref{sec:3.2}.
The  proper or  physical distance 
$r(t)$ [Eq.~\ref{GE11})], measured according to a  
standard clock at rest with
this comoving observer, at $r(t=t_0)_H=cH_0^{-1}$, is 
for redshift $z\agt 30$  according to the 
relativistic Hubble
law:
\begin{eqnarray}
r={\frac{v}{H_0}}=\Biggl[{{\frac{(1+z)^2-1}{(1+z)^2+1}}}\Biggr]
{\frac{c}{H_0}},
\label{D2} 
\end{eqnarray}
for $H_0=71~{\rm km \,s^{-1} Mpc^{-1}}$, where $z$ can be solved for:
\begin{displaymath}
z=\Biggl({{\frac{1+{{rH_0}/c}}
{1-{{rH_0}/c}}}}\Biggr)^{1/2}-1.
\end{displaymath}
Note, $r=r_H$ for $z\agt 30$, 
according to Eq.~(\ref{D2}),
appears to simply suggests that $r\le r_H$ is always the 
causally-connected observable Universe for any $z\agt 30$, which is 
relativistically true (compare Figure~2 of Ref.~\cite{Davis2004}).
 Equation~(\ref{D2}) gives the spatial separation at a common time from 
a comoving observer when the light was emitted from say a
distant galaxy at $r\leq r_H$.  
In other words---making a brief deviation 
to clarify this cosmological distance, given by 
Eq.~(\ref{metric_sph}), and our use of it---the relativistic 
Hubble law,
gives the distance measured by causally-connected observers at a 
common time $t$.  Simultaneity, means setting ${\rm d}t=0$
in the metric of  Eq.~(\ref{metric_sph}) (or in the standard 
Friedmann-Robertson-Walker metric), which implies choosing 
the local  frame of this freely falling comoving observer.  
The proper distance 
between spacetime events is then given,
 along the line-of-sight (i.e., in the radial 
direction, with ${\rm d}\phi={\rm d}\theta=0$), by
\begin{eqnarray}
r_{\rm proper}= \int_{0}^{\chi}\sqrt{\vert g_{rr}\vert}{\rm d}r,
\label{DDD2} 
\end{eqnarray}	
where $\chi $ is the comoving coordinate distance (associated 
with the proper or physical distance), mentioned in 
Sec.~\ref{sec:3.4} [compare Eq.~(\ref{GE11})]; 
and it can be shown from the metric 
component, $ g_{rr}$, of Eqs.~(\ref{metric_sph_1}), 
upon integration, that
\begin{displaymath}
r_{\rm proper}=R(t)\chi=r(t),
\end{displaymath}
which is just Eq.~(\ref{GE11}).  The proper distance measured by
a comoving observer can be understood as a spacelike separation 
using a hypothetical ruler to measure the separation at the 
time of emission, 
from say a distant galaxy, as 
opposed to a lightlike (null)
separation using light-travel time  
to measure the separation.  Both $r(t)$ and $\chi$ are spacelike
separations between events: this means that they are imaginary 
($\propto {\rm i}=\sqrt{-1}$) and cannot lie on the world line of 
any body or particle. 
From the proper distance $r(t)$ of  
Eq.~(\ref{GE11}), its derivative with respect to 
time, and Eq.~(\ref{hubble}), the Hubble law 
can be derived exactly.
The proper distance might be called the dynamical distance.
The recession velocity for the proper (or so-called dynamical) 
cosmological distance is 
always $\leq c$, at $r\leq r_H$, i.e., causally-connected regions. 
Now, since the Hubble law is 
derived exactly from the metric proper radial
distance, and since the dynamics are what we are concerned with
in this present manuscript, we appropriately
use the proper distance in these calculations.

Finally, before proceeding, for the proper distance $r$,
if we consider the
ratio of the approximated distance [Eq.~(\ref{NMR2c})] to
the more accurate relativistically
corrected distance [Eq.~(\ref{D2})],
we obtain for $z=0.03,~0.5,~ {\rm and}~1$ the ratios
$\sim 1,~1.3,~ {\rm and}~1.7$,
respectively.
In these present model calculations, to give an explanation for the
observed so-called dark energy, we need
only consider the evolution of the GM [Eq.~(\ref{NMR1})] and
GE [Eq.~(\ref{NMR2})] radial
accelerations over spacetime
points for $z\alt 1$, since, importantly, the accelerated
expansion appears to
set in, theoretically, according to this model,
 at $z\sim 0.5$
or $z\sim 0.7$,
consistent with observations,
when using either the nonrelativistic
or relativistic Hubble law, respectively,
to determine $r$, the physical distance, where both these $z$ values
have approximately the same $r$.  
 Moreover,  Eq.~(\ref{NMR1})
goes to zero,
as measured by a present epoch comoving observer, for
$z\sim 0.8$ using  Eq.~(\ref{NMR2c}), the
nonrelativistic Hubble law; and, it
 goes to zero for
$z\sim 2.3$ using Eq.~(\ref{D2}), the relativistic Hubble law
[compare Figs.~2 and~4(b)], where, again, both these $z$ values
have approximately the same $r$.
Thus, the results of this present model do not change
qualitatively when the approximate distance
[Eq.~(\ref{NMR2c})], as opposed to the relativistically
 corrected distance [Eq.~(\ref{D2})],
is used. 
Therefore, the above
and the fact that most large-scale galaxy surveys use the
nonrelativistic Hubble law appear to justify our use
of  Eq.~(\ref{NMR2c})  in this present manuscript.
Nevertheless, some relevant results using the more accurate
relativistic
Hubble law will be discussed in the last paragraph of this
section.

Proceeding, displayed in Fig.~4(b) are the magnitudes of the 
accelerations, $(g_{\rm GM})_r$  and
$(g_{\rm GE})_r $, between the interval $138~{\rm yr}\le t \le t_0$, 
measured by a comoving observer, at
the Hubble radius $r(t)=r_H(t)= ct$ 
[i.e., setting $z=1$ using Eq.~(\ref{NMR2c}) which is  
equivalent to $z  \approx 30$ according to Eq.~(\ref{D2})]
as this spacetime coordinate distance 
(or point) evolves over time, with
$t_0$ indicating the present epoch Hubble radius (causally-connected
observable Universe).
In this case, $(g_{\rm GM})_r $ falls below the magnitude
$\vert(g_{\rm GE})_r\vert$ twice, at $t\sim 2\times 10^{-6} \,t_0$
and $t\sim 5\times 10^{-5} \,t_0$, 
about  the so-called artifact (Sec.~\ref{sec:4}) 
in changing from
radiation dominance to mass dominance occurring at the
equilibrium time $t=t_{\rm eq}\sim 4\times 10^{-6}\,t_0$
($\simeq 1.7\times 10^{12}$~s),
where $n\longrightarrow  {2/ 3}$, for
$\Omega_{\rm mat}\simeq 0.3$ \cite{Liddle2003}; compare Fig~4(b).
Then,  although not shown in Fig.~4(b), 
at $t\sim 0.8\, t_0$ ($\simeq 11 \times 10^9$~yr),
$(g_{\rm GM})_r$ falls to zero,  while
$\vert(g_{\rm GE})_r\vert\sim 5.8\times 10^{-8}{\rm cm\,s^{-2}}$.
After this, $(g_{\rm GM})_r$ remains zero up to the present epoch
($t_0$) at which
$(g_{\rm GE})_r\sim -3.5\times 10^{-8}{\rm cm\,s^{-2}}$, 
consistent with a decelerating universe.
But, recall from above, as $z$ becomes less than one, 
as measured by a
present epoch observer, $(g_{\rm GM})_r$ becomes larger than
 $\vert(g_{\rm GE})_r\vert$ for $z\alt 0.5$, using Eq.~(\ref{NMR2c}),
indicating
the Universe enters into an accelerating phase
[compare Figs.~2(a) through 2(e)], consistent with recent
observations.  Note, as mentioned earlier, at $z=r=0$ and $t=t_0$,
indicating at the present epoch comoving observer,
 $(g_{\rm GM})_r\sim 4\times 10^{14}~{\rm cm\,s^{-2}}$
[compare Fig.~2(f)].

After a quantitative comparison of the calculated results,
it appears that $(g_{\rm GM})_r$ is independent of $r$ for small $r$, 
at early times, i.e.,  $(g_{\rm GM})_r$ has the same 
value for any value of $r(t)$,
changing only over time, as
measured by a comoving observer at a specific epoch up to 
some time $t\equiv t_{\rm crit}$.
For example, notice in Figs.~2 and~4(b), over the range of 
values of  $z$ measured by a present epoch 
comoving observer, the
value of  $(g_{\rm GM})_r\simeq 41.5 ~{\rm cm\,s^{-2}}$ at 
$t=10^{-8}\,t_0$ is 
the same, although the values of $r(t)$  are different. 
Analytically, this is  because the exponent of the
exponential term in Eq.~(\ref{NMR1}) goes to zero, meaning
${\rm e}^0= 1$, for small $r$ and because $(m/c)r\sin\theta$ is
$\ll 1$, where, again, $m$ is given by Eq.~(\ref{metric_2}); thence 
$(g_{\rm GM})_r$ appears independent of $r$ (or $z$); compare  
Eqs.~(\ref{NMR1}) and~(\ref{D1}).  Now, comparing
different
values of $z$ in the range $4\times 10^{-6}\le z\le 30$, 
with  $(g_{\rm GM})_r$
evolving over time, as in Figs.~2 and~4, with the scale factor being
normalized at the present epoch, it appears that
$(g_{\rm GM})_r$ begins showing dependence on $r(t)$ at the critical
time $t_{\rm crit}\simeq 1.2\times 10^{4}$~yr 
($=9\times 10^{-7}\, t_0$) after the Big Bang,
and then showing significant dependence at times 
$t\agt 2.8 \times 10^8$~yr ($ =0.02 \,t_0$).
This means that $(g_{\rm GM})_r$ begins to change significantly
over time and distance.
Before discussing below the importance of this observation,
we readily see that this validates our use of Eq.~(\ref{D1}) to
evaluate $(g_{\rm GM})_r$ in the early Universe:
it is independent of $r$ for small $r$ and it matches 
or converges to the 
value of $(g_{\rm GM})_r\simeq 41.5 ~{\rm cm\,s^{-2}}$, 
evaluated using the exact analytical expression 
[Eq.~(\ref{NMR1})],
at the so-called interface: $t=10^{-8}\,t_0=138~{\rm yr}\simeq
4.4\times 10^{9}$~s (compare Fig.~4).  

Continuing, this independence of $(g_{\rm GM})_r$  for small $r$,
at early times,
as the Universe evolves over time,
and then  becoming significantly dependent on $r$ (or $z$), for
$t\agt 0.02\, t_0$, as measured by a 
comoving observer, can be seen in Figs.~2 and~4.  Again, these
figures are given 
by Eq.~(\ref{D1}) for very small $r$ at early times and by
the exact analytical expression for $(g_{\rm GM})_r$
[Eq.~(\ref{NMR1})] for later times ($t\ge 138$~yr 
after the Big Bang).
Thus, looking at Fig.~2 and what happens at
$t\agt 0.02 \,t_0$ enable
us to see how the GM acceleration $(g_{\rm GM})_r$ 
might once again dominate
over the GE  deceleration $(g_{\rm GE})_r$,  
if it were an intrinsic part of  the equation of motion of the 
expansion (or cosmic scale factor) in the expanding and rotating 
Universe of Eq.~(\ref{Exact_1}) to produce the observed 
present-day cosmic acceleration.  This we suppose in
Secs.~\ref{sec:5.5} and~\ref{sec:5.6}, where we will 
consider, as mentioned earlier,
how the GM acceleration, and, thus, inertial spacetime frame dragging,
might be related to the torsion term (third term on the right-hand side)
in the G\"{o}del-Obukhov
equation of motion [Eq.~(\ref{Exact_1})].	

Further, and for completion, 
use of the relativistic Hubble law [Eq.~(\ref{D2})] to
determine $r$, the
physical or proper distance,  in
Eq.~(\ref{NMR1}),
appears to suggest  the following.
The GM field strength that was once
very large  in the past as shown in Fig.~4(a), dependent only on time,
independent of $r$, but becoming significantly dependent  on $r$  
around $t=0.02 \,t_0$ as described above (compare Figs.~2 and~4),
became negligibly
small ($\sim 0$) around $t=0.8 \,t_0$,  about $2.8 \times 10^9$
years ago, for  $30\agt z\agt 7$, as
shown somewhat in Fig.~4(b) for $z\approx 30$. Then,
from $z\sim 6$ the GM field gradually increased over time with
decreasing $z$ (or $r$), from $(g_{\rm GM})_r \sim 0$ at $t=0.9 \,t_0$
to a nonzero value at $z\sim 2$, as measured by a present 
epoch observer
($t=t_0$).
This is the first indication of a turning up of the curve
(compare Fig.~2 and Fig.~4), and also possibly the indication of the
coming presence of so-called
dark energy.
Finally, at  $z\sim 0.7$,
 the magnitude of $(g_{\rm GM})_r $ became greater than that of
$(g_{\rm GE})_r$ and the Universe entered into an accelerating
phase, thus, consistent with recent observations.
Further details of these results
using the relativistic Hubble law, at these large redshifts,
are presented elsewhere \cite{Williams1}.
%
\begin{figure*}
\includegraphics[width=10cm,height=16cm,angle=-90]{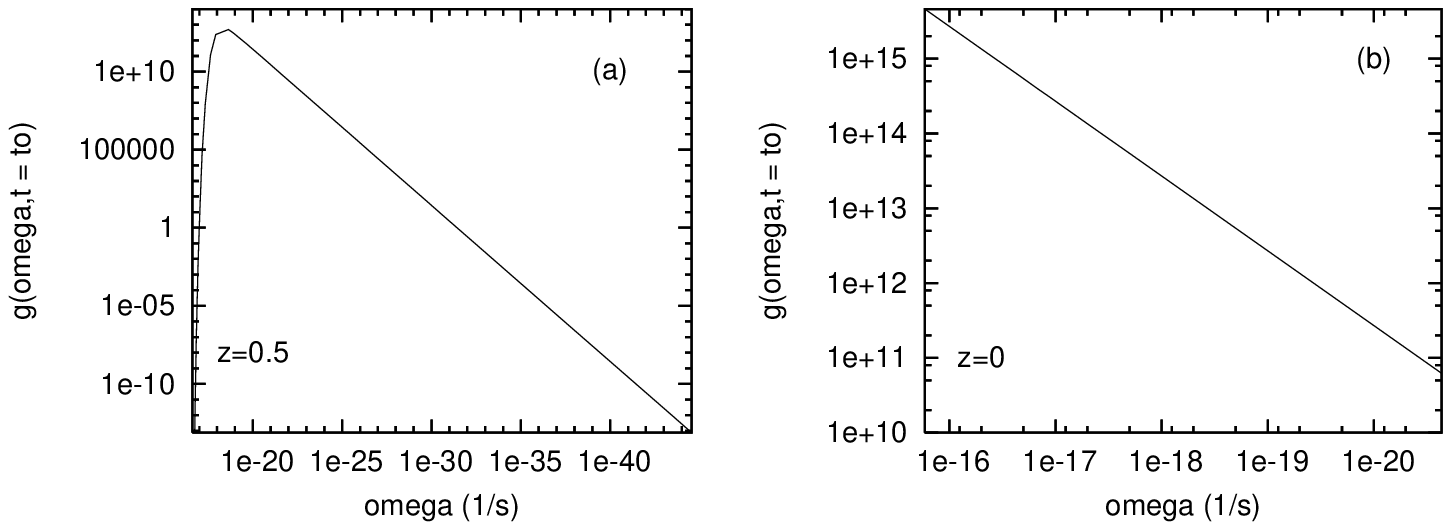}
\vspace{-4.85cm}
\caption{The GM cosmic acceleration and rotation: $(g_{\rm GM})_r$ vs.
$\omega_{\rm rot}$. (a) At redshift $z=0.5$, scale factor $R(t)=1$, at
 $t=t_0$. Note,  $\delta$ in Eq.~(\ref{D3}) decreases along the
curve, as $\omega_{\rm rot}$ decreases (see text).
(b) At redshift $z=0$, scale factor $R(t)=1$, at
 $t=t_0$.  Note, for comparison, using
$\delta=2\pi$ [see Eq.~(\ref{NMR7})], as it is done throughout these
calculations, at $t=t_0$, with $H_0=71~ {\rm km\, s^{-1} Mpc^{-1}}$,
 $(\omega_{\rm rot})_0\sim 1.5\times 10^{-17}~{\rm s^{-1}}$.
}
\label{figfive}
\end{figure*}

\subsection{The GM Acceleration, Cosmic Rotation, and the Present 
Epoch Observer}
\label{sec:5.3}

Figure~5 displays what happens to 
Eq.~(\ref{NMR1}) in a general sense 
 at
$z=0.5$ [Fig.~5(a)] and at $z=0$ [Fig.~5(b)]
when $\delta$ is allowed to vary, where 
\begin{eqnarray}
(\omega_{\rm rot})_0=\delta H_0
\label{D3} 
\end{eqnarray} 
would be the magnitude of the present-day angular velocity
[see Eqs.~(\ref{CP3}),  (\ref{CP4}), and~(\ref{CP5})], with
$H_0=71~{\rm km \,s^{-1}\, Mpc^{-1}}
\simeq 2.3\times 10^{-18}~{\rm s^{-1}}$.
  In Fig.~5(a)
$\delta$ was allowed to vary between $74\ge \delta\ge 1^{-28}$.  
We see that
$(g_{\rm GM})_r\longrightarrow 0$ exponentially (i.e., quickly), left
of the maximum, at any  
value $(\omega_{\rm rot})_0\agt 10 H_0$: falling from a 
maximum of $\sim 5\times 10^{12}~{\rm cm\,s^{-2}}$ at 
$\delta =0.1$; and $(g_{\rm GM})_r\longrightarrow 0$
linearly (i.e., slowly),  
right of the maximum, as $(\omega_{\rm rot})_0
\longrightarrow 0$ at $\delta=10^{-28}$. 
 The reason for this
behavior is that, for a present epoch observer, with $R(t=t_0)=1$, 
at large $\delta$, the exponential expression in the denominator of 
Eq.~(\ref{NMR1}) dominates, causing $(g_{\rm GM})_r$ to go 
exponentially to
zero; and at small $\delta$, the exponential expression 
goes to one and $(\omega_{\rm rot})_0$ 
in the numerator dominates [compare Eq.~(\ref{metric_2})], 
thus, causing $(g_{\rm GM})_r$ to
go to zero in a linear-like fashion.  This behavior will be important
when we analyze in Sec.~\ref{sec:5.4} 
the equation of motion describing the
expansion and deceleration of the G\"{o}del-Obukhov  spacetime
cosmology. We shall see that this 
behavior is somewhat similar to what one would
expected of the third term on the right-hand side of 
Eq.~(\ref{Exact_1}), again with $R(t=t_0)=1$.  
Such a similarity would be expected if 
 spacetime torsion and spacetime
frame dragging (producing the GM acceleration)
are indeed one in the same.  As mentioned earlier, we shall see in
Sec.~\ref{sec:5.5} that this is a valid assumption.
Figure~5(b), with 
$\delta$  allowed to vary between $74 \ge \delta\ge 1^{-3}$ at $r=z=0$,
shows what
the strength of  $(g_{\rm GM})_r$  would be 
as measured by a present epoch observer ($t=t_0$).
The reason for the linear-like decline in $(g_{\rm GM})_r$
is because with $r=0$ the exponential 
expression 
in the denominator of Eq.~(\ref{NMR1}) equals one, resulting in
$(g_{\rm GM})_r$ decreasing as  $(\omega_{\rm rot})_0$ in the 
numerator decreases.
Notice that for $(\omega_{\rm rot})_0=2\pi 
H_0\simeq 1.5\times 10^{-17}~{\rm s^{-1}}$,
 $(g_{\rm GM})_r\sim 4\times 10^{14}~{\rm cm\,s^{-2}}$, as
measured by a present epoch observer
[compare Figs.~2(f) and~5(b)].

The above general analysis of Eq.~(\ref{NMR1}),
presented in Fig.~5, showing  the general
behavior of the dependence of $(g_{\rm GM})_r$ on
$(\omega_{\rm rot})_0$ at $z=0.5$ and $z=0$,
as measured by a present epoch observer,
might have indirect
physical significance concerning  the cosmic 
time dynamical evolution presented in this paper. 
Figure~5(a)  appears to limit the proportionality
constant relating $H_0$ and $(\omega_{\rm rot})_0$
[Eq.~(\ref{D3})]; compare Eqs.~(\ref{CP3}),~(\ref{CP4}), 
and~(\ref{CP5}).
 Figure~5(b)
suggests that the value of $g_{\rm GM})_r$ measured by the present
epoch comoving observer at $z=0 $ [see also Fig.~2(f)]  will decrease
overtime, implying,  perhaps that the accelerated expansion will
decrease over time.

\subsection{Analyzing and Comparing the Standard FLRW Spacetime and the 
G\"{o}del-Obukhov Spacetime}
\label{sec:5.4}

In this section we will analyze the equations of motion of the scale
factor that contain terms that control the spacetime expansion 
of the Universe over time: these are 
Eqs.~(\ref{GE6}) and~(\ref{Exact_1}), for the
standard FLRW and the G\"{o}del-Obukhov  spacetimes, respectively.
Substitution of Eq.~(\ref{GE7}) 
into the second term on the right-hand
side of Eq.~(\ref{GE6}), and comparing 
Eq.~(\ref{GE10}), allows us to
identify this second term as the GE acceleration that decelerates the 
Universe, particularly when $\Lambda=p=0$:
\begin{eqnarray}
{\frac{\ddot R}{R}}= -q H^2,
\label{D4} 
\end{eqnarray}
or
\begin{eqnarray}
q\equiv
q_{_{\rm FLRW}}=-{\frac{\ddot R}{R H^2}}.
\label{D4a} 
\end{eqnarray}
Equation~(\ref{D4a}) is just Eq.~(\ref{dec}), 
where Eq.~(\ref{hubble}) 
has been used.  Notice, we have defined $q\equiv 
q_{_{\rm FLRW}}$ to distinguish between deceleration parameters in the 
two spacetimes we are considering: the FLRW and the G\"{o}del-Obukhov, 
where 
$q\equiv q_{_{\rm GO}}$ in the G\"{o}del-Obukhov  spacetime
(indicated by the subscript ``GO''). We will 
see below how the two may correlate.  So, we find, in general, from
Eqs.~(\ref{D4}) and~(\ref{D4a}) that
\begin{eqnarray}
q_{_{\rm FLRW}}&=&0\Rightarrow {\rm coasting,} 
\nonumber\\
q_{_{\rm FLRW}}&>&0\Rightarrow {\rm deceleration,}
\nonumber\\ 
q_{_{\rm FLRW}}&<&0\Rightarrow {\rm acceleration,}
\label{D4b} 
\end{eqnarray}
again, when $\Lambda=p=0$.
 
On the other hand, in the G\"{o}del-Obukhov
spacetime we can identify the GE acceleration as the first term on the
right-hand side of Eq.~(\ref{Exact_1}), and this term is
exactly that of the 
standard FLRW model if we set $q=1$ [see Eq.~(\ref{D4})],
as mentioned in Sec.~\ref{sec:3.4} [compare
Eq.~(\ref{GE10})]. 
Moreover, it seems reasonable to identify the fourth term on the 
right-hand side of Eq.~(\ref{Exact_1}), involving the mass density 
$\rho_t$,
as that describing the initial inertial expansion over time.  
We define this acceleration 
term, which we associate with the initial cosmic expansion, as
\begin{eqnarray}
\Biggl({\frac{\ddot R}{R}}\Biggr)_I\equiv
2\biggl({\frac{k+\sigma}{k}}\biggr)q_{_{\rm GO}} H^2,
\label{D5} 
\end{eqnarray}
where Eq.~(\ref{GE7}), the average mass density,
expressed in this case,
\begin{eqnarray}
\rho={\frac{3 q_{_{\rm GO}} H^2}{4\pi G}},
\label{D9} 
\end{eqnarray}
 has been used 
with $q\equiv q_{_{\rm GO}}$ (see Sec.~\ref{sec:3.4}
for validation of its use).  
This acceleration term being proportional
to $H^2$ and, thus, the density, will be very large in the very
early Universe.
Now, multiplying through by $r$, using Eq.~(\ref{GE11}) and 
its derivatives, we can express 
this inertial spacetime cosmic expansion
in terms of the physical separation distance $r$:
\begin{eqnarray}
{\ddot r}\equiv (a_I)_r=2\biggl({\frac{k+\sigma}{k}}\biggr)
q_{_{\rm GO}} H^2r,
\label{D6} 
\end{eqnarray}
where $r$, the proper distance, is given by Eq.~(\ref{GE11}).
So, we have identified Eq.~(\ref{D6}) as the
acceleration due to the initial cosmic expansion.
If we use the expression for the relationship between $\sigma$ and $k$
given by Eq.~(\ref{CP2}) as estimated by \cite{Obukhov2000}, 
with value
given in Sec.~\ref{sec:3.5} for constant $c_2$,
Eq.~(\ref{D6}) reduces to
\begin{eqnarray}
 (a_I)_r\simeq 2\,q_{_{\rm GO}} H^2r.
\label{D7} 
\end{eqnarray}
This is consistent with the of order expression given
near the end of Sec.~\ref{sec:2}, which was derived from insight, 
 given in the model description. 

At this point we will used the analogy of the standard FLRW cosmology,
where we set $\Lambda=p=k~(\text{spatial curvature index})=0$ 
in Eq.~(\ref{GE6}) 
to define the deceleration
parameter $q_{_{\rm FLRW}}$ and its relation to  
mass density $\rho$ or expansion rate $H$ 
[compare Eqs.~(\ref{D4}), (\ref{D4a}), and~(\ref{GE7})].
Here, in Eq.~(\ref{Exact_1}), we will set  
$\omega_{\rm rot}=B=p=0$ [compare Eq.~(\ref{D15})]
 to define the 
deceleration parameter
$q_{_{\rm GO}}$.  
Equation~(\ref{Exact_1}) reduces to     
\begin{eqnarray}
{\frac{\ddot R}{R}}=-H^2 + 
 {\frac{8\pi G}{3}} \biggl({\frac{k+\sigma}{k}}\biggr)
\rho.
\label{D8} 
\end{eqnarray}
  Using the model 
parameters of
Sec.~\ref{sec:3.5}, as was done in Eq.~(\ref{D7}), substituting 
in the mass density of Eq.~(\ref{D9}),
and dividing through by $-H^2$, Eq.~(\ref{D8}) yields
\begin{eqnarray}
q_{_{\rm GO}}\simeq {\frac{1}{2}}\Biggl({\frac{\ddot R}{RH^2}}+1\Biggr);
\label{D10} 
\end{eqnarray}
or
\begin{eqnarray} 
q_{_{\rm GO}} \simeq {\frac{1}{2}} (1-q_{_{\rm FLRW}})
\label{D11} 
\end{eqnarray}
(using Eq.~\ref{D4a}).  Then by  
Eq.~(\ref{D11}) we now  have a relationship
between the deceleration parameters of the standard FLRW and 
G\"{o}del-Obukhov 
cosmologies, which appears to be consistent with observations, as 
can be seen in the following: For
\begin{eqnarray}
q_{_{\rm GO}}\simeq {\frac{1}{2}}, ~~q_{_{\rm FLRW}}&=&0\Rightarrow 
{\rm coasting;}
\nonumber\\
q_{_{\rm GO}}<{\frac{1}{2}},~~q_{_{\rm FLRW}}&>&0\Rightarrow 
{\rm deceleration;}
\nonumber\\
q_{_{\rm GO}}>{\frac{1}{2}},~~q_{_{\rm FLRW}}&<&0\Rightarrow 
{\rm acceleration;}
\label{D12} 
\end{eqnarray}
compare Eq.~(\ref{D4b}).

Next, looking at the second and third terms on the right-hand side 
of Eq.~(\ref{Exact_1}), it appears that these
are related to the relativistic fictitious forces associated with cosmic
rotation, and at least one to 
general relativistic inertial frame dragging.  These
fictitious forces, analogous to Newtonian centrifugal and Coriolis
forces, appear in the equation of motion of an object in a 
rotating frame.  It is called a fictitious force because it does not 
appear when the motion is expressed in an inertial frame of 
reference (i.e., a frame that is not rotating nor
dragged into rotation).
The second term can be easily identified as a centrifugal-like 
acceleration, associated with the vorticity (i.e., the rotation),
 that decreases over time,  being proportional to 
$\omega_{\rm rot}^2$, and is largest in the early Universe. This 
second term appears to be associated with the initial cosmic 
rotational energy, similar to the fourth term on the right-hand side of
Eq.~(\ref{Exact_1}), involving the mass density
$\rho_t$,  in which we identified 
above [Eq.~(\ref{D5})] as 
that being associated with the initial inertial
cosmic expansional energy. Now, the  
third term, as mentioned earlier in 
Sec.~\ref{sec:3.1}, is related to torsion coupled with
spin (or rotation) and curvature
of spacetime, and appears to be 
directly related to general relativistic inertial spacetime 
frame dragging, where we 
would expect it 
to be some sort of Coriolis-like force.
This third term
behaves similar to that seen in Fig.~5(a), i.e., it goes to
zero for large $\omega_{\rm rot}$, which would be at early times;
increases as $\omega_{\rm rot}$ decreases over time; and, then,
at later times, decreases as, perhaps, the magnitude of the
cosmic magnetic field $B$ \cite{Obukhov2000}, which depends 
on $\omega_{\rm rot}$,
 decreases over time [compare Eq.~(\ref{D20})].
We shall return to this discussion of the torsion term in
Eq.~(\ref{Exact_1}) in the following sections, where
we will consider also the cosmic magnetic field $B$ and its 
relation to
$\omega_{\rm rot}$ further.

\subsection{The GM Field, Spin, and the Electromagnetic Field}
\label{sec:5.5}

In this section we will further analyze the third term on 
the right-hand side of 
Eq.~(\ref{Exact_1}), which involves the spin-torsion
cosmological coupling constants, $\lambda_1$ and $\lambda_3$,
and which also involves the 	
magnetic field, $B$, where we will look at the meaning of this
electrodynamic field.  As a vector representation this
field is the $F_{12}=B_z$ component of the electromagnetic
field tensor due to the electrodynamic characteristics of the
spacetime matter. Here we consider the Universe to be a spinning
Einstein-Cartan (source of spacetime curvature and torsion)
fluid of charge density rotating
about the global $z$-axis,
like that stated in
Sec.~\ref{sec:3.1} \cite{Obukhov1987, Obukhov2000}. We assume
that torsion is generated by the spin tensor of such a fluid.
Our goal is to relate this third term of 
Eq.~(\ref{Exact_1}) to the
GM acceleration  given by
Eq.~(\ref{NMR1}), in an effort to determine
the spin-torsion cosmological  coupling constants,
$\lambda_1$ and $\lambda_3$, whose derivations are independent 
of and will be compared to those of 
Obukhov  \cite{Obukhov2000}, in order to test the validity of the 
model presented here.  However,  we must first express
Eq.~(\ref{NMR1}) as a force per unit mass per unit length, 
in the form (i.e., units) of a component
of the scale factor equation of motion [Eq.~(\ref{Exact_1})].
Using Eq.~(\ref{GE11}), its derivatives, and dividing the vector
of Eq.~(\ref{NMR1}), 
$(g_{\rm GM})_r {\bf \hat e}_r= \ddot{\bm{r}}$,  
through by $\bm{r}$, as done in Eq.~(\ref{GE10}), 
we have the GM acceleration
expressed in the desired form, in units
of Eq.~(\ref{Exact_1}), i.e., $\rm \,s^{-2}$, expressed as
an acceleration of the scale factor (or force per unit mass per unit 
length):
\begin{widetext}
\begin{eqnarray}
\Biggl({\frac{\ddot R}{R}}\Biggr)_{\rm GM}\equiv
{\frac{(g_{\rm GM})_r}{\bm{r}}}{\bf \hat e}_r
&\sim& \biggl\lbrace{\frac{k\sigma^3(t)}
{4[k+\sigma(t)]^5}}\biggr\rbrace^{1/2}
\biggl[{\frac{(m/c) r\sin\theta+1}{R(t) {\rm e}^{(m/c) r
\sin\theta}}}
\biggr]
{\frac{\omega_{\rm rot}c}{r}},
\label{D13} 
\end{eqnarray}
\end{widetext}
where, $m$, being proportional to $\omega_{\rm rot}$, is, 
again, given by
Eq.~(\ref{metric_2}).  Notice that the
terms in the above equation are consistent with the GM force
producing a Coriolis-like
acceleration. 
Notice, also, plotted in Figs.~2, 4, and~5 
as the ordinates is
$({\ddot R/ R}) r$.  This only makes the 
scale of the ordinates larger than that 
of $({\ddot R/ R})$, but the results are
qualitatively the same.

Now, the torsion term of Obukhov \cite{Obukhov2000} [again, third term 
on right-hand side 
of Eq.~(\ref{Exact_1})], related to the spin density, 
in the Einstein-Cartan  geometrical theory of gravity, is present in
the microscopic and macroscopic regimes of spacetime.  More
familiar is the spacetime torsion, dominant at very high densities, 
due to the intrinsic angular momentum (spin) of fermions, of a 
microscopic nature, manifesting itself in the scales of typical distances 
between particles, dominant in the early Universe \cite{Ponomariev1983}.
However, at low densities
the microscopic torsion of spacetime
by the fermions is less important.
This is why the Theory of Einstein-Cartan does not compete
directly with the Theory of General Relativity \cite{Ponomariev1983}.  
It is proposed that at extremely high densities the spins
 of fermions torque spacetime producing repulsive centrifugal-like 
(i.e., fictitious-like) forces that could  possible avoid the 
initial singularity
(see, e.g., Refs.~\cite{Hehl1974,Gasperini1986}).  
On the other hand, less familiar is the ``generalized''  Einstein-Cartan 
theory of gravity
proposed by Obukhov \cite{Obukhov2000} 
and in this present manuscript
that the spin density on the macroscopic general relativistic scale
could be that of cosmic matter in a rotating universe, approximated
by a cosmological spinning fluid. The elements of cosmological fluid
approximate  particles of  intrinsic angular momentum on 
the early stage of the evolution of the 
Universe and approximate galaxies of 
global angular momentum
 on later stages, in a universe of rotating cosmic matter.  In both 
stages the spin or rotation affects the cosmic spacetime expanding 
continuum it seems.  Then,  
torsion of spacetime by the spins of particles would be  dominant in 
the early
 Universe, and torsion (or frame dragging) 
of spacetime by cosmic matter
would be dominant in the later Universe.   So, what it appears that we 
have derived in Eq.~(\ref{GM17}),~(\ref{NMR1}),  
(\ref{D1}), or~(\ref{D13}), using the metric of 
Eq.~(\ref{metric_sph}), 
is that part of 
the fictitious force (we refer to as the GM force) produced by 
torsion or frame dragging of spacetime by the global angular
momentum of the cosmic matter.
In the equation of motion for the scale factor $R$  
[see Eq.~(\ref{Exact_1})],
Obukhov \cite{Obukhov2000} derives the torsion of 
spacetime by the overall 
intrinsic angular momentum of the Universe.
The overall nontrivial angular momentum that torques the 
expanding spacetime continuum appears to be a 
combination of (1) the  intrinsic spin of the 
fermionic particles and (2) the 
intrinsic rotation of the cosmic matter about the global symmetry axis.  
Thus the spin density of item (1) would 
dominate the torsion term of Eq.~(\ref{Exact_1}) in the very early 
Universe; and the spin density of item (2) would 
dominate in the later universe [compare Eq.~(\ref{D24})].  
If this is true then the torsion 
term of Eq.~(\ref{Exact_1})
should equal the scale factor GM acceleration  of Eq.~(\ref{D13}) 
at some epoch of the evolution of the Universe, where, according to 
the above items, we would expect this intersection to be, at least,
 near the present epoch.  
Moreover, other support (in addition to the validation given below) 
 of the above proposed equality is the 
following: 
\begin{enumerate}
  \item The evolution of GM acceleration $(g_{\rm GM})_r$ 
presented in Fig.~2 is
consistent with recent observations of cosmic accelerated expansion, as
discussed in Secs.~\ref{sec:5.1} and~\ref{sec:5.2}, and 
consistent with the prediction by Obukhov \cite{Obukhov2000} that 
torsion can
either accelerate or prevent the cosmological collapse.
  \item The $(g_{\rm GM})_r$ is a Coriolis-like force derived
using the G\"{o}del-Obukhov metric [Eq.~(\ref{metric_sph})].
  \item  The behavior of the torsion term of Eq.~(\ref{Exact_1}) 
is somewhat similar to that of Fig.~5(a).
\end{enumerate}

So, if we assume that the proposed equality is true,
then, as mentioned above, at some point or epoch in time
the third term
on the right-hand side of Eq.~(\ref{Exact_1}),
associated with acceleration due to torsion,  
should equal to 
Eq.~(\ref{D13}) above, associated with 
acceleration due to so-called frame dragging 
 (in this case,  macroscopic torsion of spacetime), 
allowing us to solve
for  $4\lambda_3^2-\lambda_1^2$, a difference-of-squares 
(or ``difference'') expression 
for the 
cosmological coupling constants of spin and torsion. 
Recall, depending on the values of $\lambda_1$ and $\lambda_3$,
torsion can either accelerate or prevent cosmological collapse,
with $4\lambda_3^2 \gg\lambda_1^2$
to prevent collapse \cite{Obukhov2000}; this appears to include the 
acceleration
of the expansion as proposed here.  
Thus, we find that
\begin{eqnarray}
4\lambda_3^2-\lambda_1^2
&\sim&72
\biggl\lbrace{\frac{k^3\sigma^3(t)}
{[k+\sigma(t)]^7}}\biggr\rbrace^{1/2}
\biggl[{\frac{(m/c) r\sin\theta+1}{ {\rm e}^{(m/c) r
\sin\theta}}}
\biggr] \nonumber\\
&&\times {\frac{R^7 \omega_{\rm rot}^3 c}{B^4 r}},
\label{D14} 
\end{eqnarray}
where in cgs units, $\lambda_1$ and $\lambda_3$ have units
of $\rm {cm}\, g^{-1}$, and $B$ has units of gauss
($\rm =g^{1/2}cm^{-1/2}s^{-1}$).  
The validity of the
derivation of Eq.~(\ref{D14}) will be confirmed below.  

Hence, the frame dragging (or torsion) on the macroscopic or cosmological  
scale  appears to be caused by the inertial spacetime expansion
frame being dragged (or torqued) into rotation by the cosmic spacetime
matter (as described in Sec.~\ref{sec:2}). 
In other words, it appears that the
inertial frame of linear expansion is being torqued or dragged
into rotation by the cosmic spacetime rotating
mass density $\rho_t$, producing the GM force
per unit of moving mass per unit length [Eq.~(\ref{D13})] 
that accelerates
the cosmic expansion, affecting the scale factor.
 This behavior is
analogous to how a sufficiently
large mass density can warp (or curve) spacetime causing the inertial
cosmic expansion to
decelerate, clearly seen in the standard FLRW
cosmological model [Eq.~(\ref{GE6})], thus producing the
attractive GE force of the acceleration of  Eq.~(\ref{NMR2})
or~(\ref{GE10})
[compare also the first term on the right-hand side of
Eq.~(\ref{Exact_1})].

Note, direct discussion of microscopic torsion due to the
quantum-mechanical intrinsic spin of
fermions is beyond the scope of this paper and therefore will not
be discussed in any details here, mainly, because its effect is
negligible in the later stage of the Universe
\cite{Hehl1974}.

So, based on this present investigation, 
it seems safe to state, at least, that, in
general relativity the effect that frame dragging has on
moving objects in an expanding and rotating spacetime is 
described by the so-called GM force field. Again,
as mentioned in Sec.~\ref{sec:2}, this is somewhat 
similar to that experienced
by moving objects (i.e., test particles) in the gravitational potential
well of a rotating black hole (see fig.~2 in Ref.~\cite{Williams2002}),
 where the spacetime frame dragging is in the positive 
azimuthal direction, in the direction that the black hole is rotating, 
and produces a positive radial force. This cosmological spacetime 
frame dragging (or torsion) is, however, in the negative 
azimuthal direction, in the direction of the rotating cosmic matter, 
yet produces a positive radial force as well.
The sign of the frame
dragging angular velocity in both systems corresponds to the direction
of the rotating gravitational source.
 In an expanding and rotating universe the GM force acts on freely
falling comoving frame observers, such
as galaxies in the later universe; whereas,
with the rotating black hole it acts on freely falling local
inertial frame observers, such as particles of plasma.
According to Einstein's Equivalence Principle of Gravitation
and Inertia, the GM Coriolis-like force gives rise to a
gravitational acceleration.  That is,
just as the GE force is a by-product of mass warping spacetime,
the GM force is a by-product of rotating mass dragging spacetime.
Moreover, importantly, it appears that the cosmic expansion frame 
and the rotating spacetime matter are coupled, maintaining a
simple harmonic-like relationship between $H_t$, the expansion rate
[Eq.~(\ref{hubble})], and $(\omega_{\rm rot})_t$,
the rotation rate [Eq.~(\ref{NMR7})],
as suggested by observations [Eq.~(\ref{CP3})].	
It appears that the expanding inertial spacetime frame compensates to
``straighten'' the frame dragged (or torqued) spacetime, while
consistently slowing the cosmic matter rotation, in essence.

Next, the magnetic field given by Obukhov \cite{Obukhov2000} 
is from application of an
ideal fluid plasma of spin and energy-momentum, to a cosmological
model with rotation (shear-free) and expansion, in the framework of the
Einstein-Cartan theory of gravity.  Note, the  Einstein-Cartan theory of
gravity is just Einstein's theory including rotation (and its effect on
spacetime): a natural extension to describe a universe with expansion 
and rotation.  This cosmic magnetic field, seen 
in Eqs.~(\ref{Exact_1}) and~(\ref{D14}),
and mentioned above, is the $F_{12}=B_z$ component of the
electromagnetic field tensor describing the electrodynamic
characteristics of matter in the Universe.
It appears that this field strength
can be expressed over time by \cite{Obukhov2000} 
\begin{eqnarray}
B_t = [-2 R(t)\,\omega_{\rm rot}\,\tau_t]^{\frac{1}{2}},
 \label{D15} 
\end{eqnarray}
where $\tau_t$ [$=\tau(t)<0$] is the spin density, in units of 
$\rm {g \,cm^{-1} \, s^{-1}}$; $\tau_t <0$ means that the spin 
density is in the direction of the
vorticity $\bm{\omega}_{\rm rot}=- \omega_{\rm rot}{\bf\hat{e}}_z$.
Again, R(t) and the magnitude of the vorticity
$\omega_{\rm rot}=
\omega_{\rm rot}(t)\equiv
(\omega_{\rm rot})_t$ 
are given by Eqs.~(\ref{CP6}) and~(\ref{NMR7}), respectively.
The spin density by definition is the angular momentum per unit
volume.  It then seems reasonable to express the spin density of 
the cosmic matter as
\begin{eqnarray}
\tau_t\sim -\rho_t \omega_{\rm rot} r^2,
 \label{D16} 
\end{eqnarray}
where, also, recall $r=r(t)$. 
This means that $B$ could be a primordial cosmic global magnetic
field intrinsic to the rotating and expanding spacetime  matter in
the Universe. 
 Since $B$ of Eq.~(\ref{D15}) depends on the spin 
density $\tau_t$, and this spin density can be expressed 
in terms of  the average mass density $\rho_t$ by 
Eq.~(\ref{D16}), Eqs.~(\ref{D15})
and~(\ref{D16}) appear to
show a fundamental relationship  between the electromagnetic
field (with the electric field canceling assuming equal number of
positive and negative free charges) and the gravitational field, namely,
between the magnetic field and the mass density. It appears that they 
can be related through the spin density  $\tau_t$. 
Further details of this relationship and
of its importance in other astrophysical phenomena are
beyond the scope of this paper and will be discussed elsewhere
in a forthcoming paper \cite{Williams2}.

In the following we shall compare values or test the validity of
Eqs.~(\ref{D14}), (\ref{D15})
and~(\ref{D16}) in the early and/or later Universe against other
theoretical estimated values as well as observations.
Firstly, for comparison, we used the author's derived spin
density $\tau_t$ of Eq.~(\ref{D16}) and the cosmological
parameters used [$q_0=0.01$ and
$(\omega_{\rm rot})_0=0.1 H_0$]
by Obukhov \cite{Obukhov2000}, with $H_0=71~{\rm km\, s^{-1}Mpc^{-1}}$,
 to see if the values of the author and Obukhov \cite{Obukhov2000} agree
for spin density $\tau_0$, at the
present epoch.
The mass density from Eq.~(\ref{GE8}) is calculated to	
give $\rho_c\simeq 1.9\times 10^{-31}~
{\rm g\,cm^{-3}}$;  then, evaluation  of Eq.~(\ref{D16}) at
the Hubble radius  ($r_H\simeq 1.3\times 10^{28}$~cm) yields
$\tau_0\sim -7.4\times 10^6~{\rm g\, cm^{-1}s^{-1}}$.
The magnitude of this value is smaller than the magnitude of Obukhov's
\cite{Obukhov2000} estimated value
($\tau_0\sim -5\times 10^8~{\rm g\, cm^{-1}s^{-1}}$)
 when using the same
$q_0$ and $(\omega_{\rm rot})_0$ as used by Obukhov \cite{Obukhov2000},
with $r=r_H$.
However, using
the same $q_0=0.01$, but letting $(\omega_{\rm rot})_0=2\pi H_0$
[Eq.~(\ref{NMR7})]
as used by the author, Eq.~(\ref{D16}) then yields
$\tau_0\sim -4.7\times 10^8~{\rm g\, cm^{-1}s^{-1}}$.
As we can see, this value is approximately equal to the average
value estimated by Obukhov \cite{Obukhov2000}, given above.
Since Obukhov \cite{Obukhov2000} does not use the best 
estimates for physical
and geometrical parameters to calculate $\tau_0$, this is  
more than likely the reason why our values do not agree.  
However, it can be stated with confidence
that there is consistency of
the author's $\tau_0$ of Eq.~(\ref{D16}), as we shall see  below.
Moreover, this also appears to be the reason
Obukhov \cite{Obukhov2000} calculates the modern-day magnetic field
strength to be $\sim 10^3$ times larger than the established upper
limit from astrophysical observations (more on this below).

Secondly,  the validity of the derivation of
Eq.~(\ref{D14}) is given
by the equation of state [Eq.~(\ref{press2})],
as derived from Obukhov \cite{Obukhov2000}.  
We shall consider two cases.
Case one: As measured by a present epoch observer, for matter
dominance, with $p\approx 0$, for
$\vert 4\lambda_3\vert \gg \vert\lambda_1\vert$, 
Eq.~(\ref{press2}) yields
\begin{equation}
\lambda_3\sim -{\frac{3}{4{\tau^2_0}}}\bigl[4 (\omega_{\rm rot})_0
\,\tau_0 + c^2 \rho_0\bigr],
\label{press3}
\end{equation}
where we have used Eq.~(\ref{D15}).  When the model
values used to calculate the GM acceleration displayed in
Fig.~2(c),  for $z=0.5$, with $B_0\sim 4\times 10^{-4}$~gauss
[Eq.~(\ref{D15}) using Eq.~(\ref{D16})],
are substituted into Eq.~(\ref{D14}), 
for $4\lambda_3^2 \gg\lambda_1^2$, we find that
$\lambda_3\sim \pm 10^{-27}~{\rm {cm}\, g^{-1}}$, as measured by
a present epoch observer, where $R(t=t_0)=1$. Now, when the present
epoch mass density, $\rho_0\approx 9.5\times 10^{-30}~{\rm g\,cm^{-3}}$
 [Eq.~(\ref{GE9})], spin density
$\tau_0\sim -6\times 10^9 ~{\rm g\, cm^{-1}s^{-1}}$
[Eq.~(\ref{D16})], and $(\omega_{\rm rot})_0\sim 2\pi H_0$
[see Eqs.~(\ref{CP3}), (\ref{NMR3}), and~(\ref{NMR7})], with
$H_0\simeq 71~{\rm km\,s^{-1}\,Mpc^{-1}}$, are substituted into
Eq.~(\ref{press3}), we find that
$\lambda_3\sim - 10^{-27}~{\rm {cm}\, g^{-1}}$, the same
as that above, yet calculated independently of Eq.~(\ref{D14})!
Case two: Similarly, for radiation dominance,
as a function of time, with $p\approx c^2\rho/3$, for
$\vert 4\lambda_3\vert \gg \vert\lambda_1\vert$,
Eq.~(\ref{press2}) yields
\begin{equation}
\lambda_3\sim -{\frac{R^6}{{\tau^2}}}
\biggl({\frac{3\omega_{\rm rot}\,\tau}{R^3}}+ c^2 \rho\biggr);
\label{press4}
\end{equation}
again we have used Eq.~(\ref{D15}).
When, the evolved parameters for $z=0.5$, at
$138$~yr after the Big Bang, as measured by a comoving observer:
$r(t=138~{\rm yr})\simeq 6.5\times 10^{19}$~cm, 
$\omega_{\rm rot}\simeq 1.5\times 10^{-9}$~$s^{-1}$, 
$B_t \sim 580.6$~gauss, $R(t=138~{\rm yr})= 10^{-4}$,
as of Fig.~2(c), are substituted into Eq.~(\ref{D14}) we find 
that $\lambda_3\sim \pm 10^{-39}~{\rm {cm}\, g^{-1}}$. 
Now, when the mass density  
$\rho_c\approx 1.9\times 10^{-13}~{\rm g\,cm^{-3}}$
 [Eq.~(\ref{GE7})], spin density
$\tau\sim -1.2\times 10^{18} ~{\rm g\, cm^{-1}s^{-1}}$
[Eq.~(\ref{D16})], $(\omega_{\rm rot})$
and $R(t)$ from above, at $t=138$~yr,  are substituted into
Eq.~(\ref{press4}), we find that
$\lambda_3\sim - 10^{-39}~{\rm {cm}\, g^{-1}}$, again, the same
as that above, yet calculated independently of Eq.~(\ref{D14})!
Thus, the consistency of the above two cases, one at the present epoch 
($t=t_0$) and the other at an earlier time ($t=138$~yr), at a 
particular spacetime coordinate point (or spacetime separation), 
associated with
$z=0.5$ as measured  by a present epoch observer,
validates the proposal and
the assumption that
spacetime torsion and spacetime inertial frame dragging in a
rotating and expanding universe are
one in the same.  That is, the  spin-torsion cosmological  coupling 
constant $\lambda_3$ of the G\"{o}del-Obukhov spacetime derived from 
Obukhov \cite{Obukhov2000} [Eqs.~(\ref{press3}) and~(\ref{press4})], 
when considering torsion of spacetime,
is derived also in this present paper [Eq.~(\ref{D14})], where 
we assume torsion of 
spacetime  and Einstein's general
relativistic frame dragging are one in the same.  So, finding the 
$\lambda_3$'s nearly equal (i.e., being of the same order) in the 
cases above validates this
assumption.   
This finding also validates Eq.~(\ref{D16}), our definition
of  $\sigma(t)$ [Eq.~(\ref{CP1}], and our
choice of $c_2=-115$ (Sec.~\ref{sec:3.5}),
which was  based on observations
of the recently occurring cosmic accelerated expansion and the proposal
that it may occur when the magnitude of the
repulsive GM acceleration overtakes the magnitude of the negative
GE acceleration [compare Fig.~2(c)].

Next, the cosmic magnetic field of Eq.~(\ref{D15}) can now be 
expressed as
\begin{eqnarray}
B_t \sim [2 R(t)\omega_{\rm rot}^2 \rho_t  r^2]^{\frac{1}{2}},
 \label{D20} 
\end{eqnarray}
where we have used Eq.~(\ref{D16}). It appears that $B_t$ above is
a frozen-in cosmic primordial magnetic field modified only through 
the expansion process, such that the magnetic flux is conserved, 
consistent with what observations suggest \cite{Asseo1987}.  This 
$B_t$ would  be that measured by a comoving observer at the 
center of the metric of Eq.~(\ref{metric_sph}).  As the physical 
distance $r$ from this observer is increased, 
the observer is looking back in time, 
as usual, because of the finite speed of light. So, with this in mind one
would expect $B_t$ to be larger at large $r$, as measured by a 
present-day observer  [i.e., where the scale factor of Eq.~(\ref{CP6}) 
is normalized to one], and smaller as $r$ gets smaller, until it reaches 
the value measured locally, between  galaxies. This $B_t=B_0$ would 
be the present-day
value of the cosmic magnetic field that has been somewhat dissipated
 due to  the spacetime expansion of the Universe.  Below we
will calculate values measured by this comoving observer.  

For the very early Universe, this comoving observer measures the 
following as spacetime evolves. At the Planck scale 
(indicated by subscript $P$),
$t=t_P=\sqrt{({hG}/{2\pi c^5})}\simeq 5.4\times 10^{-44}$~s,
with $\rho_P\sim 6\times 10^{92}~{\rm g\,cm^{-3}}$ 
[Eq.~(\ref{GE7})] for $n=2/3$ [see Eqs.~(\ref{CP6}) and~(\ref{CP6_2})],
$B_P\sim 3\times 10^{37}$~gauss, according to Eq.~(\ref{D20}), at 
the Hubble radius ($r_H\equiv r_P=ct_P\simeq 1.6\times 10^{-33}$~cm or 
the so-called Planck length), where $h$ is the Planck constant.

The best way to express $B_t$, it appears, at least during
the inflationary phase, and   
to see clearly how the magnetic field falls off over
time, is to use the familiar solution to the energy conservation fluid 
equation (see, e.g., Ref.~\cite{Liddle2003}):
\begin{eqnarray}
 \rho_t  ={\frac{\rho_0}{R^3(t)}};
 \label{D21} 
\end{eqnarray}
$\rho_0$ is given by Eq.~(\ref{GE8}) with $q_0=1/2$, 
where, here,  we are ignoring the radiation contribution to
the magnetic field $B_t$, assuming it to be negligible (at least
after inflation; see below); 
then, upon substitution, Eq.~(\ref{D20}) can
be expressed as
\begin{eqnarray}
B_t \sim \Biggl[{\frac{2 \omega_{\rm rot}^2 \rho_0  r^2}
{R^2(t)}}\Biggr]^{\frac{1}{2}}.
 \label{D22} 
\end{eqnarray}
That is, since the state of matter before inflation is open to 
speculation, for simplicity we are assuming the following:
The cosmic matter is mass dominated, in thermal equilibrium,
and has negligible radiation pressure before inflation, 
such that the expansion
rate before inflation is governed by the scale factor 
[Eq.~(\ref{CP6})] at  $t\sim 0$, of  the Einstein-Lema\^{i}tre 
\cite{1931ab_Lang1980} expanding
cosmological model, with
$n=2/3$.
But after inflation
the cosmic matter becomes radiation dominated,
with $n=1/2$,
consistent with the standard version of inflation.
In the standard version, during inflation the energy stored in
the vacuum-like state is then transformed into thermal energy, 
and the universe becomes
extremely hot, and from that point onward, its evolution is described 
by the standard hot universe theory \cite{Linde1990}.  
After this hot phase
the cosmic matter returns to its so-called stable state of mass 
dominance past the radiation-mass equilibrium time
$t=t_{\rm eq}\sim 1.7\times 10^{12}$~s 
(mentioned in Sec.~\ref{sec:5.2}),  with $n=2/3$, 
like before inflation.
Now,
usually, we assume that inflation occurs at the
characteristic times between
$10^{-36}~{\rm s}\alt t_{\rm infl}\alt 
10^{-34}~{\rm s}$,  where
during inflation $H_t\approx\,\text{constant}$, since
$\Delta t\simeq 9.9\times 10^{-35}$~s $\ll 1$
(see Sec.~\ref{sec:3.5}).
The Hubble radii ($r_H= ct$) 
corresponding
to the time interval above are
$3\times 10^{-26}~{\rm cm}\alt r_H \alt 
3\times 10^{-24}~{\rm cm}$.
From Eq.~(\ref{D20}) or
Eq.~(\ref{D22})
we  calculate the magnetic field at the beginning of inflation to be
$B_t\sim 5\times 10^{32}~{\rm gauss}$, with $n={2/ 3}$.
Since we assumed that $H_t$ is approximately 
constant during inflation, it 
seems reasonable to assume that $\omega_{\rm rot}$ ($\sim 2\pi H$)
is also approximately constant.  We 
found in Sec.~\ref{sec:3.5} that the
initial and final scale factors at the beginning and after inflation
are related by 
$R_{\rm f}\sim 10^{43}R_{\rm in}$ 
(or $R_{\rm in}\sim 10^{-43}R_{\rm f}$).  It can be shown 
from Eq.~(\ref{GE11}) that the physical radius after inflation is
given by $r_{\rm f}\sim 10^{43}r_{\rm in}$ 
(or $r_{\rm in}\sim 10^{-43}r_{\rm f}$).  Thus we
can see that upon substitution  into Eq.~(\ref{D22}), the
$10^{\pm 43}$ factors in the numerator and denominator will
cancel. 
Therefore, for 
mass dominance before inflation, 
one can
continue to use  the general expression
[Eq.~(\ref{D20})]	
to express the cosmic
primordial magnetic field after inflation. 
Note, for completion, for relativistic
matter (or radiation dominance), the familiar 
solution to the energy conservation fluid
equation is 
\begin{eqnarray}
 \rho_t  ={\frac{\rho_0}{R^4(t)}},
 \label{DD21} 
\end{eqnarray}
where, in this case, $\rho_0$ is given by Eq.~(\ref{GE8}) 
with $q_0=1$; compare Eqs.~(\ref{D20}),~(\ref{D21}), 
and~(\ref{D22}).  
Consequently and importantly, concerning the assumption of 
negligibility above, 
it can be shown that the magnetic field due to radiation
before inflation, decreases by a factor
$\sim 10^{-22}$ after inflation.
\begin{figure}
\centerline{\includegraphics[width=86mm]{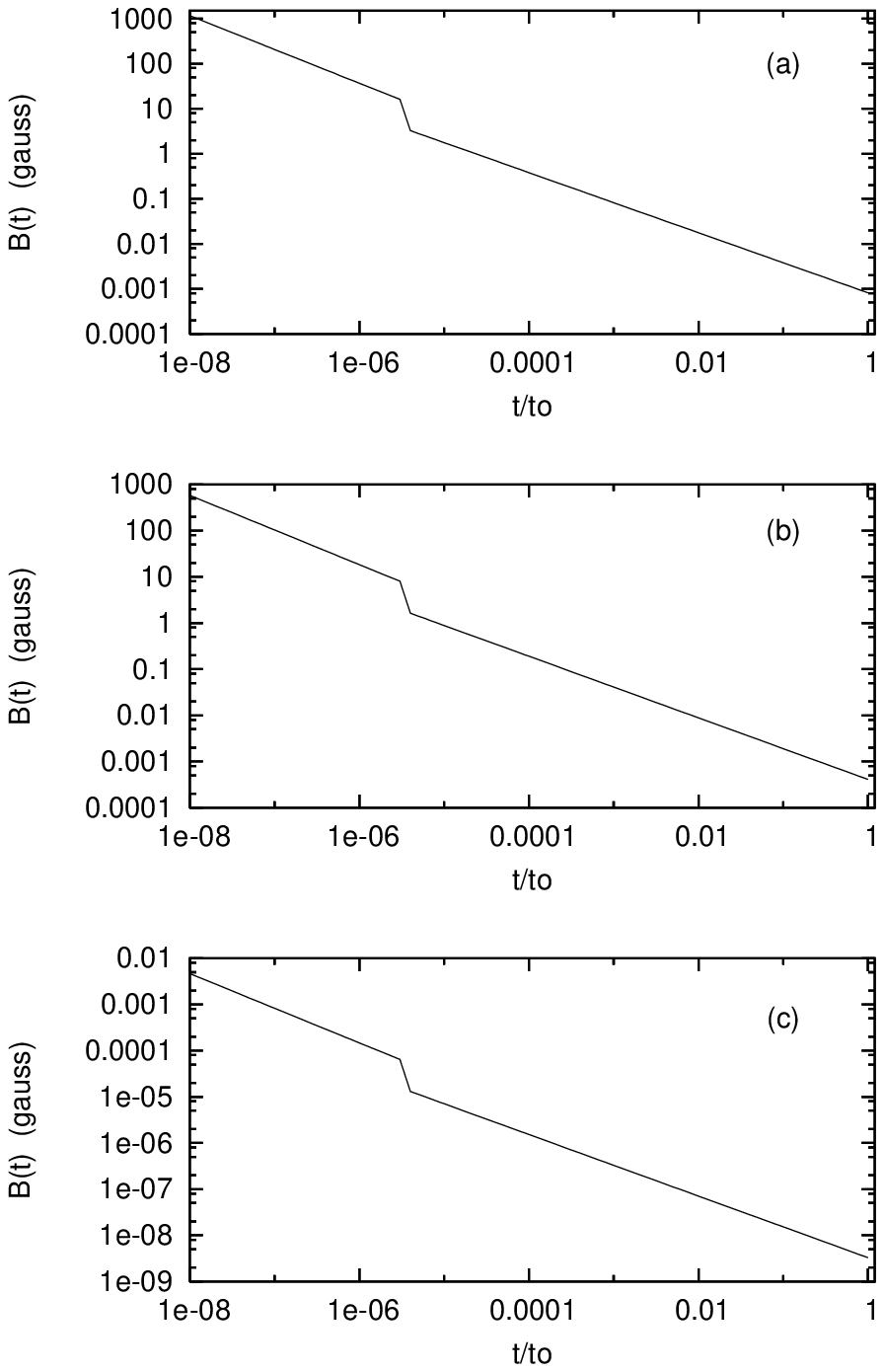}}
\caption{Evolution of the cosmic primordial universal magnetic field:
(a) $B_t$ versus $t$ at $z\approx 30$ (at the Hubble radius,
$cH_0^{-1}\approx ct_0$).
(b) $B_t$ versus $t$ at  $z=0.5$ ($\approx 2\times 10^3$~Mpc).
(c) $B_t$ versus $t$ at  $z=4\times 10^{-6}$ ($\simeq 17$~kpc).
(Note, the step-like feature is an artifact; see note on Fig.~2.)
}
\label{figsix}
\end{figure}

Using Eq.~(\ref{D20}) 
we now determine the cosmic magnetic field at
recombination ($B_{\rm rec}$) and
at the present epoch ($B_0$) as measured by a comoving 
observer at a particular proper distance $r$.  We will assume as
suggested by particle physics
that recombination occurs $\sim 350,000$~years
after the Big Bang \cite{Liddle2003}.
In  Fig.~6,  the 
magnetic field of
Eq.~(\ref{D20}) is evolved over time from when
the Universe was
$138$ years old to the present.   Figures~6(a) and~6(b) display
the evolution of the field strength $B_t$
measured at the Hubble radius [$r_H$, at $z\sim 30$ according
to Eq.~(\ref{D2})] and at $z=0.5$,
respectively,
 by a comoving observer.
At the present epoch, as measured by this comoving observer,
with $n={2/ 3}$, $q={1/ 2}$, using
Eqs.~(\ref{CP6}) and~(\ref{CP6_2}), normalized at
$t=t_0$,  for
$H_0=71~{\rm km\, s^{-1} Mpc^{-1}}=2.3\times 10^{-18}~{\rm s^{-1}}$,
$\omega_0\simeq 1.5\times 10^{-17}~{\rm s^{-1}}$
[Eq.~(\ref{NMR7})],
 $t_0=H_0^{-1}$, and $r_H =c H_0^{-1}$,
we find that
$B_t=B_0\sim 8\times 10^{-4}$~gauss and 
$B_t=B_{\rm rec}\sim 1$~gauss, at the present epoch
 and at the time of
recombination, respectively,  as can be seen
in Fig.~6(a).  Similarly, we find that 
$B_t=B_0\sim 4\times 10^{-4}$~gauss and 
$B_t=B_{\rm rec}\sim 0.5$~gauss, at $z=0.5$, as can been
seen in Fig.~6(b).
Figures~6(a) and~6(b) are to be compared
with those of Figs.~4(b) and~2(c), respectively, which plot the GM
acceleration over time.

Now, again using Eq.~(\ref{D20}), 
we can find values for the primordial 
apparently frozen-in intergalactic (as related to spaces
between galaxies) universal magnetic field
$B_t$, as measured by
a comoving observer, at the present-day, normalized
[see Eqs.~(\ref{GE11}) and~(\ref{CP6})], spacetime proper
coordinate distance
$r(t=t_0)\sim 17$~kpc
(or $z \sim 4.6\times10^{-6}$).
Recall that this observer is located at the center of the
G\"{o}del-Obukhov metric
[Eq.~(\ref{metric_sph})], which could be centered  on any
 comoving galaxy.
Figure~6(c), shows how this cosmic $B_t$ has
evolved over time, at the above
proper coordinate distance, from the spacetime separation at
$r(t=138~{\rm yr})$ to the present $r(t=t_0)$.
From this figure we see that
$B_t=B_{\rm rec}\sim 4\times 10^{-6}$~gauss and
$B_t=B_0\sim 3\times 10^{-9}$~gauss
at recombination and at the
present epoch, respectively.
Importantly, this value for the intergalactic
$B_0$ is
consistent with the upper limit constraint placed on the present
strength of any primordial homogeneous magnetic field,
which is $B_0\alt  4\times 10^{-9}$~gauss, for
$\Omega=1$ and  $H_0= 71~{\rm km\,s^{-1}\,Mpc^{-1}}$ \cite{Barrow1997}.
  The Cosmic Background Explorer (COBE)
measurements provide this constraint: set by how the amplitude of the
magnetic field is related to amplitude of the microwave background
anisotropies on large scales.  Pasquale, Scott, \& Olinto 
\cite{Pasquale1999} study the effect of inhomogeneity 
on the Faraday rotation of light from
distant quasi-stellar objects  to find a consistent limit of
$B_0\alt  10^{-9}$~gauss.  

Moreover, the strength of the magnetic field 
$B_0\sim 2.5\times 10^{-7}$~gauss at $r\simeq 1.3$~Mpc for $z=0.0003$,
with $H_0\simeq 71~{\rm km\,s^{-1}\,Mpc^{-1}}$, given by these 
present model calculations is consistent with the constraint placed by
the Planck 2015 results \cite{PLANCK2015_2}:  When effects of Faraday 
rotation on the primary CMB polarization anisotropies are considered, 
the resulting constraint is $B_{1Mpc}< 1.38\times 10^{-6}$~gauss.

Note, displayed in Fig.~6 are the evolved magnetic field strengths
$B_t$ of  Eq.~(\ref{D20}) over cosmic time, as measured by a
comoving observer at the present proper distance $r(t=t_0)$.
The
smaller this measured $r(t=t_0)$ is, as this coordinate point
evolves over time, 
from $138$ years after the Big Bang to the
present, the smaller the measured magnetic field strength will be
relative to the larger $z$ (or larger $r$) values.
This can be seen and understood by comparing the figures of
Fig.~6 and Eq.~(\ref{D20}).  This means that as
the distance $r(t=t_0)$ become smaller as measured by this comoving
observer, the magnetic field strength measured over time is smaller.
That is, the so-called hypersphere  of the 
 G\"{o}del-Obukhov spacetime metric of
Eq.~(\ref{metric_sph}),  surrounding the comoving observer,
enclosing cosmic
spacetime matter, is smaller. This
behavior of $B_t$ at  distance $r$ is similar to the behavior
of the strength of the so-called mutual universal gravitational field of
attraction  [Eq.~(\ref{NMR2})],
as measured by a comoving observer, assuming
homogeneity.  
Equations~~(\ref{D20})  and~(\ref{NMR2}) have similar
behaviors  because both $B_t$ and $(g_{\rm GE})_r$ 
are proportional
to $r$, the proper radial distance away from a comoving observer
(or between comoving observers).

The above present model calculated values of the cosmic magnetic field 
appear to be consistent with those
that would allow the Universe to evolve into its present state, 
particularly like
the cosmological model of Zeldovich (see \cite{Asseo1987}, and 
references therein).  Such cosmological models depend on the choice 
of the metric, of the equation of state, and of the cosmological constant
$\Lambda$, where most often taken to be $\Lambda=0$, until 
recently with the advent of dark energy.  Such models are characterized 
by the expansion factors, the density of constituents, and the magnetic 
field,  which evolve according to the Einstein-Maxwell field equations.
The magnetic field energy is assumed to be modified only through the
expansion process, and not through an exchange of energy with other
constituents: matter or radiation, which means that the magnetic flux 
is conserved and a uniform magnetic field evolves like $1/R_x R_y$
(for a magnetic field in the $z$-direction), where 
$R_x$ and  $R_y$  are scale factors.  This 
results from the possibility for a uniform magnetic field to exist even
 in the absence of an electric current 
($\bm{\nabla}\bm{\times} \bm{B}=0$).
This can be interpreted as the evolution of the magnetic field
compatible with the Einstein equations, as characteristic of a 
magnetic field geometrically frozen-in independently of the 
conductivity of the matter.    

\subsection{Rotation, Torsion, and Spin Density}
\label{sec:5.6}

For completion and further investigation, the torsion term of
Eq.~(\ref{Exact_1}), again third term on right-hand side,
 can be expressed in terms of the spin density $\tau_t$.   This is 
done by substituting $B$ from Eq.~(\ref{D15})
 into Eq.~(\ref{Exact_1}).
Thus we find that the Einstein-Cartan torsion term of 
Eq.~(\ref{Exact_1}) becomes
\begin{eqnarray}
\Biggl({\frac{\ddot R}{R}}\Biggr)_{\rm tor}&\equiv&
{\frac{1}{\omega_{\rm rot}^2}}
\biggl({\frac{k+\sigma}{144k}} \biggr)
(4\lambda_3^2 - \lambda_1^2){\frac{B^4}{R^8}}
\nonumber\\
         &= &\biggl({\frac{k+\sigma}{36k}} \biggr)
(4\lambda_3^2 - \lambda_1^2){\frac{\tau^2}{R^6}},
\label{D24}
\end{eqnarray}
as expressed by Obukhov \cite{Obukhov2000}.   
In Obukhov's \cite{Obukhov2000}  
generalized Einstein-Cartan theory of gravity, it appears that
$\tau$ is the total
intrinsic angular momentum per unit volume of the Universe, 
comprising, mainly, if
not exclusively, the spin of fermions at the
microscopic level and 
the global rotation of cosmic
matter at the macroscopic level, as mentioned in the above section.
Assuming the GM acceleration of Eq.~(\ref{D1}) to be valid 
at the microscopic level for very small $r$ as $t\longrightarrow 0$
[recalling that Eq.~(\ref{NMR1}) is independent of $r$ for small
$r$ at early times ($t<1.2 \times 10^4$~yr) as 
discussed in Sec.~\ref{sec:5.2}], Eq.~(\ref{D1})
might be the repulsive ``Coriolis-like'' gravitational acceleration  
experienced by the fermions, due to inertial spacetime 
frame dragging (or torsion) that causes their intrinsic spin 
vectors to precess, being coupled with the angular momentum of
the rotating Universe. 
Moreover, if we could find an equivalent microscopic GM field
like that given by Eq.~(\ref{GM1}) and subsequently like
that of Eq.~(\ref{D1}), for the fermions, due
to their intrinsic spin density $\tau$ torqueing (or frame dragging)
spacetime,
this might yield a fundamental short range gravitational
acceleration, acting
prominently at high densities, predominantly in the
early Universe.
 In addition, if we consider the equality of the GM field like
 that of Eq.~(\ref{D1}), mentioned above, and Eq.~(\ref{D24}), 
and use the equation of state given by Eq.~(\ref{press1}),
 along with the spin density $\tau$ of the fermions
(see, e.g. Ref.~\cite{Ponomariev1983}), one might be able to
understand better the initial state of the Universe at the time of the
Big Bang, or at least understand better
torsion or frame dragging of
spacetime at the microscopic level.
This is consistent with the Einstein-Cartan
theory providing a description of gravitational properties of
matter at the microphysical level  \cite{Hehl1974, Kuchowicz1978}.
The above however needs to be investigated
further. 

Also, and importantly,  recall, we found above in 
Sec.~\ref{sec:5.5} that the spin density $\tau$ 
of the rotating cosmic matter
links the magnetic field and mass density $\rho$ [compare 
Eqs.~(\ref{D15}) and~(\ref{D16})], where $\rho$ gives 
rise to the so-called GE field [Eq.~(\ref{NMR2})].  
Here, according to 
Eqs.~(\ref{D15}) 
and~(\ref{D24}), the spin density links the magnetic 
field and torsion (or frame dragging) 
of spacetime, where torsion gives rise to the so-called GM field.   
Thus, the common
link is the spin density $\tau$, which appears  to tie the 
 magnetic field to gravity, in general.

\subsection{\label{sec:5.7}Summary of Discussion}

The above discussion is summarized as follows:
\begin{enumerate} 
    \item In Sec.~\ref{sec:5.1}, we begin with 
the results of 
analyzing the GM [Eq.~(\ref{NMR1})] and the GE [Eq.~(\ref{NMR2})]
accelerations
over cosmic time from 138~years after the Big Bang to the present.  
The evolution of the GM and GE accelerations over time at	
a spacetime proper coordinate distance $r(t)$
as measured by a  present epoch comoving
observer for a specific $z$ [see Eq.~(\ref{NMR2c})]
shows that after a period of decelerating cosmic expansion,
the Universe enters into an accelerating expansion 
phase at $z\sim 0.5$, as can be seen in Fig.~2.   This is 
consistent with recent observations that suggest the presence 
of dark energy.  
     \item In Sec.~\ref{sec:5.2},
we evolve the GM and the GE
accelerations at the Hubble radius:
from the Planck  time to inflation
to the present.  We derive an
approximate analytical
expression for the GM cosmic acceleration [Eq.~(\ref{D1})] that 
appears to be valid at
early times in the Universe as well as later times (for 
small $r\sim 0$).  We use Eq.~(\ref{D1}) to evaluate the GM
acceleration in the early Universe and Eq.~(\ref{NMR1}) to evaluate
it in the later Universe.  Figure~4 displays the evolved
GM and GE accelerations from the Planck time to 138~years after
the Big Bang [Fig.~4(a)], then from 138~years to the present 
[Fig.~4(b)], evaluated at the Hubble radius ($r_H=cH^{-1}$), 
where before inflation this radius was $10^{-43} r_H$.
The magnitudes of Fig.~4(a) suggest that the GM acceleration 
does not a play a direct role in inflation, but spin and torsion
by way of the pressure might. Figure~4(b) shows that the GM
acceleration is negligible at this $z$~($\sim 30$), as measured by a 
present-day comoving observer.    
Also, in Sec.~\ref{sec:5.2}, a brief deviation was made 
to clarify the meaning of the cosmological proper distance $r$ given by
Eq.~(\ref{metric_sph}), which might be called the dynamical
distance, and is always $r\leq r_H$  by definition relativistically.
Moreover, also, it is found that the GM acceleration, being dependent
on time, appears to 
be independent of
the proper distant $r$ in the early Universe, but gradually 
becomes dependent  in the later Universe causing it
to enter into an accelerating phase.  
    \item In Sec.~\ref{sec:5.3}, we look 
at the general behavior of the GM acceleration as measured by a 
present-day 
comoving observer at $z=0.5$ [Fig.~5(a)] and  $z=0$ [Fig.~5(b)], 
with $R(t=t_0)=1$, while
$(\omega_{\rm rot})_0$ is allowed to vary from large values to very small
values.   We find that Fig.~5(a) is somewhat similar  to what one would
expected of the third term on the right-hand side of
Eq.~(\ref{Exact_1}), as discussed further in 
Sec.~\ref{sec:5.4}.  Figure~5(b) shows the strength of the GM
acceleration as measured at the comoving observer ($z=r=0$) for various
values of  $(\omega_{\rm rot})_0$.  
     \item In Sec.~\ref{sec:5.4}, we 
analyze and compare the FLRW  and the G\"{o}del-Obukhov  spacetimes
of Eqs.~(\ref{GE6}) and~(\ref{Exact_1}), respectively.  
We derive what appears to be a correlation between the two spacetimes
[Eqs.~(\ref{D11}) and~(\ref{D12})].
We also identify the terms of Eq.~(\ref{Exact_1}) that are associated
with the deceleration,  expansion, rotation,  and torsion of spacetime.
     \item In Sec.~\ref{sec:5.5}, we
 further analyze the third term on
the right-hand side of
Eq.~(\ref{Exact_1}), where we relate this term 
to the
GM acceleration  given by
Eq.~(\ref{NMR1}), to determine 
the difference expression of
the spin-torsion cosmological coupling constants,
$4\lambda_3^2-\lambda_1^2$ 
[Eq.~(\ref{D14})], and compare with 
that of Obukhov \cite{Obukhov2000}
[i.e., the equation of state of Eq.~(\ref{press2}), 
which yields Eqs.~(\ref{press3}) and~(\ref{press4})].
With the assumption that $\vert 4\lambda_3\vert 
\gg \vert\lambda_1\vert$, we find that the $\lambda_3$'s 
of Obukhov (2000) and of the author are of the same order, 
although they were derived independently,
thus, confirming the validity of the model
presented in this present paper, in essence, that frame dragging and 
torsion of spacetime are one in the same. 
Next,  we analyze the expression for the cosmic magnetic field
strength $B$,
relating it to the spin density 
(angular momentum per unit volume of the cosmic matter) 
and mass density, in 
Eqs.~(\ref{D15}), (\ref{D16}), and~(\ref{D20}).  
We find that not only is $B$ consistent with observations, but it 
yields an apparently fundamental relationship between itself and the 
mass density $\rho_t$, and, thus, indirectly, a relationship
between  electromagnetism and 
gravity [of Eq.~(\ref{NMR2})].  It appears that they 
are related (or coupled) through
the spin density $\tau$. 
       \item In Sec.~\ref{sec:5.6}, we express the torsion term of
 Eq.~(\ref{Exact_1}) in terms of the spin density $\tau$ 
(using Eq.~\ref{D15}), 
suggesting that spin density links the magnetic
field to torsion (or frame dragging)
of spacetime (that gives rise to the so-called GM field). So, 
it appears that the common
link is the spin density $\tau$, which links the magnetic field to 
torsion or 
frame dragging (i.e., the GM field)  and also links the
 magnetic field to the mass density (i.e., the GE field), 
as found in Sec.~\ref{sec:5.5}.
\end{enumerate}

Thus, overall, the main focus of this discussion is the finding
that the
recently observed acceleration of the cosmic expansion  may
possibly be explained by
considering the effect
of the GM force, due to inertial spacetime frame dragging, in a rotating
and inertially expanding universe using the G\"{o}del-Obukhov
spacetime metric of Eq.~(\ref{metric_sph}).
Moreover, it appears that spacetime frame dragging and torsion
of spacetime are one in the same, at least at the macroscopic level.
Further, and importantly, we see that the
pressure $p$ need not be negative to
explain the  recently observed cosmic acceleration of the expansion;
and the mystery surrounding Einstein's
cosmological constant
[compare Eq.~(\ref{GE6})] might be solved in the context of a
rotating universe [compare Eq.~(\ref{Exact_1})].
Nevertheless, a negative pressure,
which could possibly be produced in
the G\"{o}del-Obukhov very early Universe,
might, however,   play an important role in inflation [compare
Eqs.~(\ref{Exact_1}) and~(\ref{press1}) or~(\ref{press2})],
with the assumption that $\vert 4\lambda_3\vert
\gg \vert\lambda_1\vert$, and provided that $\lambda_3>0$.

\section{Conclusions}
\label{sec:6}
With the recent discovery of so-called dark energy (appearing to
comprise $\sim 68$ percent of
the Universe), it
seems we know little about the Universe we live in, save the
$\sim 5$ percent mass-energy we can see and the expected effects
of gravity such mass-energy displays.  With our already lack
of knowledge of what composes so-called dark matter
(appearing to comprise $\sim 27$ per cent of the Universe), this
new finding of dark energy limits our understanding even more.
The above
percentages are based on the most recent observations \cite{Planck_data}
which are almost a perfect fit to the predicted material 
 content of the Universe
by the standard FLRW cosmology, but
there are some unexplained anomalies that suggest that we should
perhaps seek further an
understanding of the
underlining force: gravity.  Such understanding could possible solve
the problem of our
recently, and somewhat embarrassingly,
increased lack of knowledge, i.e., of the nature of dark energy.
In this paper, we have adhered to
the above suggestion by seeking an understanding of dark energy by
considering it to be a component of gravity, a by product, arising in
a general
relativistically rotating and expanding cosmological spacetime.

In this manuscript is presented a general relativistic model to 
describe the dynamical evolution of the Universe.  
This model appears to answered the question,
``Could dark energy be a manifestation of gravity?'' 
and the answer it seems  
is yes.  In this model, the recently observed 
cosmic acceleration of the expansion 
may possibly be explained by considering the 
effect of the gravitomagnetic (GM)
field due to spacetime frame dragging by rotating 
cosmic matter in an inertially expanding spacetime 
 universe.  These
model calculations seem to show that application of the 
G\"{o}del-Obukhov spacetime metric
of Eq.~(\ref{metric_cc}), or Eq.~(\ref{metric_sph}),
in an Einstein-Cartan general relativistic spacetime, leading to the
derivation of Eq.~(\ref{Exact_1}), 
the equation of motion of the scale 
factor $R$,  
 yields a dynamical 
description of  how our Universe has evolved 
over time.  
Nonvanishing torsion of the Riemann-Cartan spacetime 
\cite{Mao2007} and the GM field of inertial frame dragging
appear to be one in the same in this description.
Importantly, in this model, we see that the pressure $p$ need not be 
negative, nor do we have to mysteriously resurrect 
the cosmological constant 
$\Lambda$ to explain so-called dark energy; this is contrary to 
models constrained by the standard FLRW cosmology 
[compare Eq.~(\ref{GE6})].

Not only does the model presented here appear to 
describe the dynamical evolution of the 
Universe to a large degree, 
showing how acceleration of the expansion might arise,
but it also appears to show a 
relationship between the cosmic magnetic field and the mass density, 
and a relationship between the 
cosmic magnetic field and the GM
field.  Both relationships are through the spin density of 
cosmic matter, which 
could possibly lead to a link between electromagnetism and gravity.  
These apparent relationships need to be investigated 
further, not only on
the astronomical level but the atomic level as well.
Namely, when the spin contributions of the cosmic matter are included
in the gravitational field equations according to the
Einstein-Cartan theory, the application of the spin density 
tensor can range from the microscopic case of the quantum-mechanical
intrinsic angular momentum (spin) of elementary particles, 
dominant at extremely high densities but negligible at normal 
matter densities \cite{Hehl1974},  to the 
macroscopic case of the rotating cosmic plasma  \cite{Obukhov2000},
as presented in this present paper.   In both 
cases torsion (or frame dragging) of spacetime is produced.  
Some authors 
(e.g., Refs.~\cite{Hehl1974,Kuchowicz1978,Nurgaliev1983,
Gasperini1986,Poplawski2010}) propose that the intrinsic spin and
spacetime torsion of fermions can avert the Big Bang singularity. 
Gasperini \cite{Gasperini1986} points out that inflation might
be driven by a centrifugal-like force, due to the 
spin-density tensor of matter sources when dominated 
by the  intrinsic spin of elementary 
particles, occurring in the extremely high density very early 
Universe ($t<10^{-23}$~s).  So, since it appears that both 
the GM field and the magnetic field are related to spacetime torsion 
by way of  the spin density,
perhaps the inclusion of this knowledge
will put us closer to a theory of quantum 
gravity.

As a future calculation in cosmology, one might  use the results 
of the model presented in this manuscript to
evaluate the Friedmann-like equation, describing the evolution of
the scale factor, derived from the gravitational	
field equations by Obukhov \cite{Obukhov2000}, used in 
the derivation  of Eq.~(\ref{Exact_1}) and, thus,
Eq.~(\ref{D14}).  
It appears that Eq.~(5.42) 
of Obukhov \cite{Obukhov2000} 
can be used  to see what 
fraction of the critical density the GM acceleration, due to
frame dragging (or torsion), contributes to making the observed
density parameter $\Omega\simeq 1$.  This might also shed light
on the true nature of dark matter.

Finally, it remains to be seen whether or not all the 
concepts presented in this manuscript are fully
valid (as they appear).
Nevertheless, it is a
fact that the G\"{o}del-Obukhov 
cosmology of a rotating and expanding 
universe, of Eq.~(\ref{Exact_1}), 
has the potential to exhibit cosmic acceleration in the 
equation of motion of the scale 
factor. This yields a possibility of explaining the recently 
observed cosmic 
accelerated expansion. 
If the Universe is indeed rotating,  Eq.~(\ref{NMR1}) or~(\ref{D13}),
along with Fig.~2,  shows that 
this cosmic acceleration became important in recent times,  
agreeing with observations.

\begin{acknowledgments}

I first thank God for the knowledge he has given me of
the Universe.  Next, I thank The University of Toledo for their
hospitality while this work was being completed; and I am grateful to
Dr.~Jon Bjorkman for helpful discussions.  
I am grateful also to Dr. Yuri
Obukhov for helpful comments. Finally, I thank the Ohio
Supercomputer Center. Support for this
research was
provided by an  American
Astronomical Society Small Research Grant and in part by  National
Science Foundation grant AST-0909098. 
\end{acknowledgments}

----------------------------------------------------------------------
\appendix*
\section{}
In this appendix, we use the Einstein gravitational field
equations  and  the G\"{o}del-Obukhov spacetime metric of
cosmological expansion and rotation,
given by Eq.~(\ref{metric_cc}) \cite{Obukhov2000},
used to derive the equation of motion of the scale
factor [Eq.~(\ref{Exact_1})], to show that the unknown parameter
$\sigma$, when defined in terms of the
parameters used in this present
manuscript, can be considered  as a constant or
a function of cosmological time without qualitatively
changing the gravitational field equations including
the energy-momentum tensor (sometimes
referred to as the stress-energy tensor), which does
not change qualitatively nor quantitatively.
 Specifically, as we shall see,
when $\sigma$ is defined by Eq.~(\ref{CP1})
and $k$ by Eq.~(\ref{CP2}), the terms involving the time
derivatives of $\sigma(t)$ will reduce to
a constant in the local Lorentz connections that
goes to zero in the Ricci curvature
tensor $R_{\mu\nu}$ when the time derivatives are taken.   
The Ricci curvature
tensor yields the Einstein
gravitational field equations, and, thus,
the equation of motion of the scale factor.  
Note, in this appendix, to avoid confusion of the notations,
the scale factor $R=R(t)$ is defined as $a=a(t)$.

In the framework  of Poincar\'{e}  gauge theory of gravity,
the gravitational field is
described by the tetrad $h_{\mu}^a$ and the local Lorentz connection
$ \tilde{\Gamma}_{b\mu}^a$ \cite{Obukhov2000}.  
The gravitational
Lagrangian is constructed as an invariant contraction from the 
curvature tensor
\begin{eqnarray}
R^a_{b\mu\nu}=&\partial_\mu\tilde{\Gamma}_{b\nu}^a
-\partial_\nu\tilde{\Gamma}_{b\mu}^a 
                      + \tilde{\Gamma}_{c\mu}^a\tilde{\Gamma}_{b\nu}^c
                       -\tilde{\Gamma}_{c\nu}^a\tilde{\Gamma}_{b\mu}^c, 
\label{Riemann} 
\end{eqnarray}
and the torsion tensor
\begin{eqnarray}
T^a_{\mu\nu}=&\partial_\mu h_{\nu}^a
-\partial_\nu h_{\mu}^a
                             + \tilde{\Gamma}_{b\mu}^a h_{\nu}^b
                              -\tilde{\Gamma}_{b\nu}^a h_{\mu}^b.
\label{torsion} 
\end{eqnarray}
%
Here, a Latin alphabet is used for 
the local Lorentz frame.
The independent variation of the corresponding Lagrangian (or action) 
\cite{Obukhov2000} with respect to $h_{\mu}^a$ and 
$ \tilde{\Gamma}_{b\mu}^a$ yields the gravitational field 
equations [Eqs.~(5.4) and~(5.5) of Ref.~\cite{Obukhov2000}] 
with sources 
[Eqs.~(5.19) and~(5.20) of Ref.~\cite{Obukhov2000}].  
In general,
and equivalently, introducing the asymmetric energy-momentum
tensor and the spin density, one can write the 
the gravitational field equations in the following form given by 
Sciama and Kibble in 1961 (see \cite{Trautman2006}, 
and references therein): 
\begin{equation}
R_{\mu\nu}-\frac{1}{2}g_{\mu\nu}R=8\pi G\Sigma_{\mu\nu},
\label{Einstein}
\end{equation}
which give the Einstein field equations, and, thus, the equation
of motion
of the scale factor [Eq.~(\ref{Exact_1})], and
\begin{equation}
T_{\mu\nu}^\alpha+\delta_\mu^\alpha T_{\nu\beta}^\beta
-\delta_\nu^\alpha T_{\mu\beta}^\beta=8\pi G \tau_{\mu\nu}^\alpha,
\label{Cartan}
\end{equation}
with
\begin{eqnarray}
T^\alpha_{\mu\nu}=8\pi G (\tau^\alpha_{\mu\nu}
+\frac{1}{2}\delta_\mu^\alpha\tau_{\nu\beta}^\beta
+\frac{1}{2}\delta_\nu^\alpha\tau_{\beta\mu}^\beta)
\label{torsion_spin} 
\end{eqnarray}
which gives the Cartan field equations,
where, from Eq.~(\ref{Riemann}),
\begin{eqnarray}
R_{\mu\nu}&=&-R_{\mu a \nu}^a
\label{Ricci_tensor}
\end{eqnarray}
and \cite{Obukhov2000}
\begin{eqnarray}
R=h_a^\mu h^{\nu b}R_{b \mu\nu}^a
\label{Ricci_scalar}
\end{eqnarray}
are the Riemann-Cartan-Ricci tensor and scalar, respectively, which
give the curvature of spacetime; $\Sigma_{\mu\nu}$ and
$\tau_{\mu\nu}^\alpha$   are the canonical tensors
of energy-momentum and spin, respectively: in this case, of the
cosmic matter, modeled as the Weyssenhoff spin fluid in Riemann-Cartan
spacetime \cite{Obukhov1987} for the exact solution \cite{Obukhov2000}.
Note, $\Gamma_{\alpha\beta}^\lambda=-\Gamma_{\beta\alpha}^\lambda$
and $R_{\mu\nu}=-R_{\nu\mu}$ are convenient antisymmetries of the
Christoffel symbols and the Ricci tensor in the Riemann-Cartan
spacetime \cite{Gronwald_Hehl1996}; $\delta_\eta^\alpha$ in general 
is the Kronecker delta ($=1$ for $\alpha=\eta$ and
$=0$ for $\alpha\neq\eta$).

The exact solution for the G\"{o}del-Obukhov spacetime metric 
gives two independent Einstein field equations: (1) the equation of 
motion of the 
scale factor [Eq.~(\ref{Exact_1})], and (2), upon integration, the 
Friedmann-like equation (see Ref.~\cite{Obukhov2000} for further 
details).
Now, referring back to the first paragraph of this appendix, we show 
specifically 
that the Einstein field equations depend very little, if any,
 on whether 
$\sigma$ is a constant or a function of time.  
Since we are considering here
that $\sigma$ is a function of time as opposed to a constant like  
considered by \cite{Obukhov2000}, the addition terms that could 
possibly change the Einstein field equations including 
the evolution of the scale factor
[Eq.~(\ref{Exact_1})] and the energy-momentum tensor, would be 
those involving the
time derivatives of the spacetime metric components $g_{\mu\nu}$ of 
Eq.~(\ref{metric_cc}).  
First of all, we notice that the 
energy-momentum tensor
$\Sigma_{\mu\nu}$ on the right-hand side of  
Eq.~(\ref{Einstein}), given by Eq.~(5.19) of  \cite{Obukhov2000}, does
not contain any time derivatives of the spacetime metric components
of  Eq.~(\ref{metric_cc}); therefore, considering $\sigma$ to be a 
function of
time, instead of a constant, would not add additional terms to 
the energy-momentum tensor that would change the field equations
from that of \cite{Obukhov2000}.  
Next, we point out that in the exact solution of the 
Einstein field equations
[Eq.~(\ref{Einstein})],  \cite{Obukhov2000}
assumes that the Ricci (or Riemann-Cartan curvature) scalar  is
constant:
\begin{eqnarray}
R&=&{\rm constant} \nonumber\\
    &=&-\frac{1}{2b},
\label{scalar}
\end{eqnarray}
i.e., a constant over the hyperspace ($t=\rm{constant}$), like, 
for example, the Hubble constant $H$, the scale 
factor $a$, the average mass density $\rho$, etc., for a 
specific epoch, where it can be shown from the equations that
describe the state of the matter that
\begin{equation}
b^{-1}=32\pi G \biggl(\rho-p-\frac{B^2}{a^4}\biggr),
\label{scalar_constant}
\end{equation}
with $c=1$.

So, now, we need to focus only on the Riemann-Christoffel curvature
tensor [Eq.~(\ref{Riemann})] and its contracted (Ricci) curvature tensor
[Eq.~(\ref{Ricci_tensor})] of Eq.~(\ref{Einstein}).  
The uniqueness of the curvature tensor states that
it is the only tensor that can be constructed from the metric tensor
and its first and second derivatives, and is linear in the second
derivatives \cite{Weinberg1972}.  The fact that the curvature tensor
is constructed from the metric tensor and its first and second
derivatives can be seen by comparing the general expressions
of Eqs.~(\ref{Riemann}) and~(\ref{Ricci_tensor}) with the expression for
the Christoffel symbol given by Eq.~(\ref{GM13}).

Now, considering the the Riemann-Christoffel curvature
tensor [Eq.~(\ref{Riemann})], which leads to the Einstein field 
equations [Eq.~(\ref{Einstein})], we must 
evaluate the  local Lorentz connection $ \tilde{\Gamma}_{b\mu}^a$.
We will follow the same method as \cite{Obukhov2000} except
now we will consider $\sigma$ of the spacetime 
metric [Eq.~(\ref{metric_cc})]
to be a function of time instead of a constant as assumed by  
\cite{Obukhov2000}.

Direct calculation of the local Lorentz connection \cite{ Obukhov2000}:
\begin{equation}
\tilde{\Gamma}_{b \mu}^a=h_\alpha^a h_b^\beta
\tilde{\Gamma}_{\beta\mu}^\alpha+h_\alpha^a \partial_\mu h_b^\alpha
\label{lorentz_connection}
\end{equation}
(with $\tilde{\Gamma}_{\beta\mu}^\alpha$ as the Christoffel symbols),
using the local orthonormal (Lorentz) tetrad $h_\mu^a$:
\begin{eqnarray}
h_0^{\hat{0}}&=&1, \nonumber\\
h_2^{\hat{0}}&=&-a\sqrt{\sigma}\,{\rm e}^{mx}=h_0^{\hat{2}}, \nonumber\\
h_1^{\hat{1}}&=&h_3^{\hat{3}}=a, \nonumber\\
h_2^{\hat{2}}&=&a{\rm e}^{mx}\sqrt{k+\sigma};
\label{tetrad}
\end{eqnarray}
and its inverse $h_a^\mu$:
\begin{eqnarray}
h_{\hat{0}}^0&=&1, \nonumber\\
h_{\hat{2}}^0&=&\sqrt{\frac{\sigma}{k+\sigma}}
=h_{\hat{0}}^2, \nonumber\\
h_{\hat{1}}^1&=&h_{\hat{3}}^3=\frac{1}{a}, \nonumber\\
h_{\hat{2}}^2&=&\frac{1}{a{\rm e}^{mx}\sqrt{k+\sigma}};
\label{tetrad_inverse}
\end{eqnarray}
yields for the metric of Eq.~(\ref{metric_cc}) the following
nonzero local Lorentz
connections:
\begin{subequations}
\label{allequations_1}
\begin{eqnarray}
\tilde{\Gamma}_{\hat{2}\hat{0}}^{\hat{0}}
&=&\tilde{\Gamma}_{\hat{2}\hat{1}}^{\hat{1}}
=\tilde{\Gamma}_{\hat{2}\hat{3}}^{\hat{3}}
={\frac{\dot{a}}{a}}\sqrt{\frac{\sigma}{k+\sigma}},
\label{local_connection_a}
\\
\tilde{\Gamma}_{\hat{2}\hat{2}}^{\hat{1}}
&=&-\frac{m}{a},\label{local_connection_b}
\\
\tilde{\Gamma}_{\hat{0}\hat{1}}^{\hat{2}}
&=&\tilde{\Gamma}_{\hat{1}\hat{0}}^{\hat{2}}
=-\tilde{\Gamma}_{\hat{1}\hat{2}}^{\hat{0}}
={\frac{m}{2a}}\sqrt{\frac{\sigma}{k+\sigma}},
\label{local_connection_c}
\end{eqnarray}
\end{subequations}
when $\sigma$ is considered to be a constant \cite{Obukhov2000},
where, in the local orthonormal (Lorentz) tetrad, 
\begin{eqnarray}
\tilde{\Gamma}_{bc}^a&=&\tilde{\Gamma}_{b\mu}^a h_c^\nu
\delta_\nu^\mu.
\\ \nonumber
\label{local_connection}
\end{eqnarray}
A  caret denotes tetrad indices; and, recall, a Latin alphabet 
is used for
the local Lorentz frame, i.e.,
$a, b,\ldots =  \hat{0}, ~\hat{1},~ \hat{2},~ \hat{3}$.

Since only the Lorentz connections of Eq.~(\ref{local_connection_a})
 involve the time 
derivative of the scale factor $a(t)$, and since $a(t)$ and $\sigma$
occur only as multiplying factors in Eq.~(\ref{metric_cc}), and the 
connections of 
Eq.~(\ref{local_connection_a}) are all equal, we need only look at, say, 
$\tilde{\Gamma}_{\hat{2}\hat{0}}^{\hat{0}}$ to see the change, 
if any, in the field equations if $\sigma\equiv\sigma(t)$.
So, with free index $\mu=0$ in  Eq.~(\ref{lorentz_connection})
along with $a=\hat{0}$ and $b=\hat{2}$, and with $c=\hat{0}$
substituted into Eq.~(A.14),
we  evaluate $\tilde{\Gamma}_{\hat{2}\hat{0}}^{\hat{0}}$.
Therefore,
\begin{widetext}
\begin{eqnarray}
\tilde{\Gamma}_{\hat{2}\hat{0}}^{\hat{0}}
&=&\tilde{\Gamma}_{\hat{2}{0}}^{\hat{0}} h_{\hat{0}}^0 \nonumber\\
&=&h_\alpha^{\hat{0}} h_{\hat{2}}^\beta
\tilde{\Gamma}_{\beta 0}^\alpha+h_\alpha^{\hat{0}}
\partial_{{0}} h_{\hat{2}}^\alpha \nonumber\\
&=&h_\alpha^{\hat{0}}h_{\hat{2}}^0 \tilde{\Gamma}_{00}^\alpha
+ h_\alpha^{\hat{0}}h_{\hat{2}}^1 \tilde{\Gamma}_{10}^\alpha
+ h_\alpha^{\hat{0}}h_{\hat{2}}^2 \tilde{\Gamma}_{20}^\alpha	
+ h_\alpha^{\hat{0}}h_{\hat{2}}^3  \tilde{\Gamma}_{30}^\alpha
+h_\alpha^{\hat{0}}\partial_{{0}} h_{\hat{2}}^\alpha \nonumber\\
&=&h_0^{\hat{0}}h_{\hat{2}}^0 \tilde{\Gamma}_{00}^0
+ h_0^{\hat{0}}h_{\hat{2}}^1 \tilde{\Gamma}_{10}^0
+ h_0^{\hat{0}}h_{\hat{2}}^2 \tilde{\Gamma}_{20}^0
+ h_0^{\hat{0}}h_{\hat{2}}^3  \tilde{\Gamma}_{30}^0
+h_0^{\hat{0}}\partial_{{0}} h_{\hat{2}}^0 \nonumber\\
&&+h_1^{\hat{0}}h_{\hat{2}}^0 \tilde{\Gamma}_{00}^1
+ h_1^{\hat{0}}h_{\hat{2}}^1 \tilde{\Gamma}_{10}^1
+ h_1^{\hat{0}}h_{\hat{2}}^2 \tilde{\Gamma}_{20}^1
+ h_1^{\hat{0}}h_{\hat{2}}^3  \tilde{\Gamma}_{30}^1
+h_1^{\hat{0}}\partial_{{0}} h_{\hat{2}}^1 \nonumber\\
&&+h_2^{\hat{0}}h_{\hat{2}}^0 \tilde{\Gamma}_{00}^2
+ h_2^{\hat{0}}h_{\hat{2}}^1 \tilde{\Gamma}_{10}^2
+ h_2^{\hat{0}}h_{\hat{2}}^2 \tilde{\Gamma}_{20}^2
+ h_2^{\hat{0}}h_{\hat{2}}^3  \tilde{\Gamma}_{30}^2
+h_2^{\hat{0}}\partial_{{0}} h_{\hat{2}}^2 \nonumber\\
&&+h_3^{\hat{0}}h_{\hat{2}}^0 \tilde{\Gamma}_{00}^3
+ h_3^{\hat{0}}h_{\hat{2}}^1 \tilde{\Gamma}_{10}^3
+ h_3^{\hat{0}}h_{\hat{2}}^2 \tilde{\Gamma}_{20}^3
+ h_3^{\hat{0}}h_{\hat{2}}^3  \tilde{\Gamma}_{30}^3
+h_3^{\hat{0}}\partial_{{0}} h_{\hat{2}}^3,
\label{lorentz_connection_2}
\end{eqnarray}
\end{widetext}
where we have summed over $\beta=0,~1,~2,~3$,
then summed over $\alpha=0,~1,~2,~3$, and 
used  $ h_{\hat{0}}^0=1 $ of Eq.~(\ref{tetrad_inverse}).
Upon using the local orthonormal tetrad of
Eqs.~(\ref{tetrad}) and~(\ref{tetrad_inverse}),
Eq.~(\ref{lorentz_connection_2})
reduces to
\begin{eqnarray}
\tilde{\Gamma}_{\hat{2}\hat{0}}^{\hat{0}}
&=&h_0^{\hat{0}}h_{\hat{2}}^0 \tilde{\Gamma}_{00}^0
+ h_0^{\hat{0}}h_{\hat{2}}^2 \tilde{\Gamma}_{20}^0
+h_2^{\hat{0}}h_{\hat{2}}^0 \tilde{\Gamma}_{00}^2 \nonumber\\
&&+ h_2^{\hat{0}}h_{\hat{2}}^2 \tilde{\Gamma}_{20}^2
+h_2^{\hat{0}}\partial_{{0}} h_{\hat{2}}^2,
\label{lorentz_connection_3}
\end{eqnarray}
with all the other terms being zero.

Next, we evaluate the Christoffel symbols in 
Eq.~(\ref{lorentz_connection_3}), using the spacetime metric 
of Eq.~(\ref{metric_cc}), the 
corresponding (matrix) inverse metric components $g^{\mu\nu}$:
\begin{eqnarray}
g^{00}&=&{\frac{k}{k+\sigma}}, \nonumber\\
g^{02}&=& -{\frac{\sqrt{\sigma}}{a{\rm e}^{mx}
                    [k+\sigma]}}= g^{20},  \nonumber\\
g^{11}&=&-{\frac{1}{a^2}}, \nonumber\\
g^{22}&=&-{\frac{1}{a^2 {\rm e}^{2mx}
                       [k+\sigma]}}, \nonumber\\
g^{33}&=&-{\frac{1}{a^2}},  
\label{metric_cc_inverse}
\end{eqnarray}
and Eq.~(\ref{GM13}), where $a$, $\sigma$, and $k$ 
[Eq.~(\ref{CP2})] are 
now all considered to be functions of cosmological time $t$.  
Thus, we find that
\begin{subequations}
\label{allequations_2}
\begin{eqnarray}
\tilde{\Gamma}_{00}^0&=&{\frac{\sigma}{a(k+\sigma)}}
\biggl(\frac{\dot{\sigma}}{2\sigma}a+\dot{a}\biggr),
\label{christoffel_a}
\\
\tilde{\Gamma}_{20}^0&=&{\frac{\sqrt{\sigma}\,
{\rm e}^{mx}}{2(k+\sigma)}}
(2\dot{a}k+a\dot{k}),
\label{christoffel_b}
\\
\tilde{\Gamma}_{00}^2
  &=&{\frac{\sqrt{\sigma}}{a^2{\rm e}^{mx} (k+\sigma)}}
\biggl(\frac{\dot{\sigma}}{2\sigma}a+\dot{a}\biggr),
\label{christoffel_c}
\\
\tilde{\Gamma}_{20}^2&=&{\frac{1}{2a(k+\sigma)}}
(2\dot{a}k+a\dot{k}),
\label{christoffel_d}
\end{eqnarray}
\end{subequations}
where the comoving coordinate distance $x$ is by definition fixed.  

Evaluation of the terms on the right-hand side of 
Eq.~(\ref{lorentz_connection_3})
separately and consecutively using Eqs.~(\ref{tetrad}), 
~(\ref{tetrad_inverse}),
and~(\ref{allequations_2}) yields
\begin{subequations}
\label{allequations_3}
\begin{eqnarray}
h_0^{\hat{0}}h_{\hat{2}}^0 \tilde{\Gamma}_{00}^0&=&
\frac{1}{a}\biggl(\frac{\sigma}{k+\sigma}\biggr)^{3/2}
\biggl(\frac{\dot{\sigma}}{2\sigma}a+\dot{a}\biggr),
\label{lorentz_connection_a}
\\
h_0^{\hat{0}}h_{\hat{2}}^2 \tilde{\Gamma}_{20}^0&=&
\frac{1}{2a}\sqrt{\frac{\sigma}{k+\sigma}}
\biggl(\frac{1}{k+\sigma}\biggr)
(2\dot{a}k+a\dot{k}), \nonumber \\
&&
\label{lorentz_connection _b}
\\
h_2^{\hat{0}}h_{\hat{2}}^0 \tilde{\Gamma}_{00}^2&=&
-\frac{1}{a}\biggl(\frac{\sigma}{k+\sigma}\biggr)^{3/2}
\biggl(\frac{\dot{\sigma}}{2\sigma}a+\dot{a}\biggr),
\label{lorentz_connection _c}
\\
h_2^{\hat{0}}h_{\hat{2}}^2 \tilde{\Gamma}_{20}^2&=&
-\frac{1}{2a}\sqrt{\frac{\sigma}{k+\sigma}}
\biggl(\frac{1}{k+\sigma}\biggr)
(2\dot{a}k+a\dot{k}), \nonumber \\
&&
\label{lorentz_connection _d}
\\
h_2^{\hat{0}}\partial_{{0}} h_{\hat{2}}^2&=&
\sqrt{\frac{\sigma}{k+\sigma}}
\biggl(\frac{\dot{\sigma}}{2\sigma}+\frac{\dot{a}}{a}\biggr).
\\ \nonumber
\label{ lorentz_connection _e}
\end{eqnarray}
\end{subequations}
Upon substitution of the terms of Eq.~(\ref{allequations_3}) into 
Eq.~(\ref{lorentz_connection_3}) yields for the
 local Lorentz connection 
$\tilde{\Gamma}_{\hat{2}\hat{0}}^{\hat{0}}$,
\begin{eqnarray}
\tilde{\Gamma}_{\hat{2}\hat{0}}^{\hat{0}}
&=&
\biggl(\frac{\dot{a}}{a}+\frac{\dot{\sigma}}{2\sigma}\biggr)
\sqrt{\frac{\sigma}{k+\sigma}}.
\label{lorentz_connection_4}
\end{eqnarray}
Compare the above metric connection of 
Eq.~(\ref{lorentz_connection_4}) for 
$\sigma=\sigma(t)$ to that of Eq.~(\ref{local_connection_a}) 
 for $\sigma={\rm constant}$ (as given by Ref.~\cite{Obukhov2000}).
The only difference is the added term involving the time derivative 
of $\sigma$.

Finally, upon substitution of the model parameters used in this
present manuscript for $\sigma$ and $k$ [Eqs.~(\ref{CP1})
and~(\ref{CP2})] and their derivatives:
\begin{subequations}
\label{parameters}
\begin{eqnarray}
\sigma(t)&\equiv&{\rm e}^{c_1 t/t_0},
~~~ \dot{\sigma}(t)=\frac{c_1}{t_0}\sigma(t),
\label{parameter_sigma}
\\
\nonumber
\\
k&=&c_2\sigma,
~~~\dot{k}=c_2\dot{\sigma}=\frac{c_2 c_1}{t_0}\sigma,
\label{parameter_k}
\end{eqnarray}
\end{subequations}
with
\begin{equation}
{\frac{\sigma}{k+\sigma}}=\frac{1}{c_2+1},
\label{parameters_2}
\end{equation}
Eq.~(\ref{lorentz_connection_4}) reduces to a term
independent of $\sigma$, $k$,
and their derivatives: 
\begin{eqnarray}
\tilde{\Gamma}_{\hat{2}\hat{0}}^{\hat{0}}
&=&
\biggl(\frac{\dot{a}}{a}+\frac{c_1}{2t_0}\biggr)
\sqrt{\frac{1}{c_2+1}},
\label{lorentz_connection_5}
\end{eqnarray}
yet dependent mainly on $a$ and $\dot{a}$
as in the case when $\sigma$ is considered to be a constant
(compare Eq.~(\ref{local_connection_a}), but with a trivially 
 small added constant
$c_1/2t_0\approx -1.3\times 10^{-16}~{\rm s^{-1}}$, with $c_1=-115$ 
and $t_0=13.8\times 10^9~{\rm yr}$, whose absolute value is  
$<< 1~ {\rm s^{-1}}$,
and whose value goes to zero in the Riemann-Christoffel curvature
tensor [Eq.~(\ref{Riemann})] and its contracted (Ricci) 
curvature tensor
[Eq.~(\ref{Ricci_tensor})] of Eq.~(\ref{Einstein})
when the time derivatives are taken, validating the 
assumption of triviality of the additional terms in the 
gravitational field
equations, as stated in  Sec.~\ref{sec:3.5}, in these of 
order calculations 
[compare, e.g.,Eqs.~(\ref{NMR1}), (\ref{D14}), (\ref{press3}), 
and~(\ref{press4})]. 
That is, there will be no time derivatives of the parameter
$\sigma(t)$ in the gravitational field equations.

Moreover, since we are specifically using the torsion 
acceleration term
in Eq.~(\ref{Exact_1}) (third term on the right-hand side) to
compare with the gravitomagnetic acceleration
[Eq.~(\ref{NMR1})] brings
out the negligibility  or triviality in these present
calculations of the constant term $c_1/2t_0$.
In order to see the exact role of this constant term,
if any, one would have to re-evaluate the Riemann-Cartan curvature
tensor and thus Einstein-Cartan gravitational field equations
in their entirety, which is beyond the scope of
this manuscript. Nor does it seem neccessary since we
can estimate its role:
If this constant term does not cancel and appears as
a square  $ (c_1/2t_0)^2$, which would be at its maximum 
second order value, expressing a constant acceleration per unit length
in the  field equations, like the Hubble parameter
[compare Eqs.~(\ref{Exact_1}), (\ref{hubble}), and
(\ref{lorentz_connection_5})],
though not changing over time like the Hubble parameter,
it could possibly contribute to
the cosmic expansion at some point in time. However, its value
must be compared with the other accelerations in
Eq.~(\ref{Exact_1}), which change over time, to see when in time,
if ever, this constant would be important.
This possibility is investigated
elsewhere \cite{Williams2014}.
However, a preliminary investigation
further validates the assumption of triviality of such constant term 
$(c_1/2t_0)^2$, again, if it exist in the gravitational
field equations and the equation of motion of the scale 
factor [Eq.~(\ref{Exact_1})].
This  preliminary investigation 
shows that the evolution of the terms in Eq.~(\ref{Exact_1}) 
for the range of $z$ values in  Fig.~2 reveals that $ (c_1/2t_0)^2$
is much smaller than the other terms in the early
universe and would not appear to become relevant until 
near the present day where its absolute value is still
smaller than the second term on the right-hand side of
Eq.~(\ref{Exact_1}), and this relevance continues to be
diminished by the third term, i.e., the torsion or GM acceleration
as $z$ gets smaller, as measured by a present day observer.

%

\end{document}